\documentclass[%
 aip,
 amsmath,amssymb,
 reprint,%
]{revtex4-1}

\usepackage{graphicx}% Include figure files
\usepackage{dcolumn}% Align table columns on decimal point
\usepackage{bm}% bold math
\usepackage{xcolor}
\usepackage{subcaption} % For subfigure environment
\usepackage{multirow} % For multirow cells in tables
\usepackage[utf8]{inputenc}
\usepackage[T1]{fontenc}
\usepackage{mathptmx}
\usepackage{etoolbox}

\makeatletter
\def\@email#1#2{%
 \endgroup
 \patchcmd{\titleblock@produce}
  {\frontmatter@RRAPformat}
  {\frontmatter@RRAPformat{\produce@RRAP{*#1\href{mailto:#2}{#2}}}\frontmatter@RRAPformat}
  {}{}
}%
\makeatother
\begin{document}

\preprint{AIP/123-QED}

\title[]{Role of gravity on preferential clustering of microparticles in unsteady wake flows}
% Force line breaks with \\
\author{Siddhi Arya}
\author{Partha S. Goswami}%
\email{psg@iitb.ac.in}
\affiliation{ 
Department of Chemical Engineering, Indian Institute of Technology Bombay,
Mumbai - 400076 (India)%\\This line break forced with \textbackslash\textbackslash
}%

\date{\today}% It is always \today, today,

\begin{abstract}
 Direct numerical simulations are carried out for particle-laden flow over the cylinder to investigate preferential clustering of particles in an unbounded vertical channel flow. The flow is examined at Reynolds numbers Re $ =100$ and $200$ for varying particle Stokes number, particle loadings and Froude numbers to quantify the combined influence of particle inertia and gravitational settling on particle motion. The unladen flow exhibits the classical vortex shedding pattern observed in flow over bluff bodies at both Reynolds numbers. Reynolds number dependent wake width, wake recovery, and velocity-deficit evolution are observed, characterizing the coherent flow structures that govern particle dynamics. In particle-laden unsteady wake flows, the non-uniform particle distribution leads to formation of coherent voids and clusters, whose shape are directly correlated with background flow dynamics. Gravity modifies particle-fluid interaction, which leads to an increase in slip velocity, weakens vortex-induced particle clustering and promotes them to travel through vortices, resulting in a transition of the void shape from individual leaf-like structure to snake-like void zone and eventually into a nearly vertical void structure. In upstream region infront of the cylinder, inertial particles form a bow-shock-like structure whose extent increases with increase in Stokes number and finite Froude conditions. Voronoi based analysis combined with local Q values is used to investigate effect of gravity and inertia on particle distribution. The dimensionless settling velocity, $St/Fr^2$, is identified as the governing parameter controlling the evolution of void shape, normalized void cell area and the probability distribution of Voronoi cell areas. The effect of wake dynamics, particle inertia, and gravity is reported to jointly govern preferential clustering in bluff-body wakes.

\end{abstract}

\maketitle
\section{Introduction}

The flow past a stationary obstacle is classical area of study, where fluid interacts with the bluff body and undergoes flow separation over a critical Reynolds number. The unladen flow past a stationary circular cylinder is a benchmark to study boundary-layer separation, wake formation and alternate vortex pattern observed in such flows. \citet{roshko1954development} studied wake transition from two-dimensional state to flow with distinct three-dimensional instabilities. He identified a stable vortex-shedding regime ($40 \lesssim Re \lesssim 150$) with regular vortex streets, followed by a transition regime ($150 \lesssim Re \lesssim 300$) with the presence of turbulent fluctuations in the wake. \citet{Gerrard_1966} proposed that two simultaneous characteristic lengths are required to describe the wake, namely the formation-region length and the diffusion length, which governs the vortex shedding and its shedding frequency. \citet{williamson1988existence}, \citep{wiliamson1996vortex} extensively mapped these transitions, identifying Mode A shedding (at $Re \approx 180$), characterized by large-scale spanwise vortex loops, followed by Mode B shedding (for $Re > 230$), which exhibits finer-scale streamwise structures. Similar observation were reported by \citet{Barkley_Henderson_1996} and \citet{HENDERSON_1997} , who confirmed these two distinct transitions through Floquet stability analysis and direct numerical simulations. 
%\citet{roshko1955wake}.

In case of particle-laden flow over such bluff bodies, the particle motion is governed by the inertia of the particles relative to the fluid. A defining characteristic of inertial particle dynamics in multiphase flows is preferential concentration, which is the tendency of particles to accumulate to form regions of high local concentration, rather than distributing uniformly (\citet{brandt2022particle, Monchaux}). \citet{squires1991, eaton1994} showed that particles are centrifugally expelled from regions of high vorticity and preferentially accumulate in regions of high strain. The extent of this clustering is commonly characterized by the Stokes number ($St=\tau_p/\tau_f$), defined as the ratio of the particle response time to a characteristic fluid timescale. When $St=O(1)$, the particle response time becomes comparable to the characteristic timescale of the flow's coherent structures, resulting in the strongest preferential concentration. In contrast, particles with $St\ll1$ behave as passive tracers, whereas those with $St\gg1$ possess sufficient inertia to decouple from local fluid velocity fluctuations. \citet{nilsen2014mechanisms} demonstrated that tendency for particles to cluster in regions of low $Q$ (or vorticity) is far more pronounced in non-Gaussian turbulence, where low inertia particles are convected with the flow structures. At higher Stokes numbers, it observed that clustering is driven by the formation of caustics, where fast particles overtake slower ones, leading to particle accumulations in regions of low fluid kinetic energy. 
In case of flows with well-defined vortical patterns, such as the periodic von Karman street generated by a bluff body, the interaction between particles and local coherent structures yields highly nonlinear dynamics. Idealized vortex models provide a foundational understanding of the mechanisms that drive the preferential concentration. \citet{ravichandran2014attracting} demonstrated that heavy inertial particles in a frame co-rotating with a vortex pair are not only centrifuged outward, but can be captured by stable attracting fixed points, provided their Stokes number remains below a critical threshold ($St < St_{cr}$). \citet{nath2024clustering} investigated the dynamics of heavy inertial particles in flow with isolated, non-axisymmetric vortices, utilizing the elliptical Kirchhoff vortex and its strained variant(the Kida vortex) as a canonical model. They demonstrated that flow non-axisymmetry generates stable, co-rotating attractors that trap particles and enhances clustering, while external straining, modeled by the Kida vortex, leads to chaotic particle transport. They discussed that particle inertia competes with this strain, effectively suppressing chaotic advection beyond a critical Stokes number.  Macroscopic flow conditions also play a critical role in governing particle organization in bluff-body wakes. In wall-bounded turbulence, \citet{dhankarghare2025direct} showed using DNS that inertial particles preferentially accumulate within near-wall low speed streaks, where particle preferential concentration significantly influences turbulence modulation through particle-fluid interactions. \citet{harishteja2026particle} demonstrated that planar shear alters symmetry of void zones and enhances particle accumulation along the wake centerline, while Voronoi tessellation and cluster analyses revealed scale-free cluster-size distributions characterized by power-law behavior. These studies highlight that particle distribution is governed by its interaction with large-scale flow structures.

In particle laden flows under the influence of gravity, a low Froude number ($Fr$), corresponding to strong gravitational settling, limits the ability of vortices to trap particles. \citet{wang_maxey_1993} carried out direct numerical simulations to examine the settling behavior of heavy particles in homogeneous isotropic turbulence. They observed an increased mean settling velocity reaching upto $50\%$ of their terminal velocity and attributed it to inertial bias, which causes particles to preferentially sample low vorticity regions and get swept with the downward-moving fluid.  \citet{Good_Ireland_Bewley_Bodenschatz_Collins_Warhaft_2014} investigated using experiments and direct numerical simulations and showed that gravity and particle inertia reduce particle fluctuations.  \citet{DÁVILA_HUNT_2001,ayala2008effects,ROSA2016217} observed that turbulence-induced settling enhancement of inertial particles is governed primarily by the particle Froude number, with the maximum enhancement occurring for Froude number of order unity. This reflects an optimal balance between particle inertia, gravitational settling, and particle-eddy interaction timescales. While coherent wake vortices influence particle transport at downstream locations, interactions between inertial particles and the cylinder can lead to pronounced upstream spatial inhomogeneity, modulating local particle clustering. \citet{shi2020bow} carried out one-way coupled simulations to study inertial particle distributions in a flow around a wetted cylinder at $Re=100$ and observed that Stokes number governs particle distribution, forming a bow-shock-like structure near the cylinder. It was found that high Stokes particles form wider shocks, extending farther across the cross-stream direction at downstream locations with reduced influence of vortex on clustering.

Traditional methods of quantifying particle concentration, such as box counting method \cite{kulick_fessler_eaton_1994, Aliseda}, often fail to capture the multi-scale nature of particle clusters and voids. In such methods, the results are highly sensitive to the chosen grid resolution. To overcome this, contemporary studies have widely adopted Voronoi tessellation. Introduced to studies on multiphase flows by researchers like \citet{Monchaux}, Voronoi analysis divides the spatial domain into polyhedral cells, with each cell containing exactly one particle. By comparing the Probability Density Function (PDF) of the normalized Voronoi cell areas to a standard Gamma distribution (which represents a spatially random, Poisson-distributed particle field), researchers can determine cell area thresholds for clusters and voids. \citet{Tagawa_Mercado_Prakash_Calzavarini_Sun_Lohse_2012} and  \citet{fong_amili_coletti_2019}  used this method in turbulent flows, demonstrating its robustness in identifying coherent particle clusters independent of grids. \citet{shi2021clusters} performed one-way coupled simulations of inertial point particles in the wake of a circular cylinder and showed that preferential clustering in strain-dominated regions gives rise to organized cluster-void patterns surrounding the vortices shed from the cylinder. They further reported a non-monotonic variation of particle velocity along particle trajectories, linking particle dispersion to the underlying coherent vortices. \citet{shi2022scale} investigated preferential clustering of inertial particles in a three-dimensional flow over a cylinder at $Re=200$ using direct numerical simulations. By combining Voronoi tessellation with the local fluid Qvalues, they directly correlated fluid phase with the spatial non-homogeneity of the particle phase. It was observed that vortices shed from the cylinder and streamwise braids traps particles in clumps and influences the cluster formation. The voronoi-based analysis reveals scale-dependent clustering behavior for unsteady wake flows and shows the strongest clustering occurs in the case of Stokes number of order of unity. 

Despite these studies, a comprehensive understanding of how the competing mechanisms of vortex-induced clustering, gravity-driven vortex piercing, and upstream particle-cylinder elastic collisions jointly affects the downstream void-cluster dynamics in near-wake region remains incomplete. The present study helps to bridge this gap by carrying out detailed analysis on one-way coupled Direct Numerical Simulations, in order to quantify the influence of the Reynolds number, Stokes number, Froude number, and particle loading on preferential concentration of particles.

\section{Methods}

\subsection{Governing equations}
 The incompressible fluid flow is solved in an Eulerian framework using the finite-volume method implemented in the open-source software OpenFOAM. The Direct Numerical Simulations are carried out by solving the continuity and Navier-Stokes equation,

\begin{equation}
\frac{\partial u_i}{\partial x_i} = 0
\label{label_equation_continuity}
\end{equation}

\begin{equation}
 \frac{\partial u_i}{\partial t}
+ u_j \frac{\partial u_i}{\partial x_j}
=
-\frac{1}{\rho_f}\frac{\partial p}{\partial x_i}
+ \nu \frac{\partial^2 u_i}{\partial x_j \partial x_j}
+ f_i/\rho_f
\label{label_equation_momentum}
\end{equation}

Here, $u_i$, $\nu$, $\rho_f$ and $p$ are the fluid velocity, viscosity, density and pressure, respectively. The term $f_i$ is the feedback force per unit volume, which represents the momentum force exerted by the particles phase on to the fluid. The feedback force is calculated from drag force and lift force, and can be expressed as

\begin{equation}
F_i = -\sum_{I} \left( F_{d,i}^{(I)} + F_{l,i}^{(I)} \right)
\label{label_equation_feedback}
\end{equation}

where $F_{d,i}^{(I)}$ and $F_{l,i}^{(I)}$ are the drag and lift forces acting on the $I$-th particle, respectively. 

 Pressure-velocity coupling is implemented through the PISO algorithm, and second-order Gauss schemes are employed for the spatial discretizations of the pressure gradient, convective, and viscous diffusion term. Temporal discretizations are performed using a second-order Crank Nicolson scheme with a blending factor of 0.5 for time derivative.

 The dispersed phase is resolved in a Lagrangian framework using the discrete element method implemented using LIGGGHTS \citep{goniva2012open} in finite volume based opensource software CFDEM \citep{el2021theories}. The particle motion is governed by Newton’s second law,
 
\begin{equation}
m_p \frac{d v_{i,I}}{dt}
=
F_{d,i}^{(I)}
+
F_{l,i}^{(I)}
+
\sum_{J} F_{i,IJ}
+
F_{w,i}^{(I)}
+
m_p g_i.
\end{equation}

where, $m_p$,$ v_{i,I}$ is the mass of particle and the $i^{th}$ component of  velocity of $I^{th}$ particle respectively.  $F_{d,i}^{(I)} $ , $F_{l,i}$ are the drag and lift forces acting on the particle where lift force is calculated based on correlation provided by \citet{mei1992approximate} and \citet{loth2009equation},  while Schiller-Naumann correlation\cite{schiller1933drag}  are used to calculate drag force, given by

\begin{equation}
F_{d,i}^{I}
=
3\pi\mu_f d_p
\left(u_{i,I}(x,t)-v_{i,I}\right)
\left(1+0.15\,Re_p^{0.687}\right).
\end{equation}

Interparticle and particle-wall collisions are modeled using a soft-sphere spring-dashpot approach. A nonlinear Hertz-Mindlin contact law is used to compute the normal and tangential contact forces during collision between $ i $ and $j$ particles, given as

\begin{equation}
{F}_{n,ij}
=
\left(-k_n \, \delta_n^{3/2}
- \eta_{nj} \, {G}  \cdot {n} \right){n}
\end{equation}
\begin{equation}
{F}_{t,ij}
=
-k_t  {\delta_t}
- \eta_{tj} \, {G}_{ct}
\end{equation}

Here, the normal and tangential components is denoted by the terms $n$ and $t$ in subscripts.  $\delta$ , $k$ , $\eta$ are the overlap distance,  stiffness coefficients and the damping coefficients, respectively. The vector $\mathbf{G}$ is the relative  velocity between particles $i$ and $j$ and $\mathbf{n}$ is the unit vector line connecting the centers of the two particles. The tangential slip velocity at the contact point, $\mathbf{G}_{ct}$, is obtained by subtracting the normal component of the relative velocity from $\mathbf{G}$, i.e.$
\mathbf{G}_{ct}=\mathbf{G}-(\mathbf{G}\cdot\mathbf{n})\mathbf{n}$. The coefficient of restitution is set as unity and both sliding and rolling friction are neglected, so contacts are modeled as frictionless and purely elastic. The particle property is set with a Young's modulus of $10^{8}~\mathrm{Pa}$ and a Poisson's ratio of $0.3$. 

\subsection{Flow configuration}
\begin{figure}[htbp]
\centering
\includegraphics[width=0.4\textwidth]{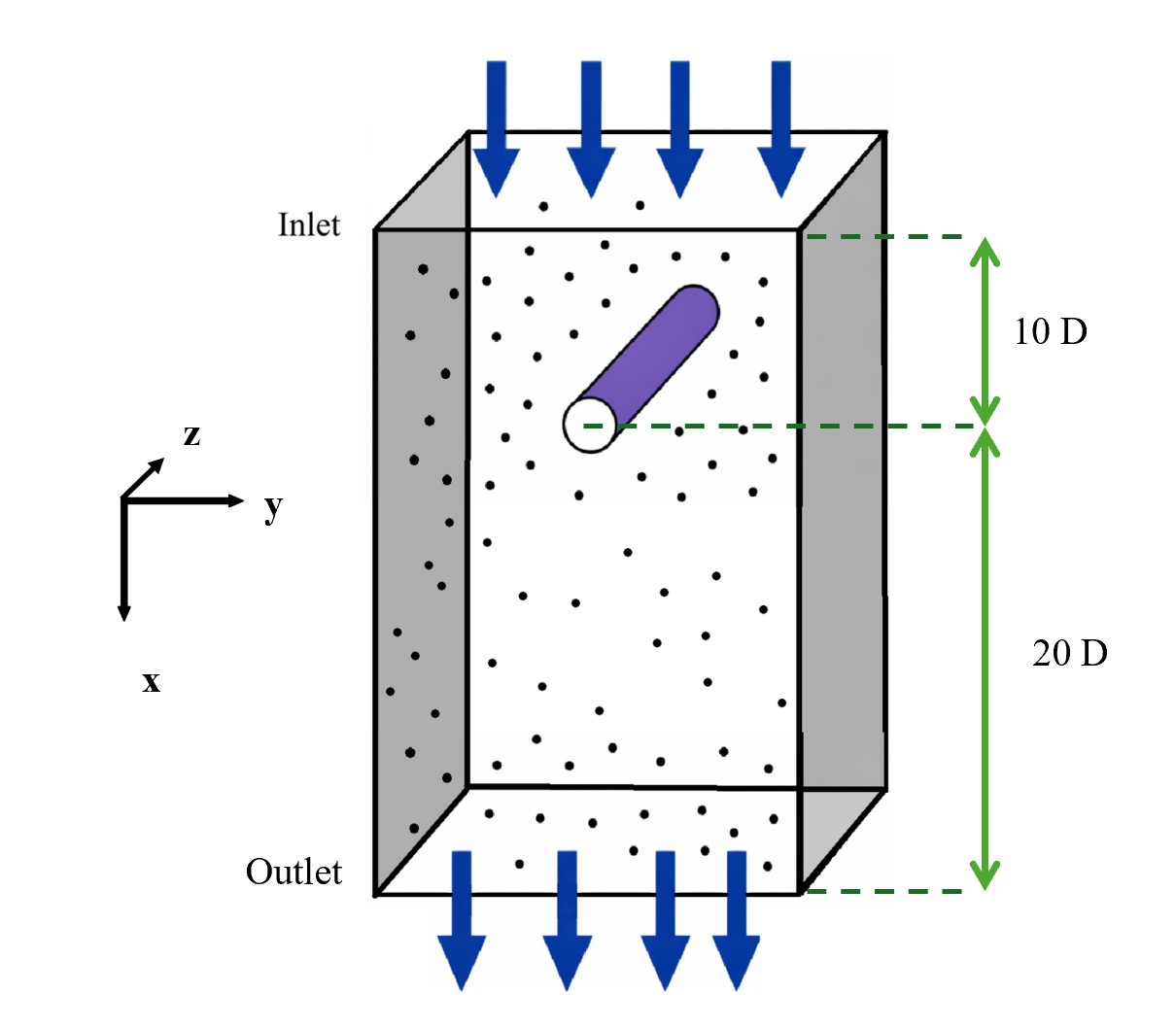}
\caption{Three-dimensional representation of an unbounded channel for flow over a circular cylinder, where $D$ is the cylinder diameter, placed at a distance of $X/D = 10$ from the inlet.}
\label{cylinder}
\end{figure}

Three-dimensional simulations of particle-laden flow over a circular cylinder are carried out at two Reynolds numbers, $Re=100$ and  $Re=200$, where $Re = U_0 D/ \nu$ and  $D$,$U_0,\nu$ are the cylinder diameter, inlet fluid velocity and kinematic viscosity, respectively. The computational domain spans $30D \times 25D \times 3.4D$ in the streamwise ($x$), cross-stream ($y$), and spanwise ($z$) directions, respectively. The cylinder is placed at a distance of $10 D$ from the inlet surface in the spanwise z direction with its length equal to the spanwise length of the domain, as shown in Figure~\ref {cylinder}. The center of the cylinder is placed at the origin $(X,Y) = (0,0)$. The computational domain is discretized using 2,193,120 non-uniform mesh cells to resolve the smaller flow length scales, particularly in the boundary layer on the cylinder surface. Finer mesh cells are placed in the vicinity of the cylinder and throughout the developing wake region, with coarser cells placed farther away from the cylinder. The analysis is carried out using one-way coupled simulations in which the fluid phase is considered to be unaffected by the presence of particles at low particle loading. A no-slip boundary condition is applied on the cylinder surface, and a slip condition is applied in the cross-stream (y) direction to enforce the unbounded channel condition, allowing the flow to develop without artificial blockage effects. The spanwise (z) boundaries are set as periodic in nature to set an infinite cylinder length.

Particles enter the unbounded vertical channel from the inlet at top, where the width of the inlet zone is kept equal to one particle diameter. The particle phase inlet velocity is kept the same as the fluid inlet velocity. The ratio of cylinder diameter to particle diameter ($d_p$)  is set at $D/d_p=60$. For the flow over a circular cylinder, the particle Stokes number is expressed as $St = \tau_p / \tau_f = \tau_p U_0 / D$, where $\tau_p$ is the particle relaxation time and $\tau_f = D / U_0$ is the characteristic flow time scale. The parameters associated with particle and fluid phase are described in Table ~\ref{foc_table}. The particle loading in the inlet region is set to particle volume fraction of  $\phi_v=2 \times 10^{-5}$(low volume fraction, LVF) and $2 \times 10^{-4}$ (high volume fraction, HVF).

\begin{table*}[t]
\caption{Flow and particle parameters used in the simulations}
\centering
\begin{tabular}{|c|c|c|c|c|}
\hline
\textbf{Cases} & \textbf{Re} & \textbf{Fr} & \textbf{St} & \textbf{Particle density (kg/m$^3$)} \\
\hline
\multirow{4}{*}{Case 1: Flow with Gravity} 
 & \multirow{2}{*}{$Re = 100$} & \multirow{2}{*}{$Fr = 1$} & $St = 1$   & 800 \\
\cline{4-5}
 &                            &                           & $St = 6.5$ & 5000 \\
\cline{2-5}
 & \multirow{2}{*}{$Re = 200$} & \multirow{2}{*}{$Fr \approx 2$} & $St = 1$   & 400 \\
\cline{4-5}
 &                            &                           & $St = 6.5$ & 2500 \\
\hline
\multirow{4}{*}{Case 2: Gravity-free Flow} 
 & \multirow{2}{*}{$Re = 100$} & \multirow{2}{*}{$Fr\rightarrow\infty$} & $St = 1$   & 800 \\
\cline{4-5}
 &                            &                               & $St = 6.5$ & 5000 \\
\cline{2-5}
 & \multirow{2}{*}{$Re = 200$} & \multirow{2}{*}{$Fr\rightarrow\infty$} & $St = 1$   & 400 \\
\cline{4-5}
 &                            &                               & $St = 6.5$ & 2500 \\
\hline
\end{tabular}
\label{foc_table}
\end{table*}

\section{Results}

\subsection{Fluid phase dynamics}

\begin{figure}[htbp]
\centering
\begin{minipage}{0.3\textwidth}
\includegraphics[width=1\textwidth]{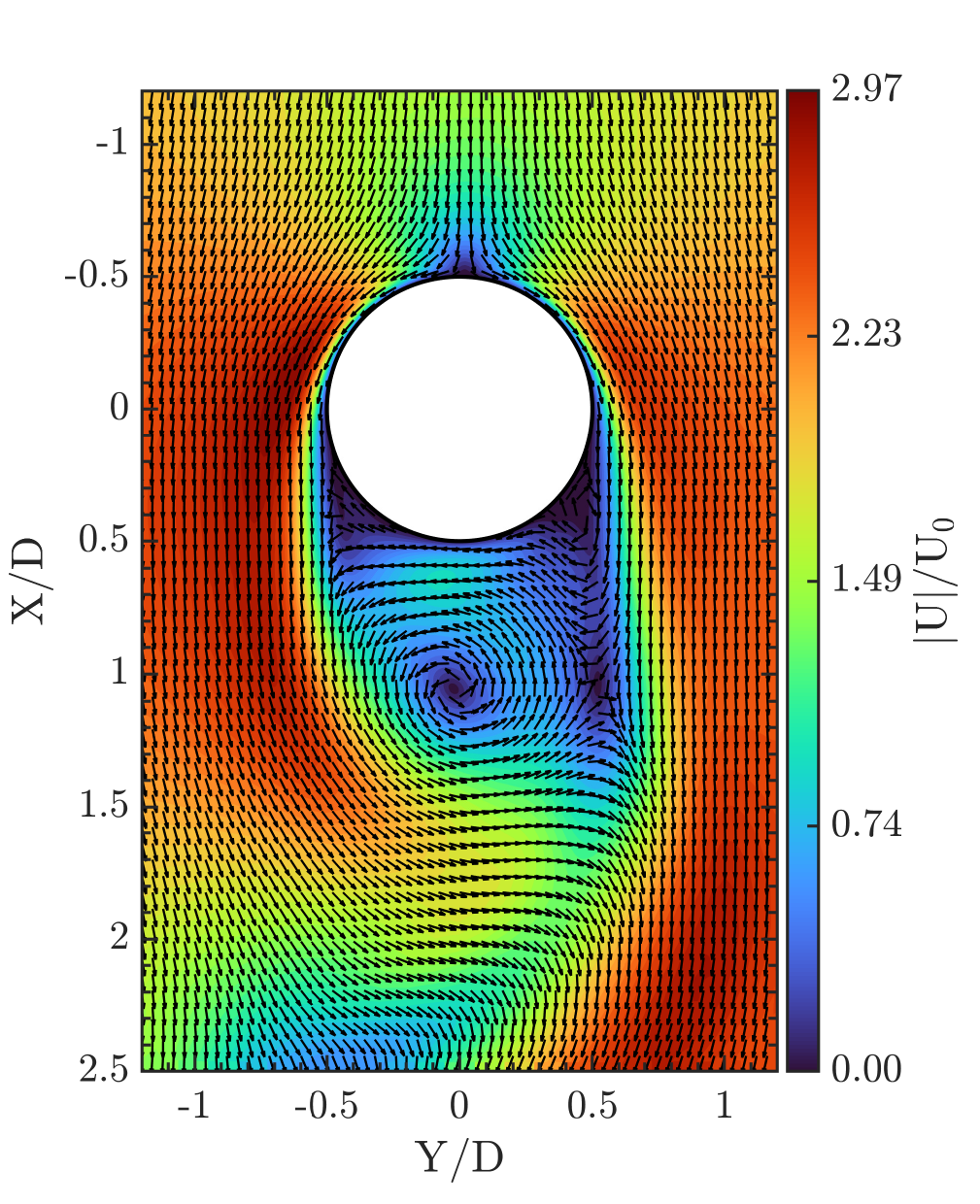}
\end{minipage}\hfill
\caption{Enlarged view of the instantaneous velocity field, normalized by the inlet velocity, $U_0$, in the near wake of a circular cylinder for the unladen flow. The contours represent the normalized velocity magnitude and the arrows indicate the local fluid velocity direction.}
\label{unladen glyph flow shedding}
\end{figure}

Figure \ref{unladen glyph flow shedding} shows the instantaneous contour of the velocity magnitude for the unladen flow over a circular cylinder, normalized by the fluid inlet velocity. The arrows in the figure show the direction of local fluid velocity. It is to be noted that the figure shown here is a part of the flow domain and provides an enlarged view of the flow near the downstream of the cylinder. Overall, the figure illustrates how the cylinder acts as an obstruction, resulting in regions of acceleration, deceleration, and local fluid recirculation. These modifications to the velocity field play a central role in governing particle dispersion and clustering in the particle-laden cases.

\begin{figure}[htbp]
\centering
\includegraphics[width=0.35\textwidth]{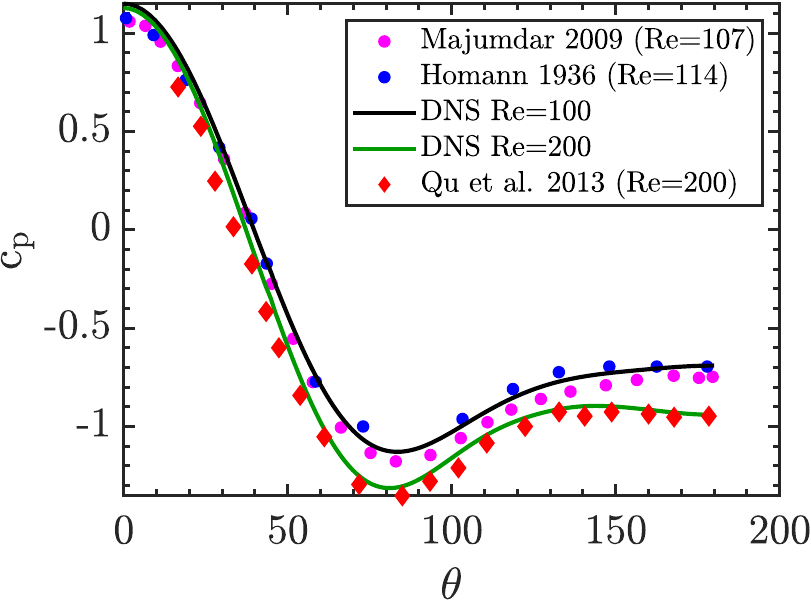}
\caption{Pressure coefficient distribution along the surface of a circular cylinder for the unladen flow, plotted as a function of the angle $\theta$. The present DNS results are compared with available experimental and numerical data from \citet{rajani2009numerical,homann1936einfluss,QU2013347}}
\label{pressure_coefficient}
\end{figure}

Figure~\ref{pressure_coefficient} presents the pressure coefficient profile around the surface of the circular cylinder for the unladen flow, plotted as a function of angle $\theta $. It is measured as the angle between the x-axis and a point on the cylinder surface in the x-y plane. Here, $\theta = 0 $ represents the upstream stagnation point. The pressure coefficient is defined as

\begin{equation}
C_p = \frac{p - p_\infty}{\tfrac{1}{2}\,\rho\,U_0^2}
\end{equation}

where p is the local pressure at the point, $p_\infty$ is the free stream pressure, $\rho$ is the fluid density, and $U_0$ is the fluid inlet velocity. It is the ratio of the difference between local and freestream static pressure to the dynamic pressure generated by the fluid flow. The pressure coefficient is maximum ($C_p \approx 1$) at the stagnation point located in front of the cylinder. The fluid decelerates as it approaches the cylinder and the local velocity reduces to zero at the stagnation point, resulting in a maximum pressure difference. The pressure at the cylinder surface decreases rapidly as the fluid gains momentum, leading to a minimum pressure coefficient at either side of the cylinder. The resultant pressure coefficient distribution shows a good agreement with classical experimental and numerical results reported in the literature for laminar vortex-shedding regimes at $Re=100$ and 200.

\begin{figure}[htbp]
\centering
\includegraphics[width=0.3\textwidth]{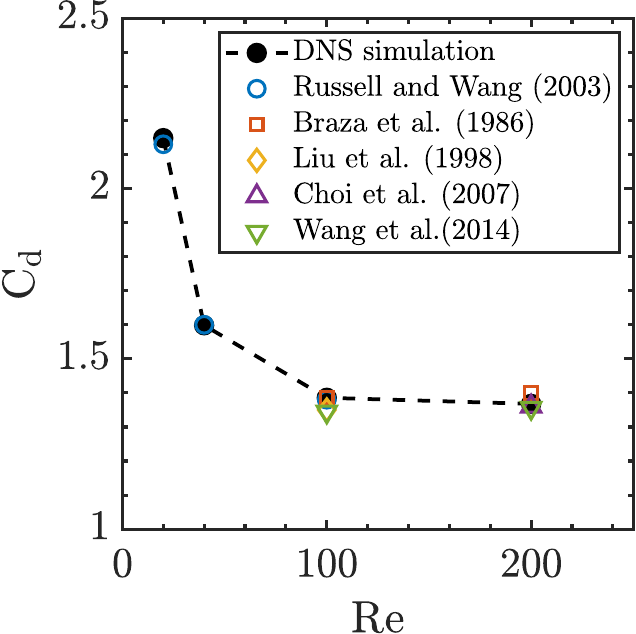}
\caption{Variation of the drag coefficient with Reynolds number for unladen flow past a circular cylinder. The present DNS results are compared with the published data of  \citet{Braza_Chassaing_Minh_1986,RUSSELL2003177,liu1998preconditioned,choi2007immersed,Wang_Shu_Wu_Yang_2014}.}
\label{drag_coefficient}
\end{figure}

Figure~\ref{drag_coefficient} shows the variation of the drag coefficient($C_d$) with Reynolds number. The drag coefficient decreases rapidly with increase in Reynolds number in the low $Re$ regime. The good agreement of DNS results with published data suggests that the present numerical setup is adequate to simulate the flow over cylinder for current Reynolds number range of Re $100$-$200$.

\begin{figure*}[!t]
\centering

\begin{minipage}{0.48\textwidth}
\includegraphics[width=0.9\textwidth]{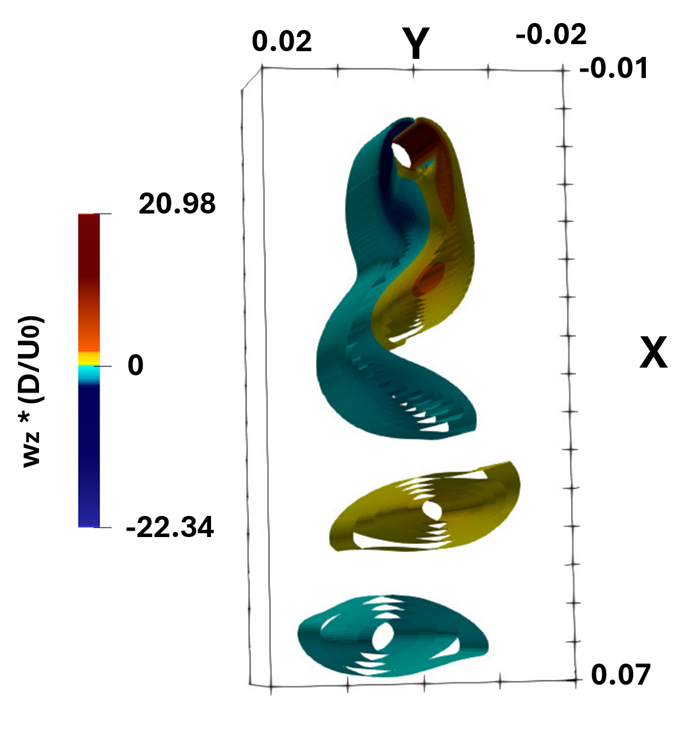}
\caption*{(a)}
\end{minipage}
\begin{minipage}{0.48\textwidth}
\includegraphics[width=0.9\textwidth]{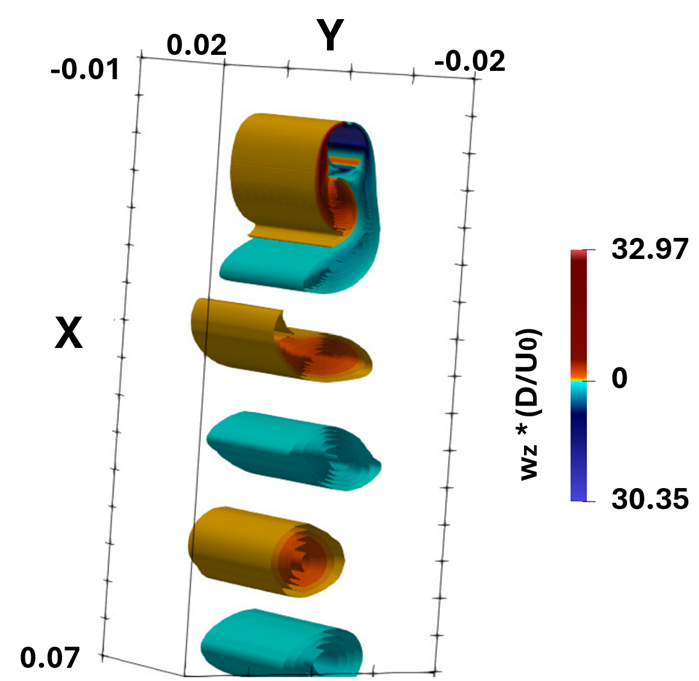}
\caption*{(b)}
\end{minipage}\hfill
\caption{Instantaneous three-dimensional visualizations of the spanwise vorticity field($w_z$), normalized by fluid timescale ($D/U_0$), in the wake of a circular cylinder for (a) $Re=100$ and (b) $Re=200$. Here, $D$ is the cylinder diameter and $U_0$ is the inlet fluid velocity.}
\label{unladen flow shedding}
\end{figure*}

Three-dimensional contours of spanwise vorticity field in the wake of a cylinder for $Re=100$ and $Re=200$ are presented in Figure \ref{unladen flow shedding}. Here, the spanwise vorticity ($ w_z \times D/U_0$) is normalized with flow timescale (ratio of cylinder diameter and inlet fluid velocity). The vortex shedding becomes much more frequent at higher Re with vortices placed closer to each other along the streamwise direction, reflecting an increased shedding frequency.

\begin{figure*}[!t]
    \centering

    \begin{subfigure}{0.45\textwidth}
        \centering
        \includegraphics[width=0.5\linewidth,trim={3cm 0cm 0cm 0mm},clip]{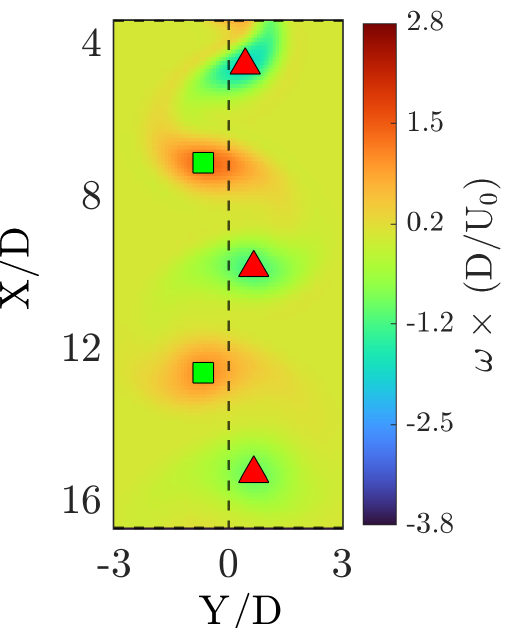}
        \caption{}
    \end{subfigure}
    \hfill
    \begin{subfigure}{0.39\textwidth}
        \centering
        \includegraphics[
            width=0.9\linewidth,
            trim={0cm 0cm 0cm 5mm},
            clip
        ]{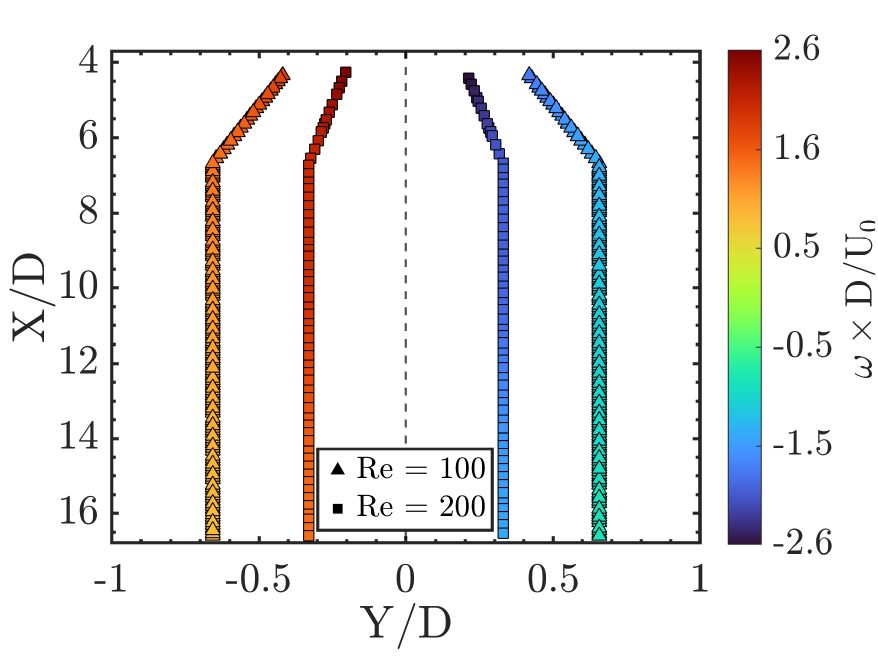}
        \caption{}
    \end{subfigure}
\caption{
(a) Instantaneous vorticity field of the unladen flow, normalized by fluid timescale ($D/U_0$), illustrating the alternating vortex-shedding pattern, with vortex cores identified by minimum ($\triangle$) and maximum ($\square$) local vorticity. Here, $D$ is the cylinder diameter and $U_0$ is the inlet fluid velocity.
(b) Downstream evolution of the vortex core trajectories in the cylinder wake for $Re=100$ ($\triangle$) and $Re=200$ ($\square$).
}
    \label{vortex_fig}
\end{figure*}

The normalized spanwise vorticity field and downstream vortex core evolution for the unladen flows at two Reynolds numbers $Re = 100$ and $Re = 200$ are presented in Figure \ref{vortex_fig}. The fluid instantaneous spanwise vorticity field with a classical Karman vortex pattern is shown in Figure \ref{vortex_fig}(a). The vortex cores are tracked using a MATLAB post-processing script by identifying extrema of the local vorticity field, which is represented by triangles and squares in the figure. The trajectories of the extrema downstream highlight the symmetric nature of the wake structure and the regular vortex shedding patterns for the Reynolds number investigated here. Figure \ref{vortex_fig} (b) displays the evolution of the downstream trajectory of vortex cores over multiple shedding cycles. The comparison between the trajectories for $Re=100$ and $Re=200$ reveals substantial effects of the Reynolds number on vortex dynamics. The vortex cores retain higher vorticity magnitudes at higher Reynolds number case, with the vortices persisting over longer downstream distances, indicating formation of more energetic vortices. In contrast, spanwise vorticity of vortex cores is observed to decrease downstream at $Re = 100$, resulting in a comparatively less intense vortex near the exit. Additionally, flow exhibit a wider wake width at $Re = 100$ with higher cross-stream displacement of vortex cores. At $Re = 200$, flow exhibits stronger vortex cores with narrower wake width, indicating presence of a more energetic and compact wake in the flow. These variations in vortex strength and wake width significantly influence fluid momentum deficit and the evolution of unsteady flow structures, thereby governing wake dynamics and influencing the fluid-particle interaction. 

\begin{figure*}[!t]
	\begin{subfigure}[b]{1\textwidth}
	\minipage{0.5\textwidth}		
	\includegraphics[width=0.75\textwidth]{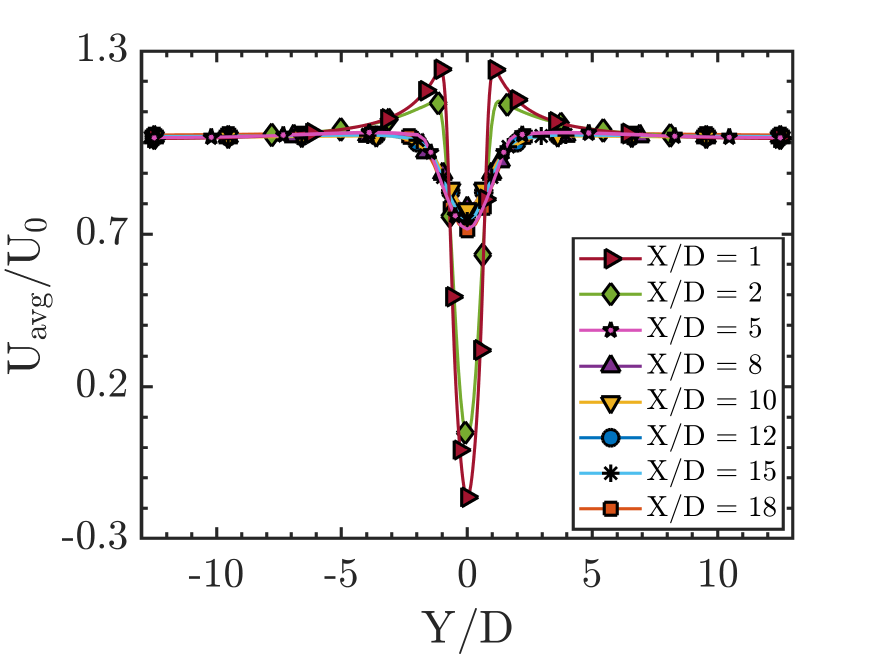}
	\caption{}
	\endminipage
	\minipage{0.5\textwidth}		
	\includegraphics[width=0.75\textwidth]{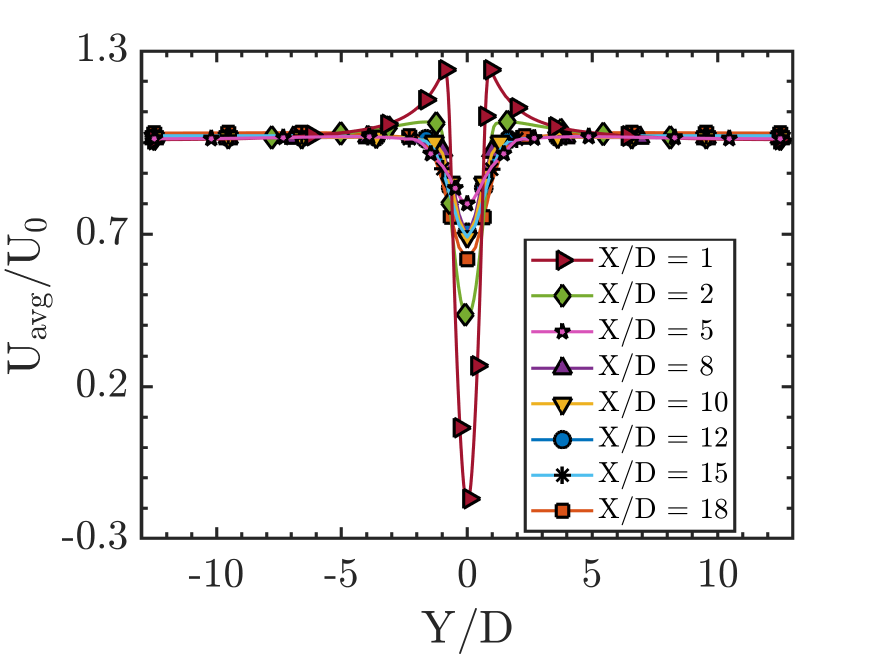}
	\caption{}
	\endminipage \\ [0.2cm]
	\minipage{0.5\textwidth}		
	\includegraphics[width=0.75\textwidth]{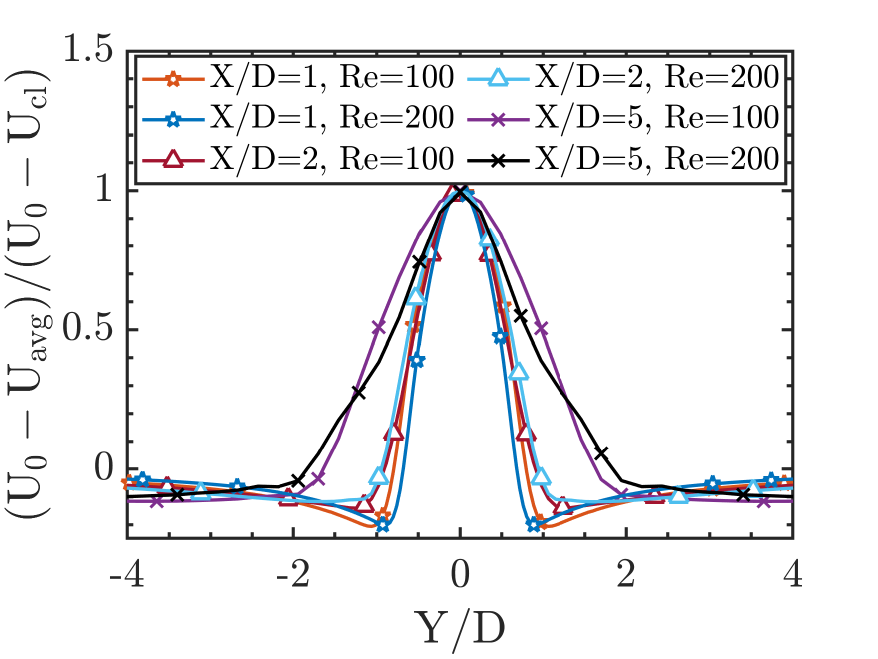}
	\caption{}
	\endminipage
	\minipage{0.5\textwidth}		
	\includegraphics[width=0.75\textwidth]{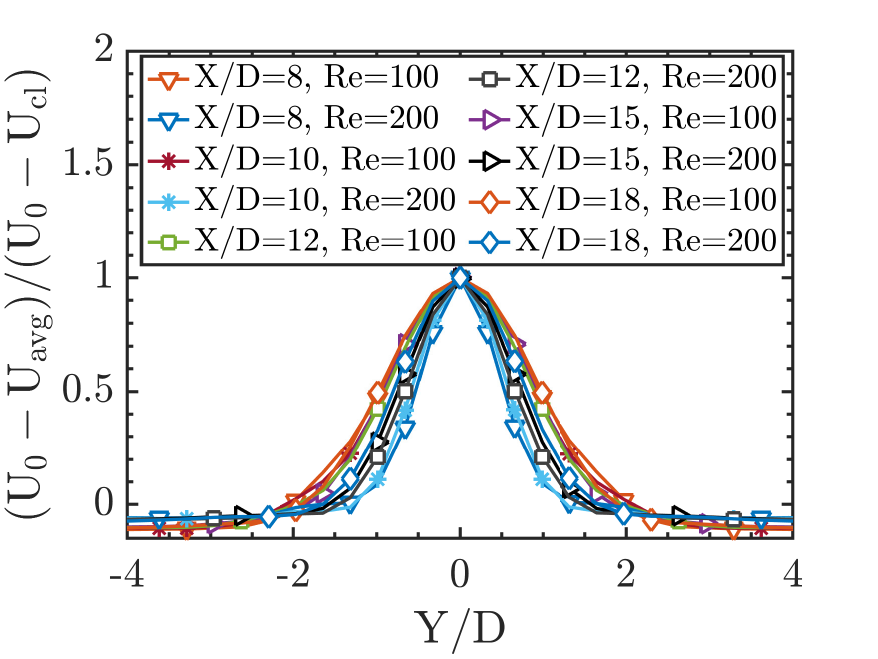}
	\caption{}
	\endminipage			
	\end{subfigure}
\caption{Normalized average streamwise velocity and velocity-deficit profiles at different downstream locations in the wake of a circular cylinder. The normalized mean streamwise velocity profiles are shown for (a) $Re=100$ and (b) $Re=200$. The normalized velocity-deficit profiles are shown for downstream locations (c) $X/D \leq 5$ and (d) $8 \leq X/D \leq 18$.}
\label{vel_fluid}
\end{figure*}

Figure \ref{vel_fluid} (a), (b) presents the mean streamwise velocity profiles in the wake of a cylinder at different downstream locations for Reynolds numbers $Re = 100$ and $200$, respectively. Here, the mean streamwise velocity profiles are normalized by the inlet velocity. The streamwise velocity is observed to decrease immediately after the cylinder in the region near the wake centerline at both Reynolds numbers, due to the fluid recirculation caused by flow separation. At the downstream, the wake gradually recovers its momentum, resulting in a more uniform velocity profile. Flow Reynolds number affects the recovery of fluid momentum within the wake region. For $Re = 200$, the wake exhibits a faster recovery of the centerline velocity, particularly in the near cylinder zone. This is caused by enhanced shear-layer instabilities where vortex are shed much closer to the cylinder, resulting in stronger fluid mixing. 

The normalized velocity deficit is the ratio of the difference between fluid inlet($U_{0}$) and local mean streamwise velocity($U_{avg}$), and the difference between fluid inlet velocity and mean streamwise velocity at the center of the wake ($U_{cl}$). We have presented fluid velocity deficit profiles to verify the self-similarity behavior of wake. Figure~\ref{vel_fluid} (c) shows profiles up to the downstream location of $X/D = 5$ to observe near-cylinder wake behavior, whereas Figure~\ref{vel_fluid} (d) shows wake behavior at further downstream locations. At larger cross-stream locations away from the centerline, the normalized velocity deficit approaches zero for both Reynolds numbers, indicating that the outer flow remains largely undisturbed by the wake. In free shear flows, self-similarity is attained when the normalized mean velocity profiles collapse onto a single curve, indicating that the wake evolution for the region is not influenced by the near-cylinder flow. Figure~\ref{vel_fluid}(c),(d) shows the normalized velocity-deficit profiles exhibits higher extent of self-similarity for same Reynolds number at downstream locations further away from the cylinder with the development of wake. The velocity-deficit profiles at downstream positions $X/D <=5$  (as shown in Figure~\ref{vel_fluid}(c)) presents a clear tendency to collapse into a common curve  for cases $Re =100 $ and $200$ at that location. Negative values of the normalized velocity deficit near $Y/D\leq\pm2$ at $X/D=1$ and 2, indicate local flow acceleration outside the shear layers on either side of wake, where the mean streamwise velocity exceeds the free-stream velocity ($U_{\mathrm{avg}}>U_0$). Here, the normalized velocity profiles remain strongly dependent on the downstream location. Further downstream ($X/D\geq8$), as shown in Figure~\ref{vel_fluid}(d), the normalized velocity-deficit profiles for both $Re=100$ and $Re=200$ collapse onto a nearly symmetric Gaussian-like distribution, indicating that the wake has reached a self-similar state. In this regime, the shape of the mean velocity deficit becomes nearly invariant with downstream distances. The collapse is more pronounced for the lower Reynolds number case, reflecting the faster wake evolution and reduced unsteadiness. In contrast, shear-layer instabilities are stronger at high Reynolds flow due to developing wake dynamics, resulting in a weaker collapse of profiles. The progression of profiles toward a single, symmetric Gaussian-like distribution at $X/D \geq 8$ is similar to findings in previous literature. A similar asymptotic Gaussian-like distribution has been reported by \citet{Wygnanski_Champagne_Marasli_1986}, who showed experimentally that although the characteristic wake scales depend on the cylinder geometry and inflow conditions, the normalized mean velocity-deficit profile approaches a universal distribution sufficiently far downstream.  However, they observed such collapse at much higher downstream distance. In the present study, a comparable collapse with slight difference is observed at considerably shorter downstream distances, as the wake, particularly for $Re=200$, has not yet reached a fully self-similar state. 

\begin{figure*}[!t]
	\begin{subfigure}[b]{1\textwidth}
	\minipage{0.5\textwidth}		
	\includegraphics[width=0.8\textwidth]{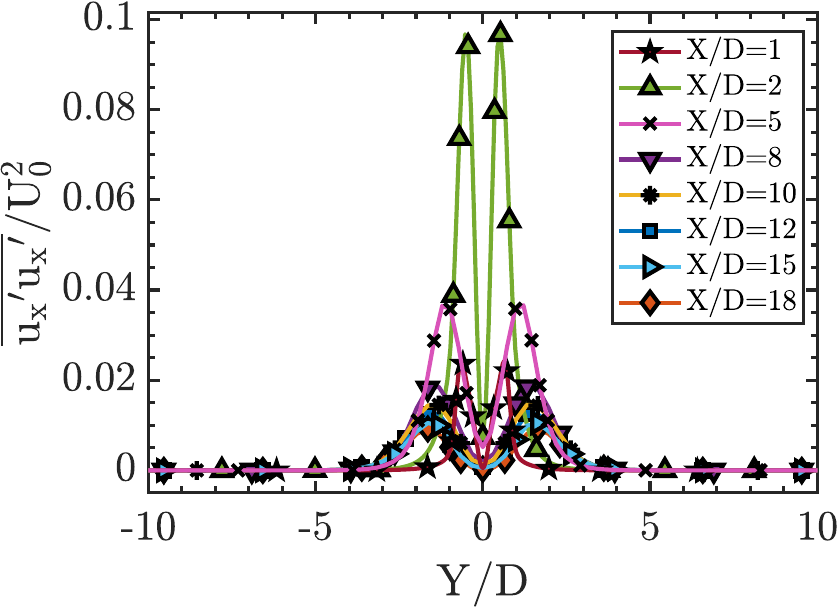}
	\caption{}
	\endminipage
	\minipage{0.5\textwidth}		
	\includegraphics[width=0.75\textwidth]{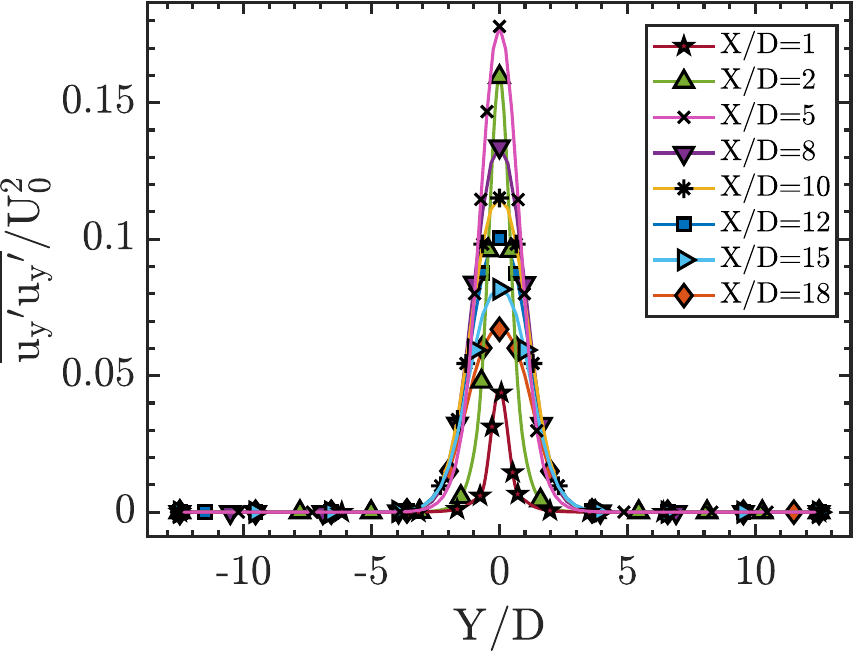}
	\caption{}
	\endminipage \\ [0.2cm]
	\minipage{0.5\textwidth}		
	\includegraphics[width=0.75\textwidth]{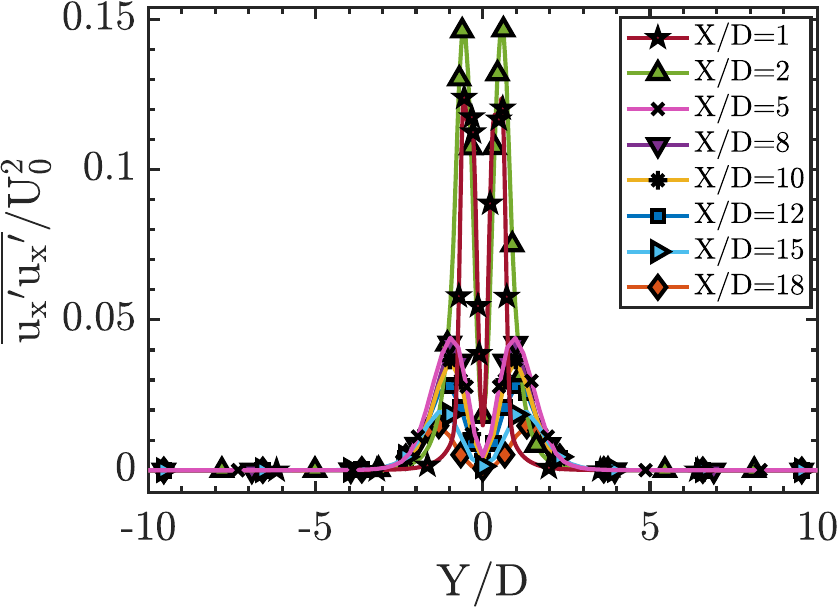}
	\caption{}
	\endminipage
	\minipage{0.5\textwidth}		
	\includegraphics[width=0.75\textwidth]{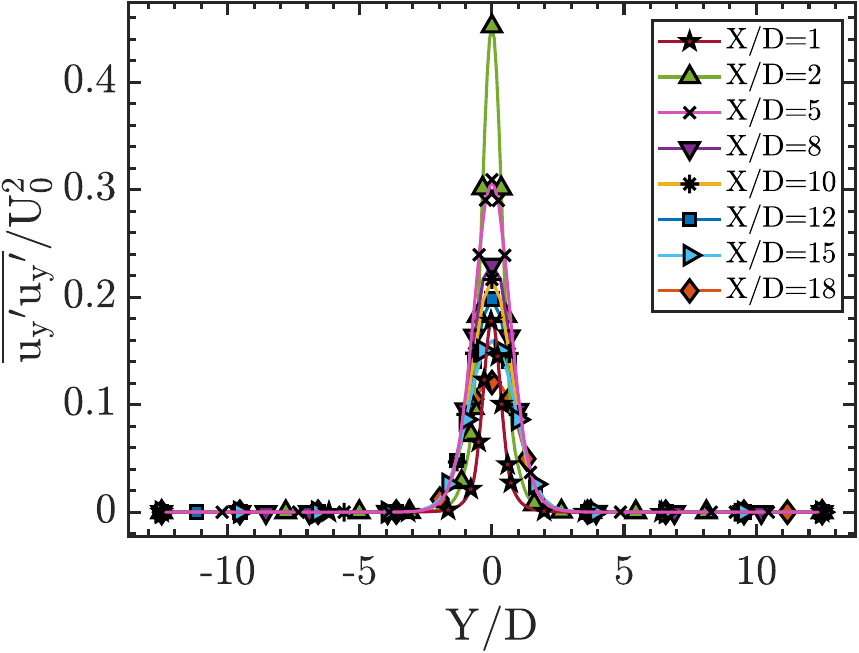}
	\caption{}
	\endminipage			
	\end{subfigure}
\caption{Profiles of the mean-square fluid velocity fluctuations, normalized by the inlet velocity($U_0$) for unladen flow at different downstream locations in the wake of a circular cylinder. Results for $Re=100$ are shown in (a,b), whereas those for $Re=200$ are shown in (c,d). The normalised streamwise and cross-stream velocity fluctuations are shown in (a,c) and (b,d), respectively.}
\label{fluc_fluid}
\end{figure*}

 The unsteady nature of the flow has been characterized by the mean square of the unsteady part of the velocity components, where fluctuations are calculated by subtracting local average velocity from the instantaneous velocity. In Figure \ref{fluc_fluid}, profiles of the normalized  mean-square streamwise fluctuations and cross-stream velocity fluctuations are plotted at different downstream locations from the cylinder. The streamwise fluctuations exhibits distinct peaks on either side of the centerline, corresponding to the shear layer present at the boundary of the wake zone, which is the region of high mean velocity gradient. The maximum fluctuation intensity occurs relatively close to the cylinder (at $X/D = 2-5$). As the flow moves further downstream ($X/D > 5$), the peaks gradually widen and diminish in magnitude as the wake diffuses and the shear layers interact.  It can be observed that the unsteady fluctuations are more intense for $Re=200$  and remains significant over a longer extent downstream. For both the Reynolds number, cross-stream fluctuations is significantly higher than the streamwise fluctuations. Increasing the Reynolds number strengthens the unsteady wake, leading to higher fluctuation intensities in both velocity components. The peak average streamwise fluctuation normalized with inlet fluid velocity ($\overline{{{u}_{x}}^{\prime}{{u}_{x}^{\prime}}} / {U}_{0}^2$) increases from approximately 0.1 at $Re=100$ to about 0.15 at $Re=200$. A more pronounced increase is observed for the normalized average cross-stream fluctuation ($\overline{{{u}_{y}}^{\prime}{{u}_{y}^{\prime}}} / {U}_{0}^2$), whose maximum value rises from approximately 0.17 to about 0.45. Furthermore, the location of the peak cross-stream fluctuation shifts from $X/D\approx5$ at $Re=100$ to $X/D\approx2$ at $Re=200$, indicating that the vortex formation region moves closer to the cylinder with increasing Reynolds number.

\subsection{Particle phase characteristics}
\begin{figure*}[!t]
\centering

\begin{minipage}{0.4\textwidth}
\includegraphics[width=\textwidth]{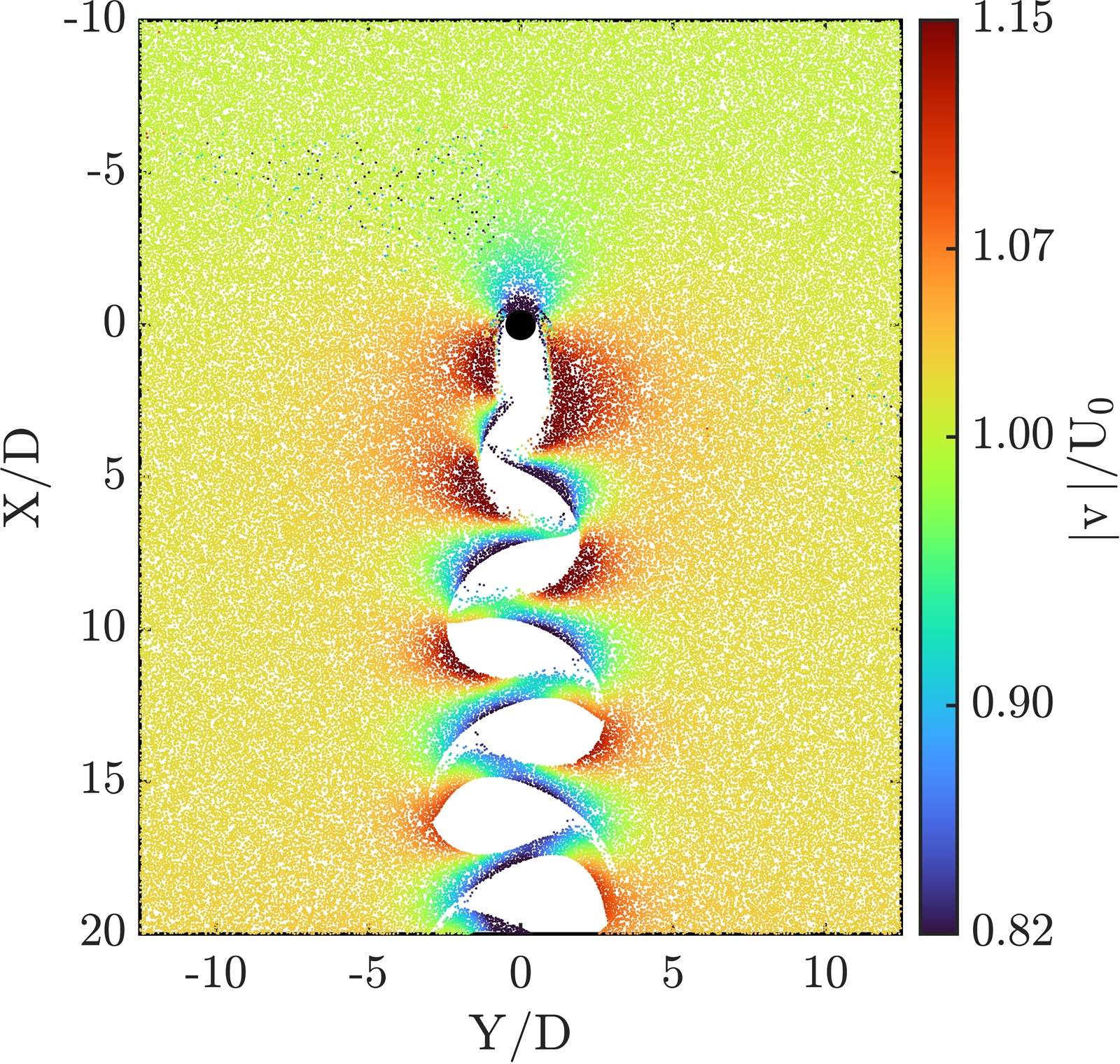}
\caption*{(a)}
\end{minipage}\hfill
\begin{minipage}{0.4\textwidth}
\includegraphics[width=\textwidth]{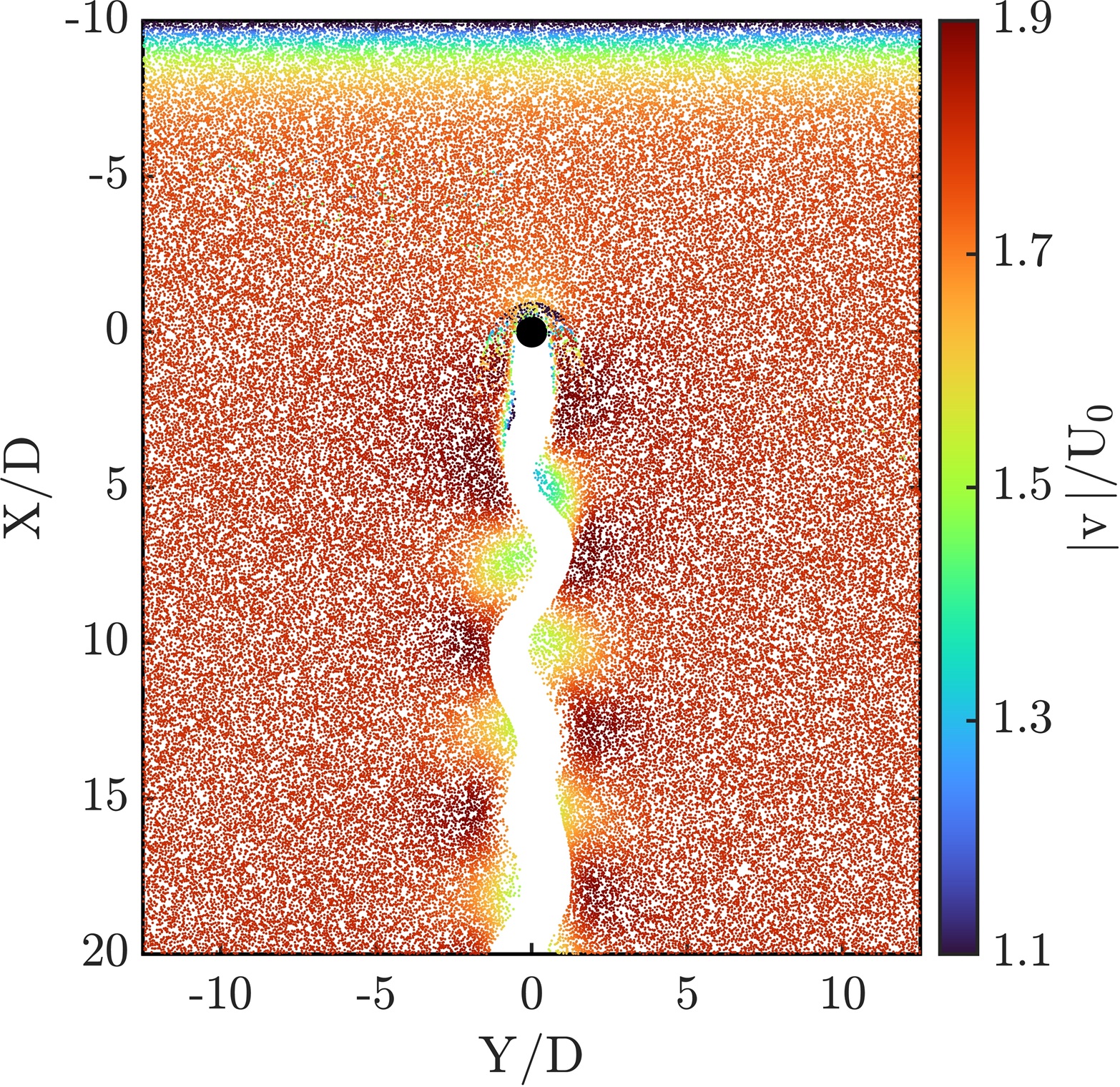}
\caption*{(b)}
\end{minipage}

\vspace{2mm}

\begin{minipage}{0.4\textwidth}
\includegraphics[width=\textwidth]{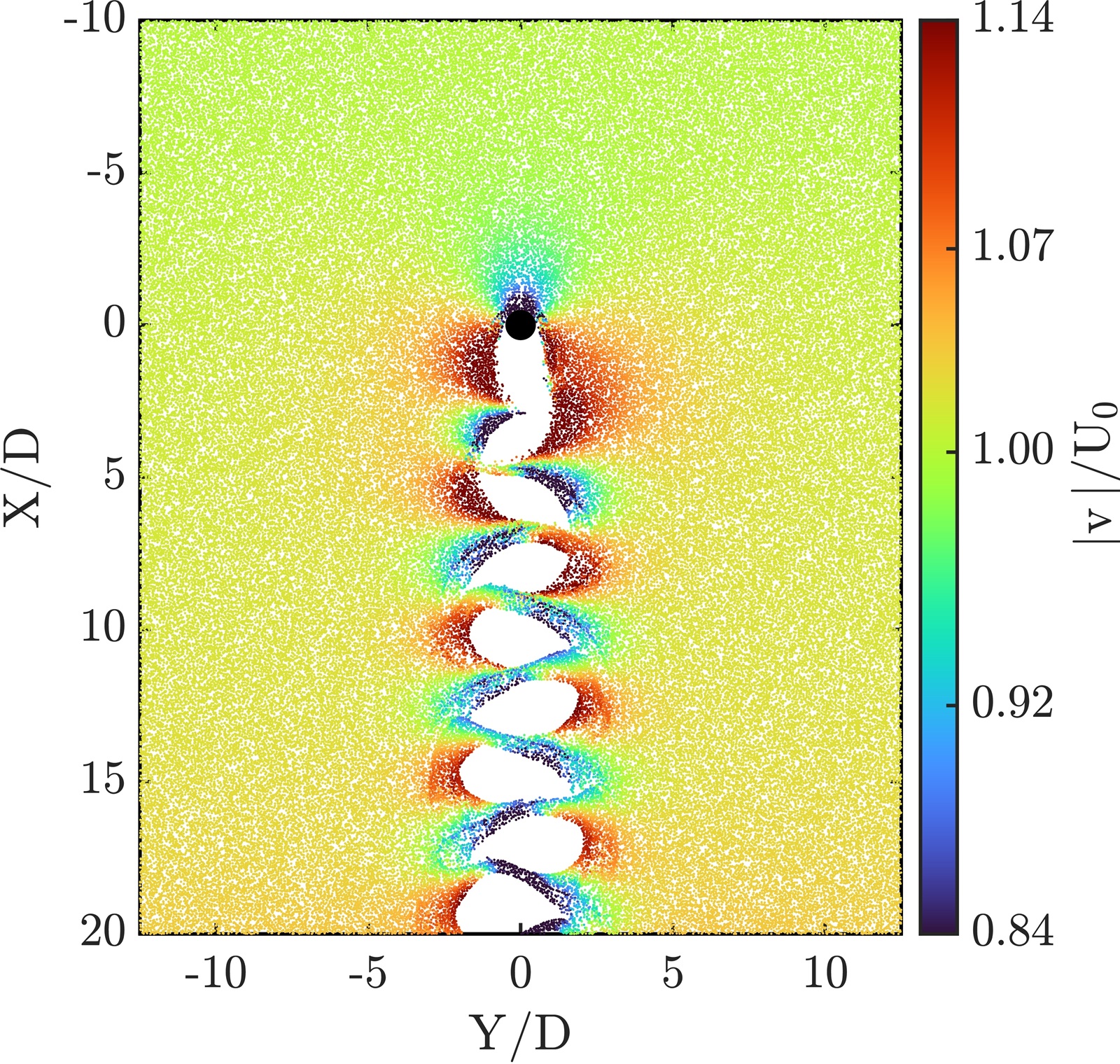}
\caption*{(c)}
\end{minipage}\hfill
\begin{minipage}{0.4\textwidth}
\includegraphics[width=\textwidth]{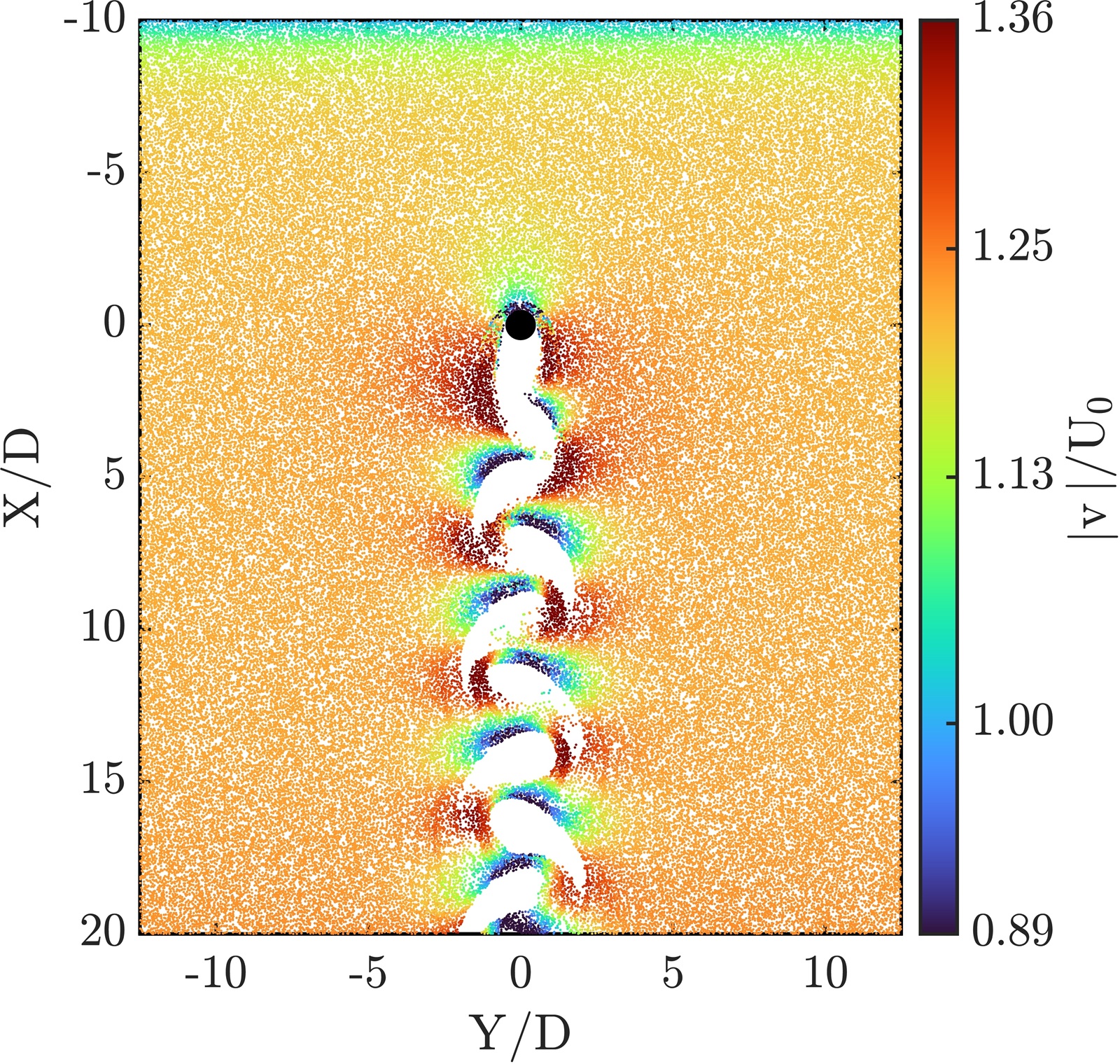}
\caption*{(d)}
\end{minipage}

\caption{Scatter plots of instantaneous particle positions, colored by the magnitude of their particle velocity normalized by the inlet velocity($U_0$), in the wake of a circular cylinder for high particle loading ($\phi_v=2\times10^{-4}$) and low inertia particle ($St=1$) cases. Results for $Re=100$ and $200$ are shown in (a,b) and (c,d), respectively. Cases without gravity ($Fr\rightarrow\infty$) are shown in (a,c), whereas cases with gravity are shown in (b) ($Fr=1$) and (d) ($Fr=2$).}
\label{Scatter_plot_vp_LSt}
\end{figure*}

\begin{figure*}[!t]
\centering

\begin{minipage}{0.4\textwidth}
\includegraphics[width=\textwidth]{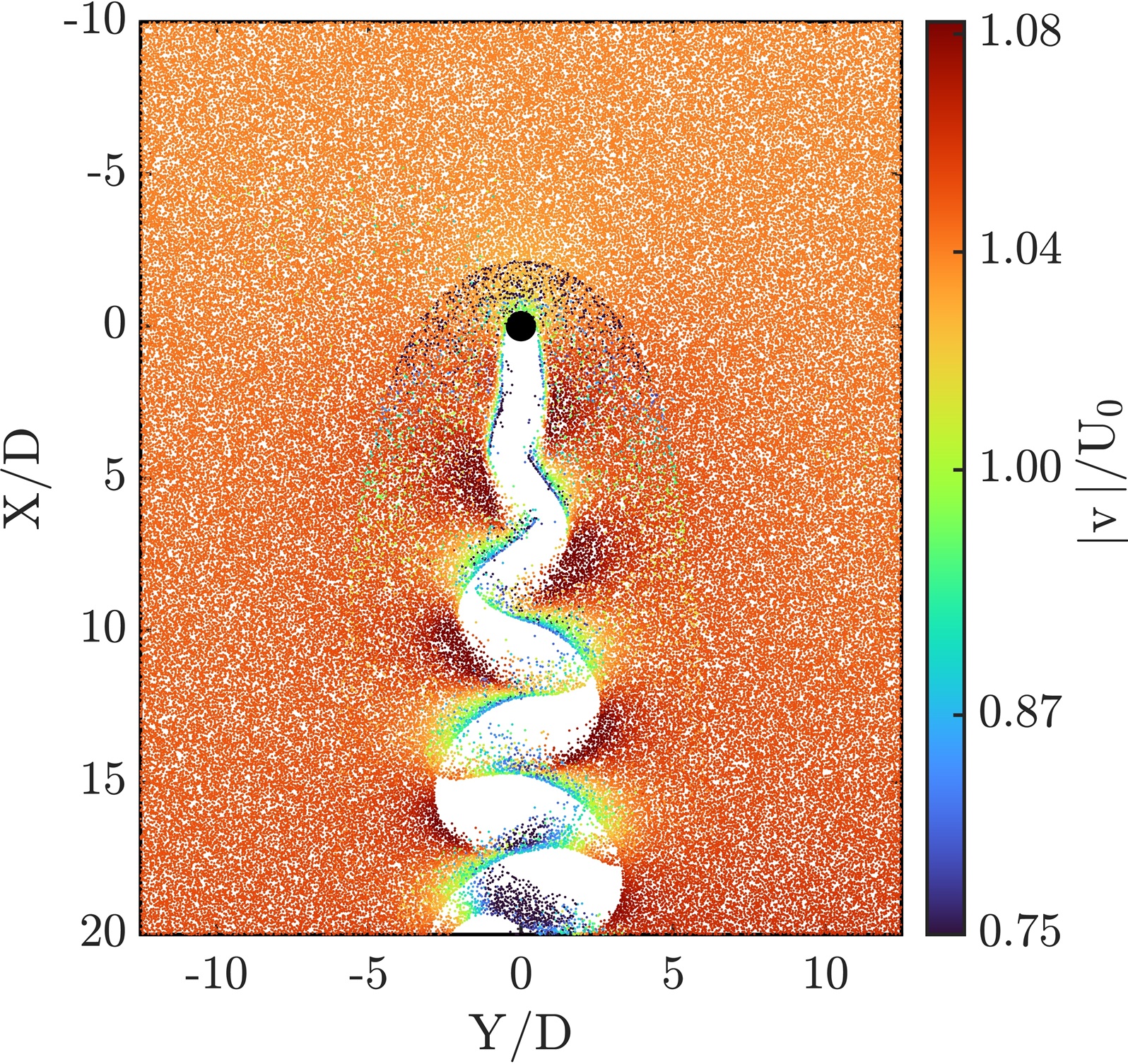}
\caption*{(a)}
\end{minipage}\hfill
\begin{minipage}{0.4\textwidth}
\includegraphics[width=\textwidth]{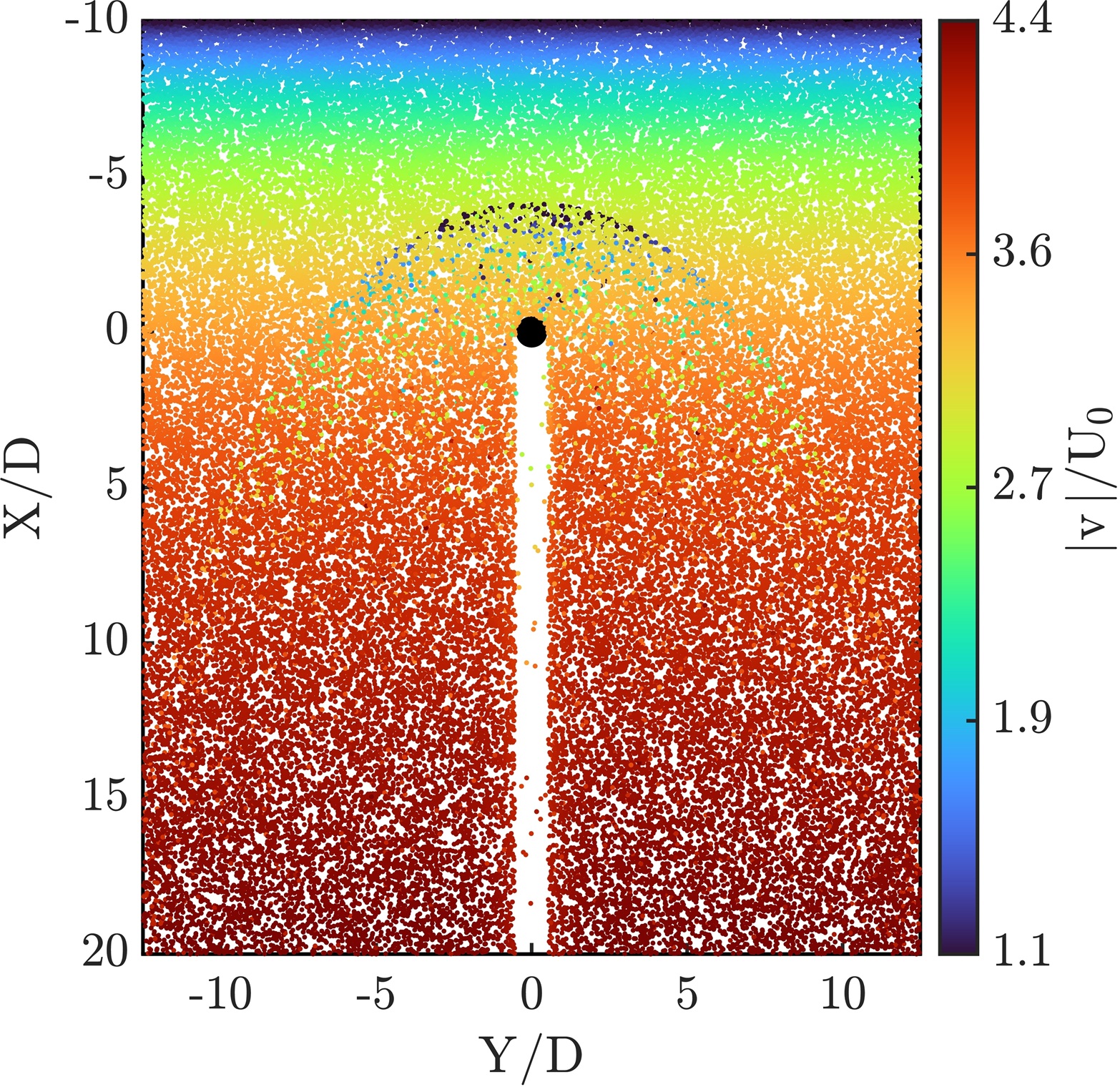}
\caption*{(b)}
\end{minipage}

\vspace{2mm}

\begin{minipage}{0.4\textwidth}
\includegraphics[width=\textwidth]{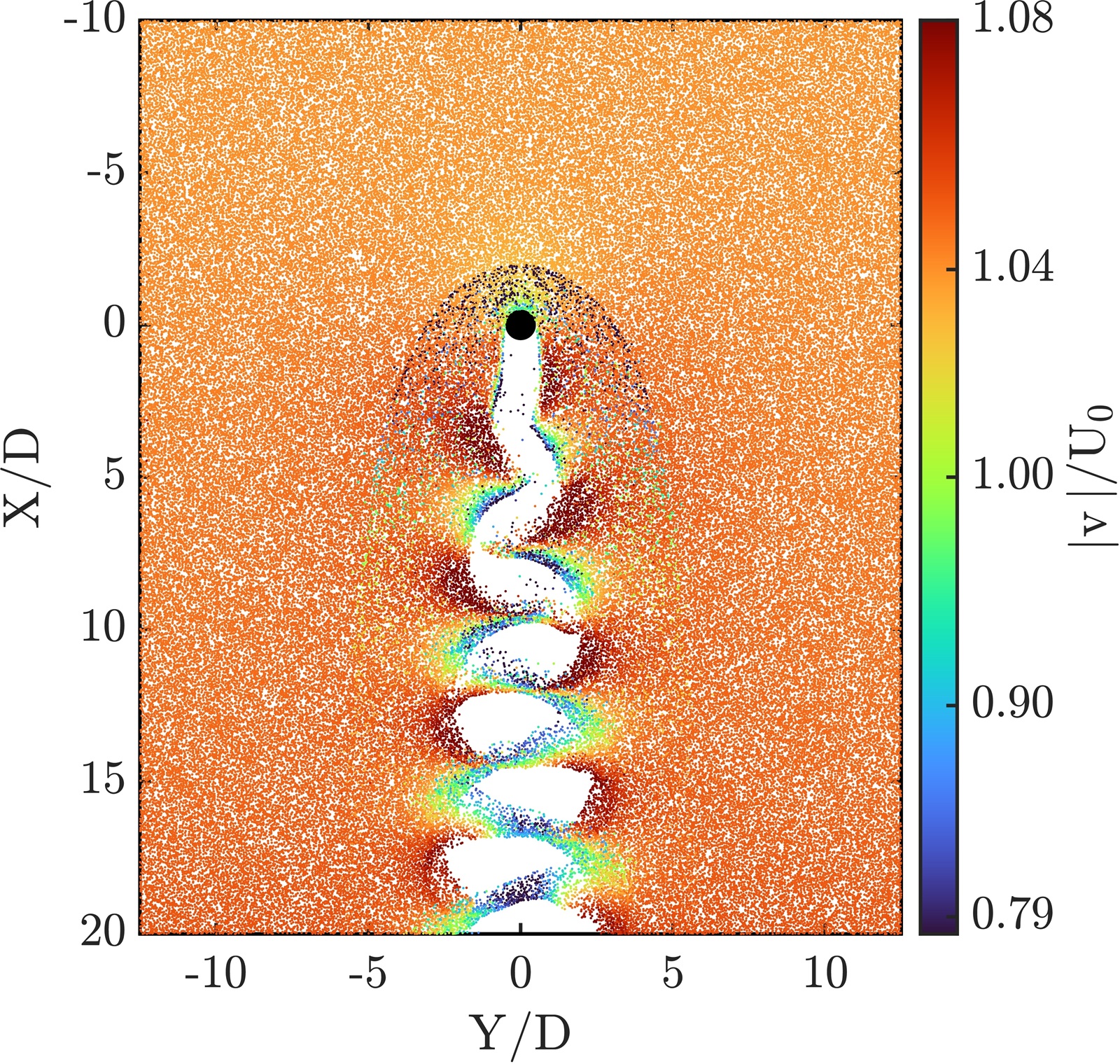}
\caption*{(c)}
\end{minipage}\hfill
\begin{minipage}{0.4\textwidth}
\includegraphics[width=\textwidth]{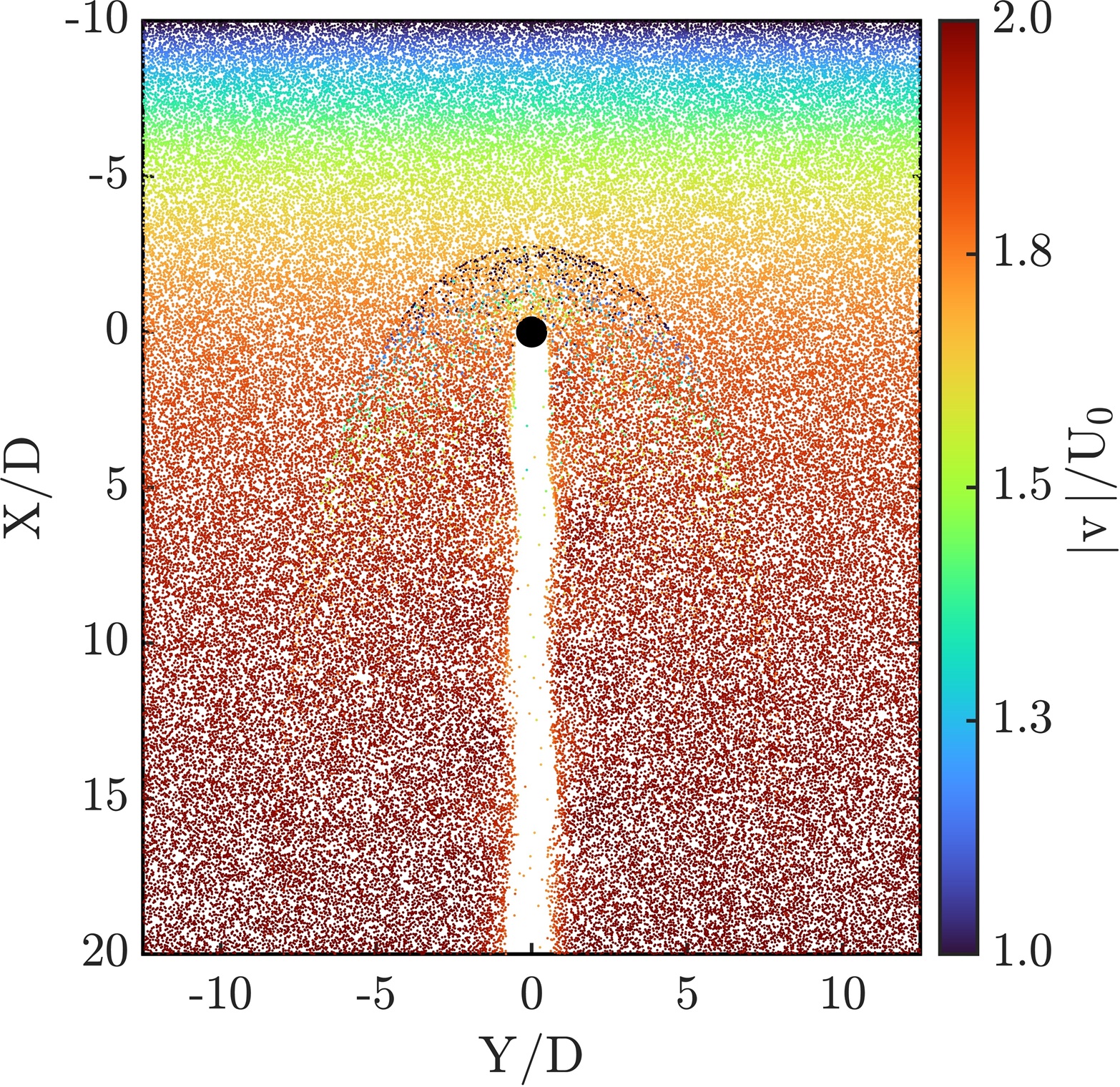}
\caption*{(d)}
\end{minipage}

\caption{Scatter plots of instantaneous particle positions, colored by the magnitude of their particle velocity normalized by the inlet velocity($U_0$), in the wake of a circular cylinder for high particle loading ($\phi_v=2\times10^{-4}$) and high inertia particle ($St=6.5$) cases. Results for $Re=100$ and $200$ are shown in (a,b) and (c,d), respectively. Cases without gravity ($Fr\rightarrow\infty$) are shown in (a,c), whereas cases with gravity are shown in (b) ($Fr=1$) and (d) ($Fr=2$).}
\label{Scatter_plot_vp_HSt}

\end{figure*}

The spatial distributions of particle position and their velocity are examined to highlight the combined effects of particle inertia and gravity on particle phase in the wake of the cylinder, as shown in Figures~\ref{Scatter_plot_vp_LSt} and~\ref{Scatter_plot_vp_HSt}. Here, particle centers are represented by scatter points, with color indicating the magnitude of the particle velocity, normalized by the inlet velocity. The larger black-filled circle represents the cylinder's position in the channel. The particles, on interaction with local fluid fluctuations, arrange themselves downstream to the cylinder, leading to the formation of coherent voids and clusters in the presence of the wake, as reported earlier \citep{shi2021clusters}. In the unsteady wake, inertial particles are centrifugally expelled from the cores of the periodically shed vortices and forms voids, which are  regions of reduced local particle concentration. 
In the absence of gravity ($Fr\rightarrow\infty$), low inertia particles closely follow the vortical structures and forms a leaf-like void pattern, as shown in Figures~\ref{Scatter_plot_vp_LSt} (a) and (c). Particles are preferentially expelled from regions of high vorticity and accumulate along strain-dominated zones between successive vortices, resulting in a well-organized, traveling pattern downstream of the cylinder. This behavior is consistent with preferential concentration mechanisms for inertial particles in unsteady flows in the absence of gravity \citep{shi2021clusters,shi2020bow}.

The influence of gravity on the particle distribution is more pronounced at the lower Reynolds number ($Re =100$), where particles penetrate vortical regions and alter particle spatial distribution. For $St = 1$, an elongated void region is observed downstream of the cylinder at $Re =100$, whereas at $Re = 200$, a leaf-like void pattern is formed, as shown in Figures~\ref{Scatter_plot_vp_LSt} (a) and (b). Figure~\ref {Scatter_plot_vp_HSt} presents the scatter plot of particle centers at high Stokes number ($St=6.5$). It is observed that gravity plays a significant role in modifying the particle distribution at both Reynolds numbers, with its influence being more pronounced for the high Stokes number case. For the high Stokes case in the absence of gravity ($Fr\rightarrow\infty$), leaf-like void structures are observed (Figure~\ref{Scatter_plot_vp_HSt} (a),(c)), whereas an elongated void region is observed for gravity-driven flows (Figure~\ref{Scatter_plot_vp_HSt} (b),(d)). Here, the figures present the high particle loading case ($\phi_v = 2 \times 10^{-4}$). Similar behavior is observed for the low particle loading case ($\phi_v = 2 \times 10^{-5}$), suggesting that the particle void behavior in the cylinder wake is relatively insensitive to particle loading compared with the other parameters investigated over the range of Froude numbers considered in the present study (Table~\ref{foc_table}). The Froude number also affects the particle velocity distributions, as shown in the color map of Figure~\ref{Scatter_plot_vp_LSt} and ~\ref{Scatter_plot_vp_HSt}, highlighting the influence of gravity and particle inertia on particle motion. In the absence of gravity ($Fr \rightarrow \infty$), particle velocities remain close to the inlet velocity ($|v| \approx U_0$), with only minor variations induced by the vortices on neighbouring particles. Particles exhibit slight deceleration within or near the leaf-like void regions and slight acceleration around the outer edges of the vortices. Particles also decelerate on rebounding after collision with cylinder. When gravity is introduced, particle velocity increases as it travels in the streamwise direction due to gravitational settling. This is evident from the nearly horizontal bands of constant particle velocity, indicating an increase in particle velocity with downstream distance. The effect becomes increasingly pronounced with an increase in particle inertia, where the particle has a higher terminal velocity. For the low inertia case ($St = 1$) at $Re = 100$ and low particle loading ($\phi_v = 2 \times 10^{-5}$), particle velocities reach a maximum magnitude of approximately $|v|/U_0 \approx 1.9$ (Figure~\ref{Scatter_plot_vp_LSt}b). In contrast, for the high inertia case ($St = 6.5$) at $Re = 100$, particle velocities reach approximately $|v|/U_0 \approx 4.4$ (Figure~\ref{Scatter_plot_vp_HSt}b), indicating that gravitational settling becomes the dominant mechanism governing particle motion and produces a nearly vertical void region. To understand the effect of gravity on particle dynamics in detail, we investigate the particle slip (relative) velocity for different parametric conditions.

\begin{figure*}[!t]
	\begin{subfigure}[b]{1\textwidth}
	\minipage{0.45\textwidth}		
	\includegraphics[width=0.9\textwidth]{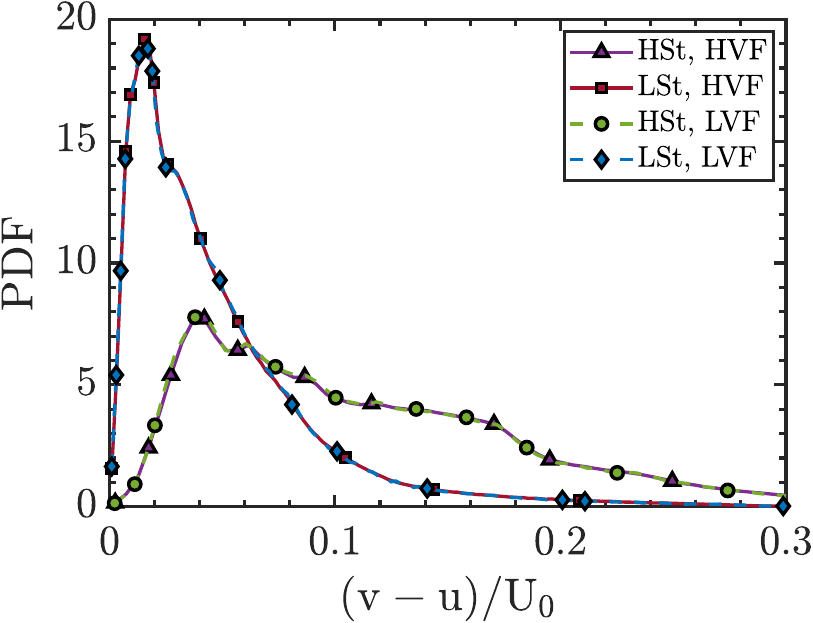}
	\caption{}
	\endminipage
	\minipage{0.45\textwidth}		
	\includegraphics[width=0.9\textwidth]{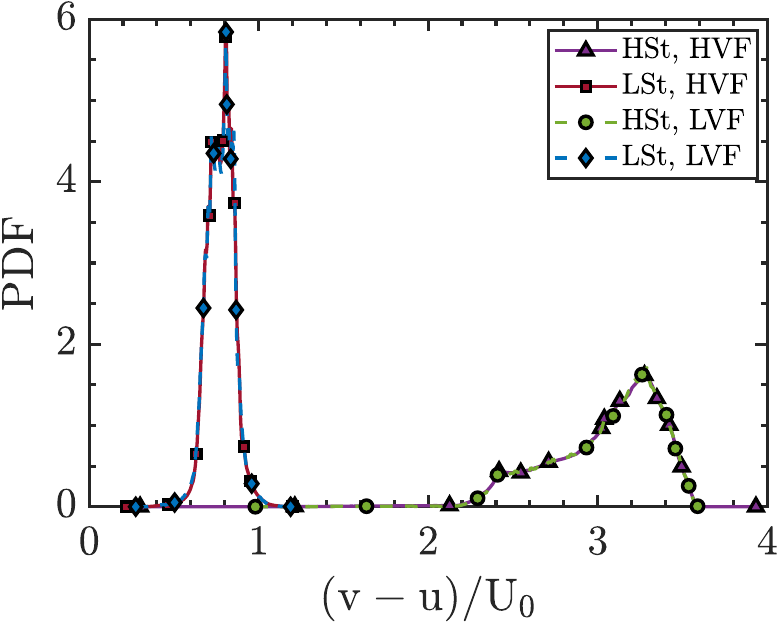}
	\caption{}
	\endminipage \\[1cm]
	\minipage{0.45\textwidth}		
	\includegraphics[width=0.9\textwidth]{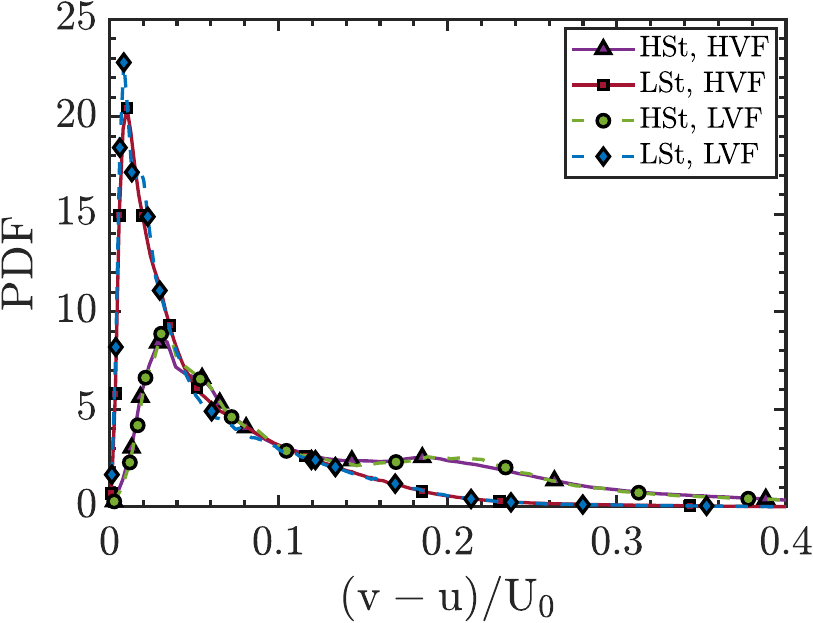}
	\caption{}
	\endminipage
	\minipage{0.45\textwidth}		
	\includegraphics[width=0.9\textwidth]{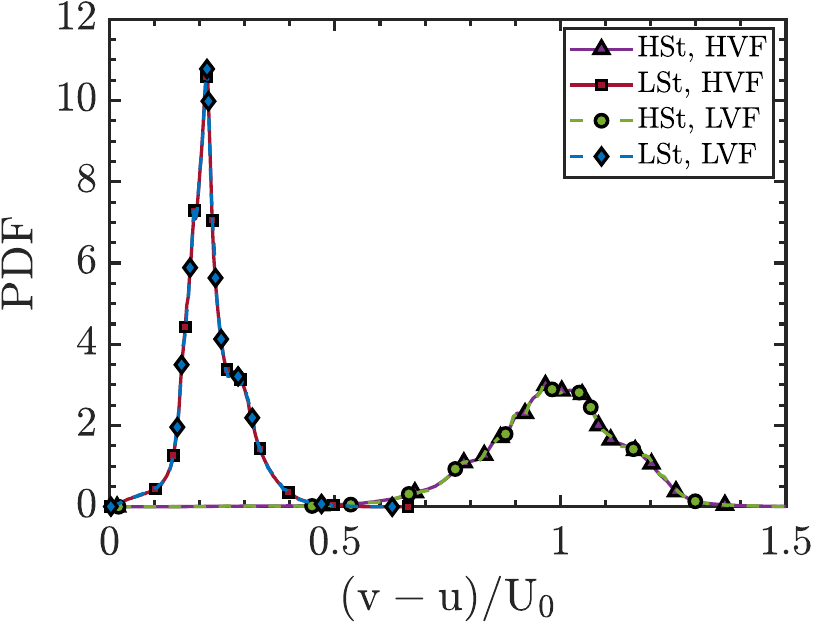}
	\caption{}
	\endminipage			
	\end{subfigure}
\caption{Probability density functions (PDFs) of the particle-fluid relative velocity, normalized by inlet fluid velocity ($U_0$), for (a,b) $Re=100$ and (c,d) $Re=200$. Cases without gravity ($Fr\rightarrow\infty$) are shown in (a,c), whereas cases with gravity are shown in (b) ($Fr=1$) and (d) ($Fr=2$).}
\label{PDF of relative velocity}
\end{figure*}

\begin{figure*}[!t]
	\begin{subfigure}[b]{1\textwidth}
	\minipage{0.45\textwidth}		
	\includegraphics[width=0.9\textwidth]{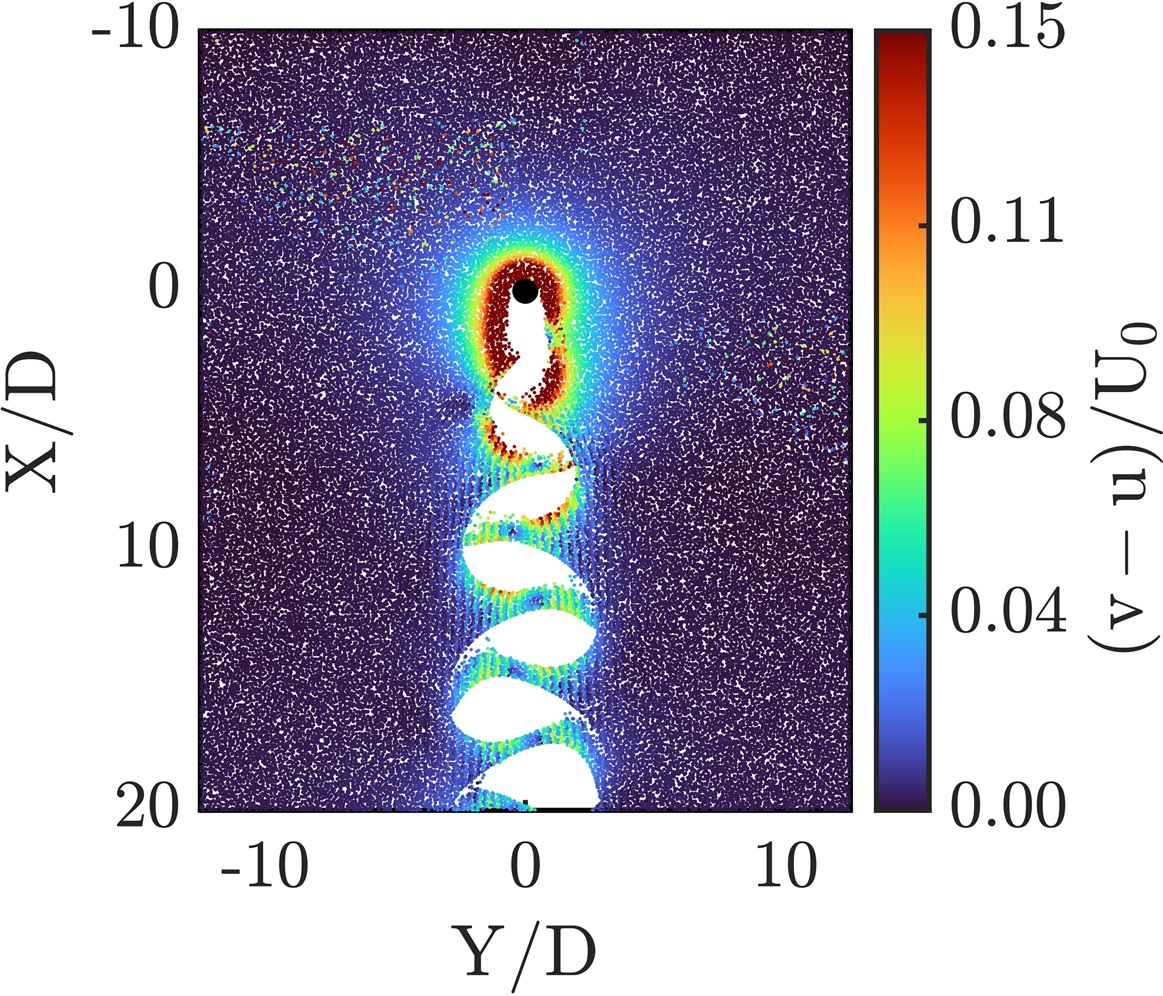}
	\caption{}
	\endminipage
	\minipage{0.45\textwidth}		
	\includegraphics[width=0.9\textwidth]{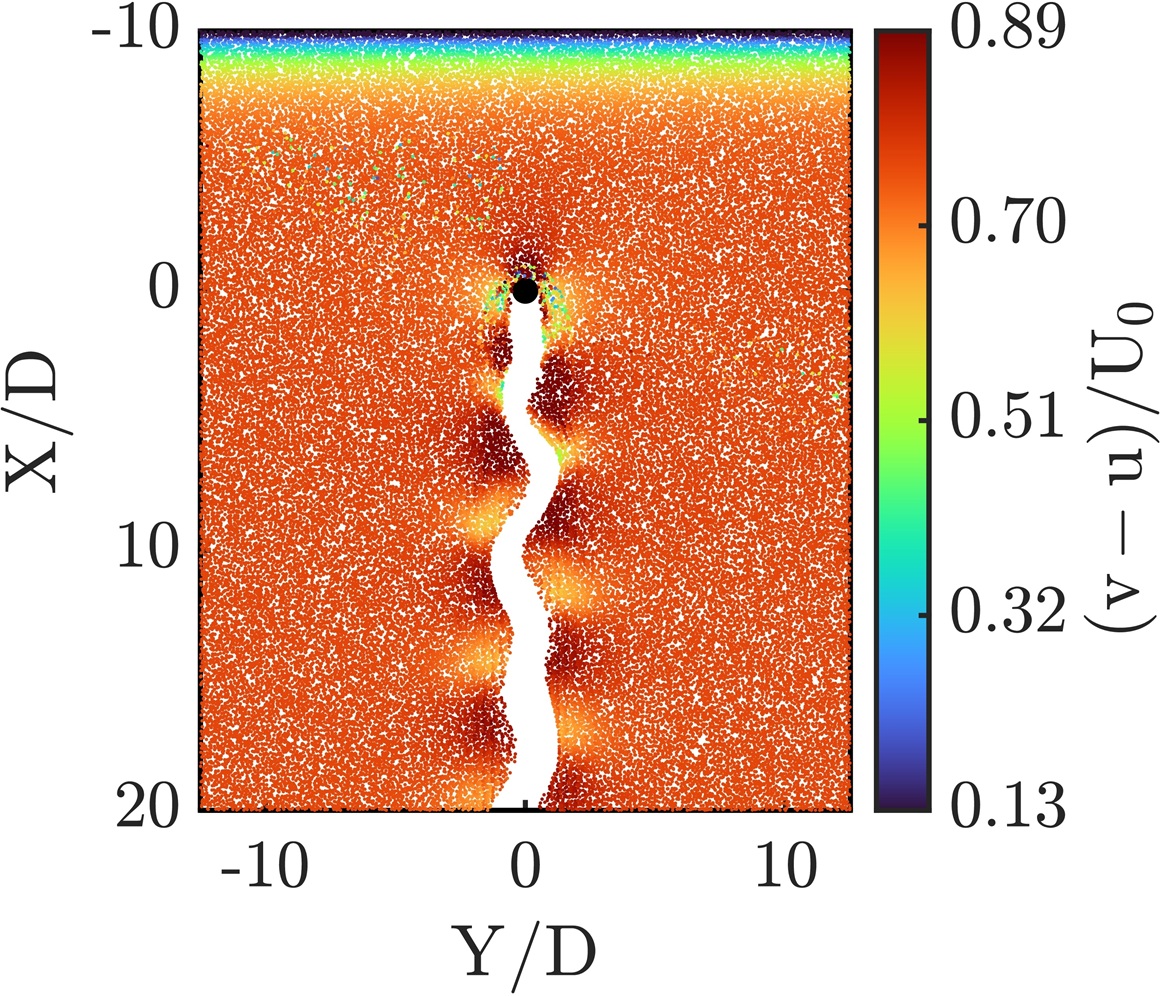}
	\caption{}
	\endminipage  
    \\[1cm]
	\minipage{0.45\textwidth}		
	\includegraphics[width=0.9\textwidth]{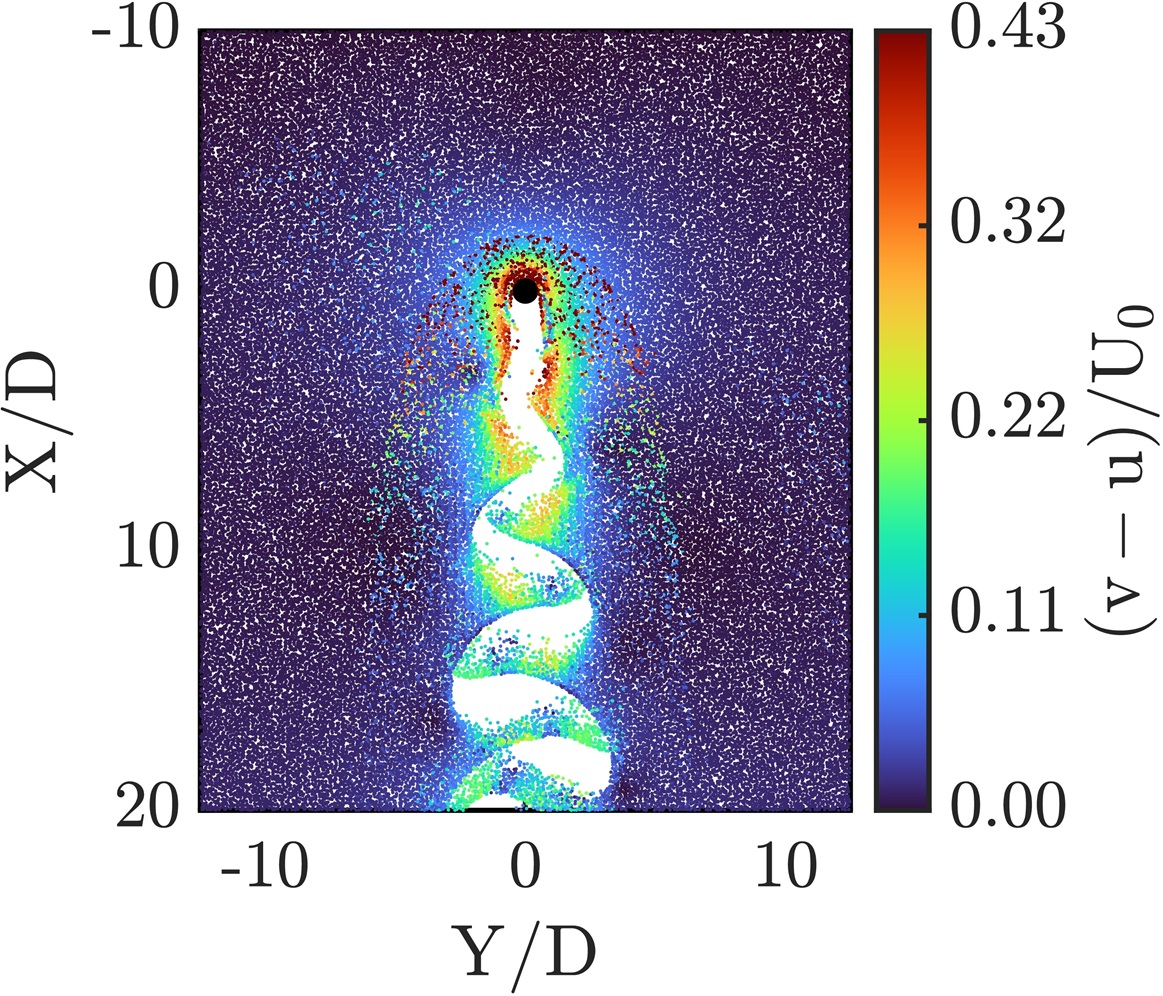}
	\caption{}
	\endminipage
	\minipage{0.45\textwidth}		
	\includegraphics[width=0.9\textwidth]{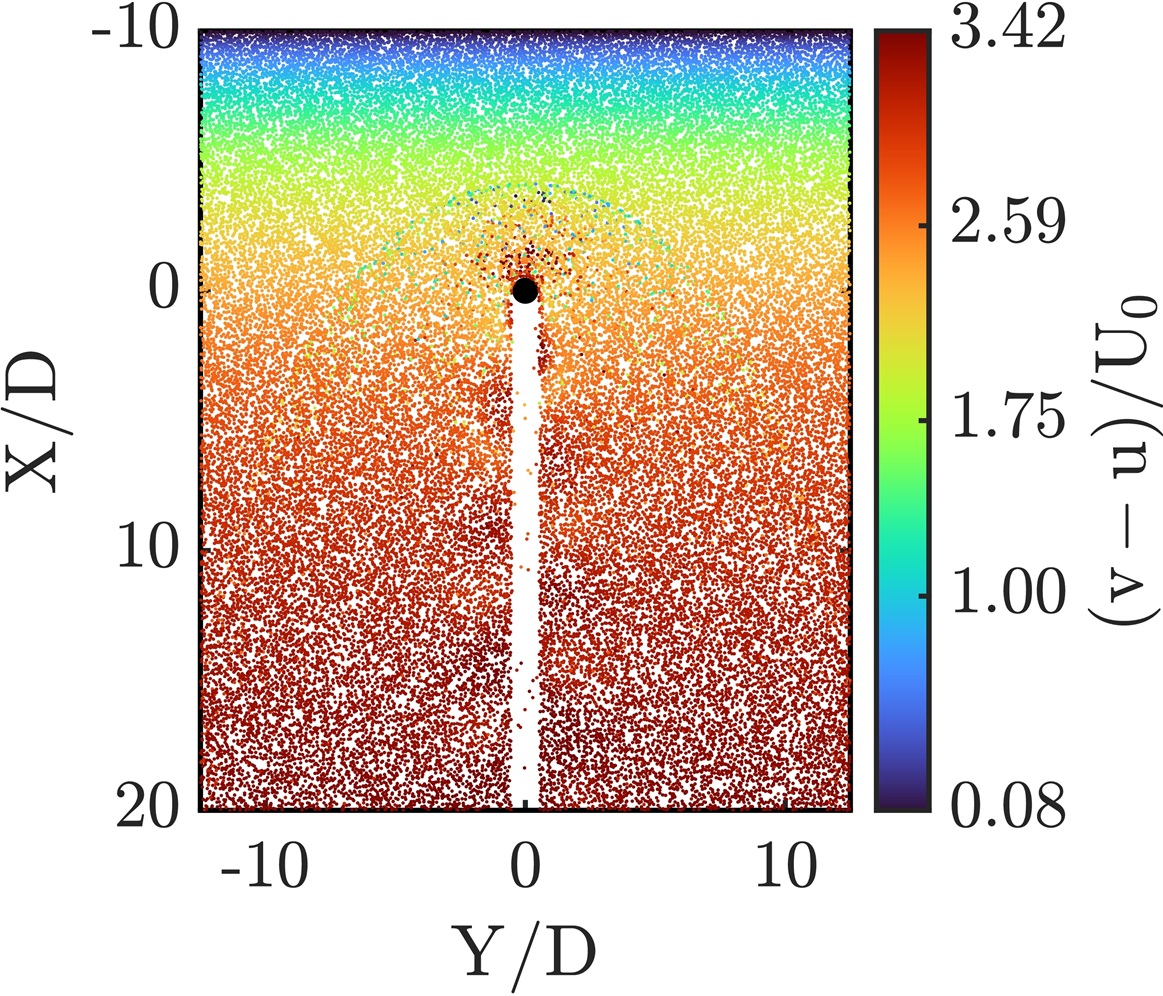}
	\caption{}
	\endminipage			
	\end{subfigure}
\caption{Instantaneous particle positions, shown as colored scatter plots of the normalized slip velocity, for $Re=100$. Results for low inertia particles ($St=1$) are shown in (a,b), whereas those for high inertia particles ($St=6.5$) are shown in (c,d). Cases without gravity ($Fr\rightarrow\infty$) are shown in (a,c), whereas cases with gravity for $Fr=1$ are shown in (b,d).}
\label{scatter relative velocity}
\end{figure*}

Figure~\ref{PDF of relative velocity} shows the probability density functions (PDFs) of the normalized particle-fluid slip velocity for the varying Reynolds numbers, for both gravity-free and gravity-driven flow configurations. The PDFs are narrowly peaked around zero slip velocity in case of $Fr\rightarrow\infty$, indicating that particles remain strongly coupled to the fluid flow. This observation is more pronounced for lower inertia particles, which closely follow the unsteady fluid flow induced by vortex shedding. The PDF peaks shift towards higher slip velocities and develop broader tails for high inertia particles having high Stokes number. This is because particles are unable to follow unsteady fluid fluctuations, leading to higher slip velocity. The  PDF of normalized slip velocity for higher inertia particles is slightly broader at the tail section in the absence of gravity. This broadness of PDF is due to larger fraction of particles having higher slip due to high inertia. At $Re=100$, increasing the Stokes number shifts the PDF peak towards a much higher normalized relative velocity as shown in Figure~\ref{PDF of relative velocity}(b). The increase in particle-fluid slip leads to decoupling of the fluid and particle phase, resulting in the formation of elongated voids. A similar peak shift is also observed for $Re=200$ in Figure~\ref{PDF of relative velocity}(d). It is to be noted that increasing particle inlet volume fraction does not have any significant effect on slip velocity. To obtain a clear picture on the location of high slip velocity, scatter plots of particle centers colored by their slip velocity are shown in Figure ~\ref{scatter relative velocity}.

As seen in Figure ~\ref{scatter relative velocity}(a),(c), in the case of $Fr\rightarrow\infty$, particles located away from the cylinder wake exhibit relatively low slip velocities, whereas particles with higher slip velocities are concentrated in the near-wake region. For the low Stokes number particles, higher slip velocities are observed in the vicinity of the cylinder. In contrast, in the high Stokes number case, higher normalized slip velocities persist over a much larger portion of the wake, up to the much greater downstream vortical region. Thus, a larger fraction of particles shows moderate-to-high slip velocities, resulting in the broader PDF and the shift of the distribution towards higher normalized slip velocities. Additionally, high inertia particles rebound farther upstream on impact with the cylinder, being less responsive to the underlying unsteady fluid fluctuations. These factors contribute to an increase in slip velocity and a broader slip velocity PDF in the high Stokes number case.

At finite Froude number (Figure~\ref{scatter relative velocity}(b,d)), particles accelerate in the presence of gravity, leading to an enhanced normalized slip velocity. Therefore, the minimum slip velocity is observed at the inlet region at top. For the low Stokes number case, localized regions of relatively lower slip velocity are still observed at near wake, indicating that the particles remain coupled to the underlying flow structures. In contrast, at higher Stokes numbers, slip velocities are increased as vertical sections over almost the entire domain, unaffected by the vortices. Also, the particles, after impact on the cylinder, travel farther upstream due to their inertia. However, they rapidly recover their gravity-driven settling velocity downstream of the cylinder, resulting in uniformly high normalized slip velocities across most of the cross-section. Consequently, the PDF is primarily governed by the gravity influence on particle velocity rather than the localized bow-shock-like region and vortices. Here, particles exhibits high slip velocities, leading to the shift of the PDF towards higher normalized slip velocities. These scatter plots reflect the combined influence of particle inertia and gravity, which enhances particle-fluid decoupling and weakens the effect of vortices on particle distribution at higher Stokes number.

Another interesting phenomenon observed is the formation of dense particle zone in the shape of bow-shock-like structure upstream to the cylinder. Previous studies have reported the formation of such structures, leading to the inhomogeneous distribution of inertial particles in the presence of unsteady wake  \citep{shi2020bow, shi2021clusters, SCHUSTER2023104487}. However, those studies focused on gravity-free cases in which particle-cylinder collisions are neglected. The present work carries out investigations while accounting for particle-cylinder collisions, assuming perfectly elastic collisions in both infinite and finite Froude conditions.

\begin{figure*}[!t]
	\begin{subfigure}[b]{1\textwidth}
	\minipage{0.45\textwidth}		
	\includegraphics[width=0.9\textwidth]{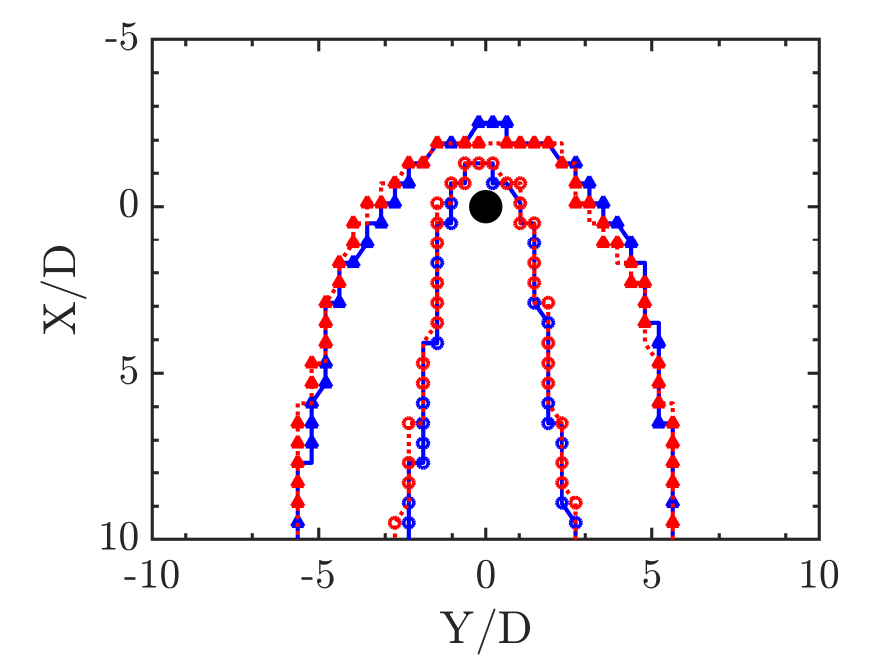}
	\caption{}
	\endminipage
	\minipage{0.45\textwidth}		
	\includegraphics[width=0.9\textwidth]{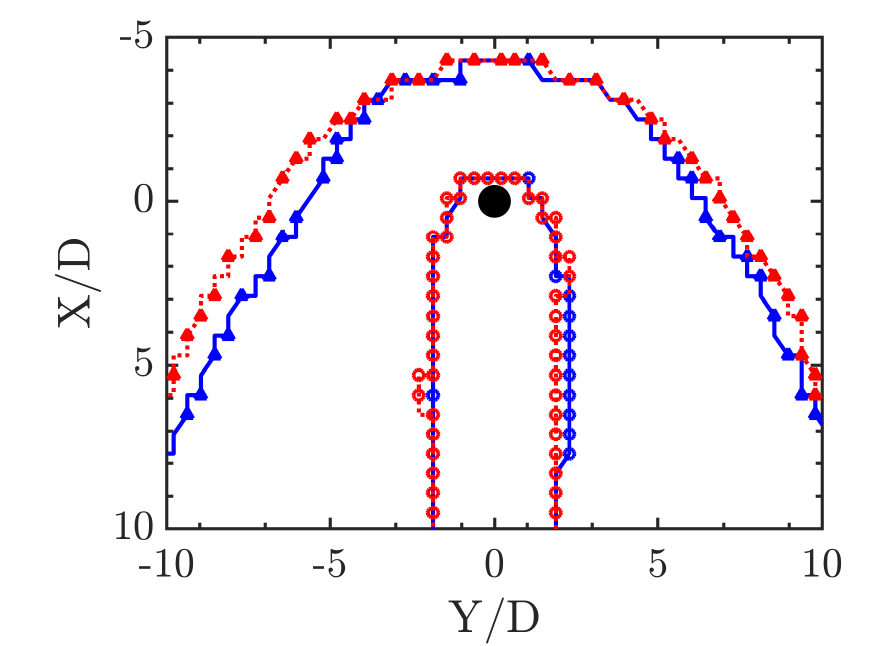}
	\caption{}
	\endminipage \\
	\minipage{0.45\textwidth}		
	\includegraphics[width=0.9\textwidth]{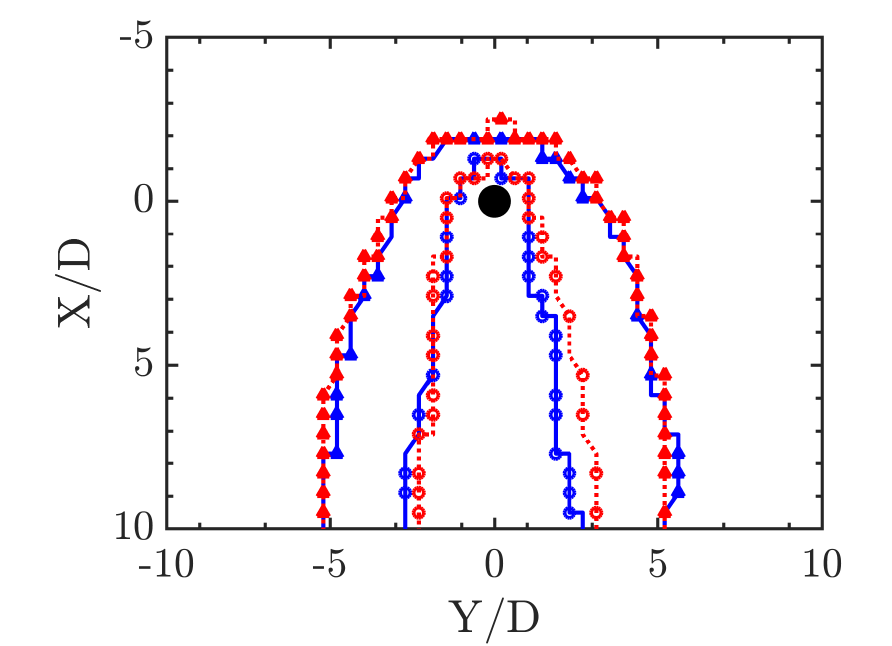}
	\caption{}
	\endminipage
	\minipage{0.45\textwidth}		
	\includegraphics[width=0.9\textwidth]{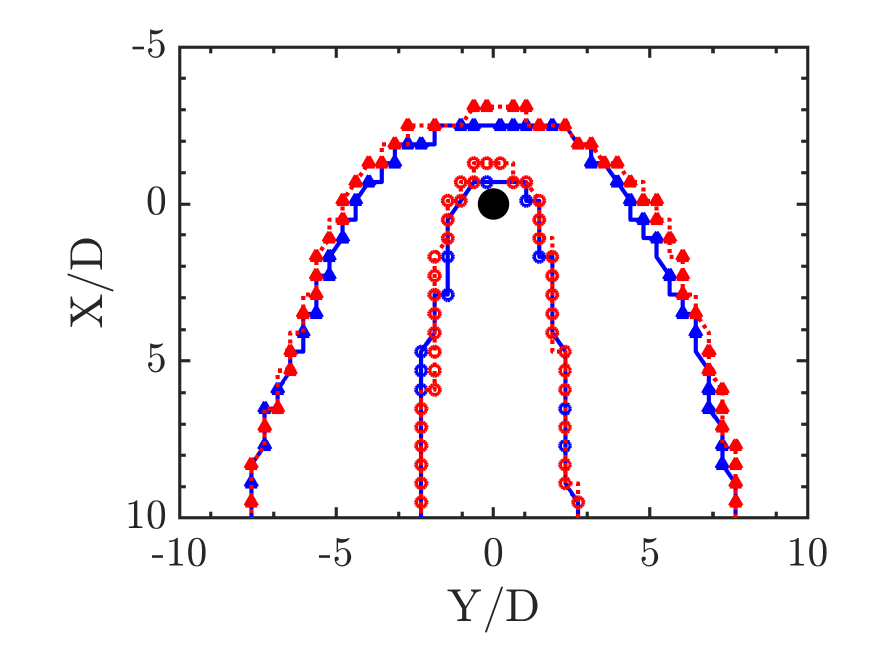}
	\caption{}
	\endminipage			
	\end{subfigure}
\caption{Visualization of the bow-shock-like structure, obtained from the time-averaged particle concentration field, is shown in the figure. Results for $Re=100$ and $200$ are shown in (a,b) and (c,d), respectively. Cases without gravity ($Fr\rightarrow\infty$) are shown in (a,c), whereas cases with gravity are shown in (b) ($Fr=1$) and (d) ($Fr=2$). Red and blue symbols denote high particle loadings ($\phi_v = 2 \times 10^{-4}$) and low particle loadings ($\phi_v = 2 \times 10^{-5}$), respectively, while circles ($\circ$) and triangles ($\triangle$) represent $St=1$ and $St=6.5$, respectively.}
\label{bow shock}
\end{figure*}

Figure~\ref{bow shock} shows the bow-shock-like structure formed upstream of the cylinder for flows with varying Reynolds numbers. In the gravity-free cases ($Fr\rightarrow\infty$), low inertia particles form bow-shock-like boundaries that remain closer to the centerline irrespective of the Reynolds number, while high inertia particles form bow shocks that extend to larger upstream heights. This indicates that they do not closely follow the curved fluid streamlines, leading to a wider bow-shock-like structure in comparison. The bow-shock boundaries for low and high particle volume fractions nearly overlap, suggesting that particle loading has a negligible influence on the bow-shock structure and is majorly driven by the impact of inertial particles on the cylinder. In the finite Froude case, particles tend to accelerate under the effect of gravity, leading to high slip velocity. The particles rebound on collision with the cylinder and follow trajectories defined by their angle of impact on the cylinder surface. Here, high inertia particles form a more pronounced and spatially extended bow-shock structure in the finite Froude case.

\begin{figure*}[!t]
    \centering

    \begin{subfigure}[b]{0.49\textwidth}
        \centering
        \includegraphics[width=0.95\linewidth,
                         trim={6cm 0cm 6cm 0cm},clip]
        {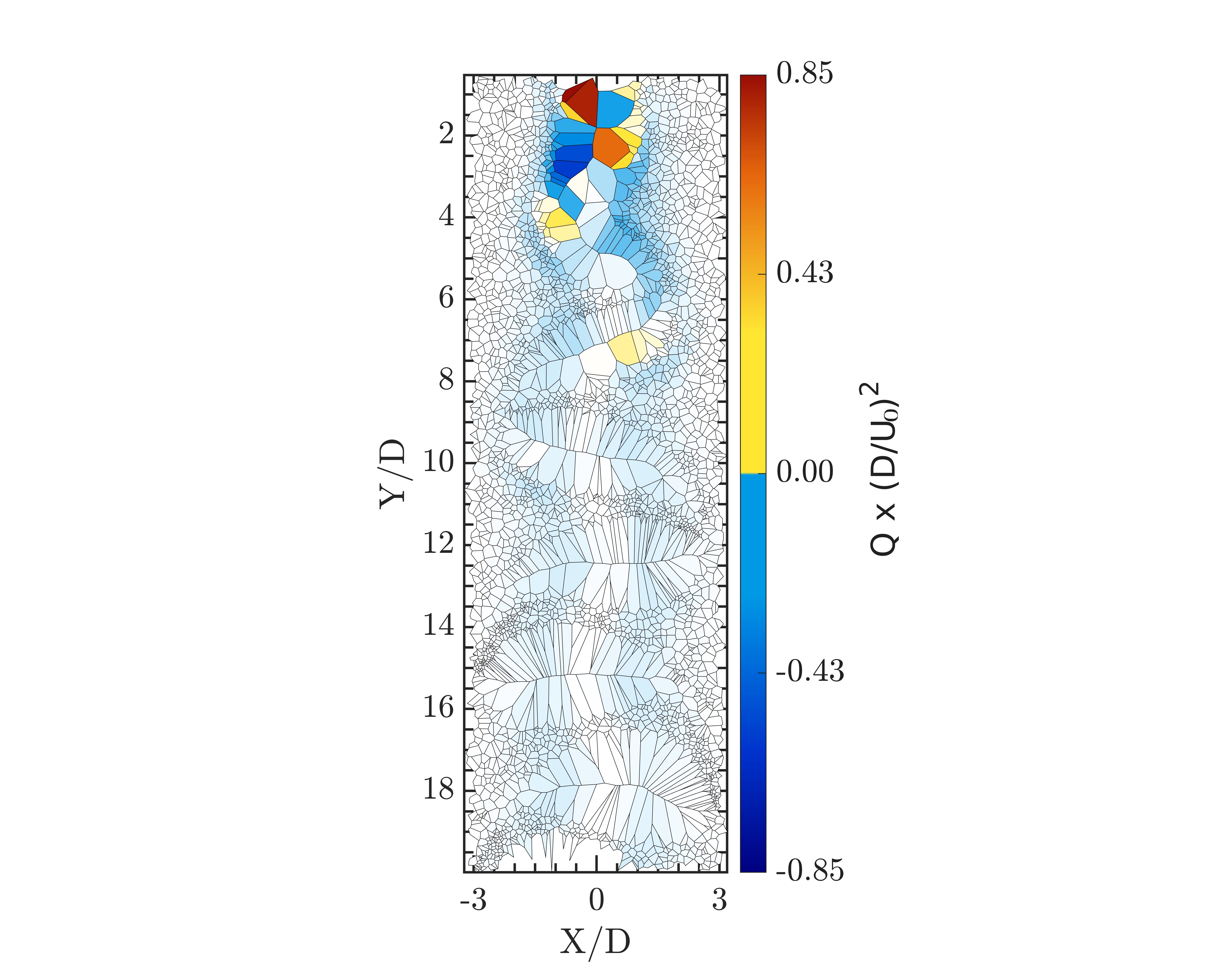}
        \caption{}
    \end{subfigure}
    \hfill
    \begin{subfigure}[b]{0.49\textwidth}
        \centering
        \includegraphics[width=0.95\linewidth,
                         trim={6cm 0cm 6cm 0cm},clip]
        {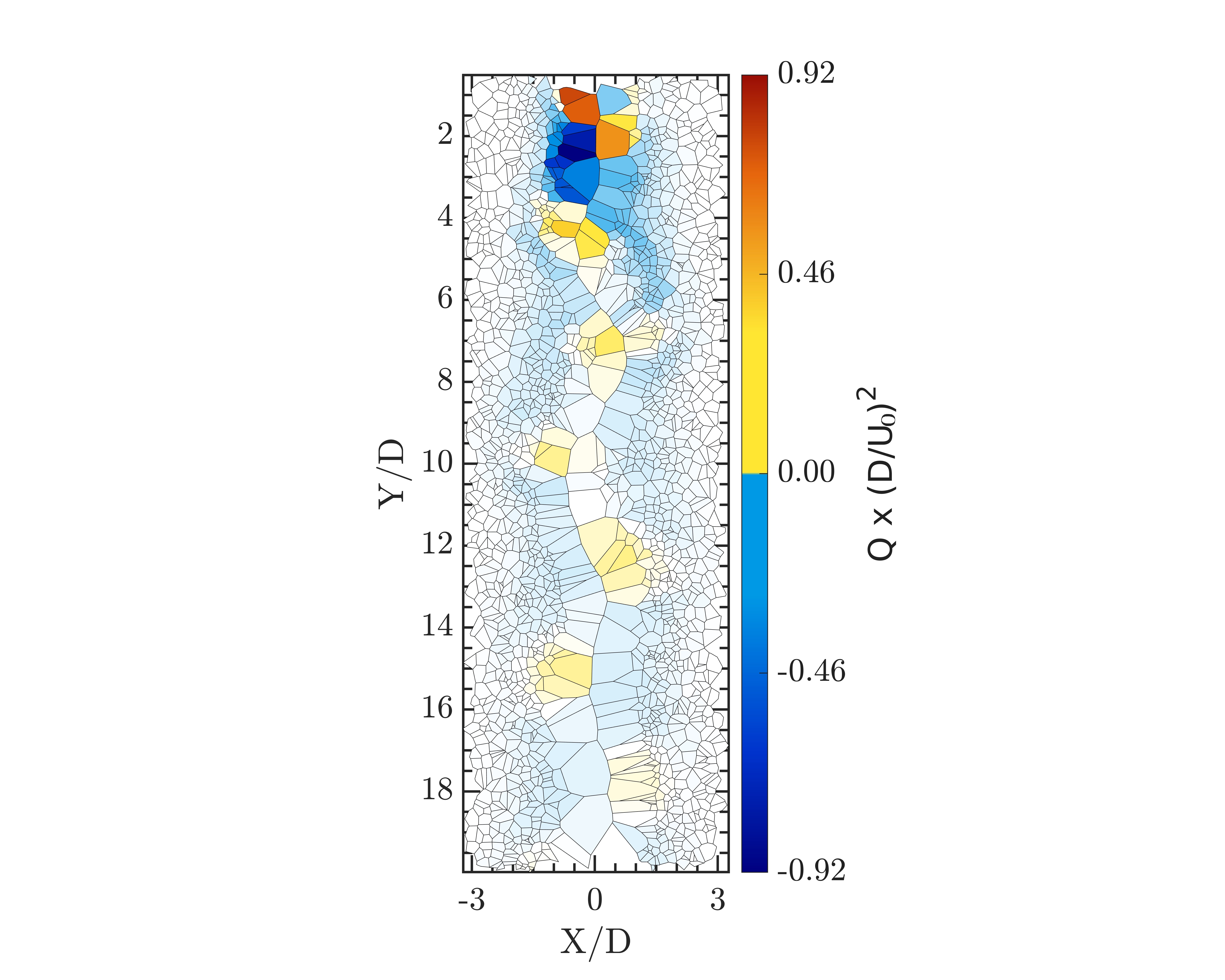}
        \caption{}
    \end{subfigure}

    \caption{Instantaneous snapshot of voronoi tessellation for low inertia particles (St$=1$) at low particle loading( $\phi_v=2\times10^{-5}$) and  $Re=100$ in case of (a) without gravity ($Fr\rightarrow\infty$) (b) with gravity ($Fr=1$).}
    \label{cylinder void leaf}
\end{figure*}

While the bow-shock-like structure characterizes the large-scale particle distribution upstream of the cylinder and the subsequent particle transport in the downstream region, voronoi tessellation is used to analyze and quantify the unsteady wake-driven particle distribution. It partitions the flow domain into discrete polygonal cells, each cell containing exactly one particle, such that any spatial point within a given cell is closer to its enclosed particle than to any other. The local particle concentration is inversely proportional to the individual Voronoi cell area, $A$. This method characterizes the particle clusters and voids formed by the coherent vortices, where small voronoi cells correspond to densely clustered particles, whereas large cells indicate void regions. Figure~\ref{cylinder void leaf} shows the voronoi cells colored by the local second invariant of the velocity-gradient tensor, $Q$, evaluated at the particle centers for the gravity-free ($Fr\rightarrow\infty$) and gravity-driven ($Fr=1$) cases. The $Q$-criterion is defined as $Q=\frac{1}{2}(\|\boldsymbol{\Omega}\|^{2}-\|\boldsymbol{S}\|^{2})$, where $\boldsymbol{\Omega}=\frac{1}{2}(\nabla\mathbf{u}-(\nabla\mathbf{u})^{T})$ and $\boldsymbol{S}=\frac{1}{2}(\nabla\mathbf{u}+(\nabla\mathbf{u})^{T})$ are the rotation-rate and strain-rate tensors, respectively. Positive values of $Q$ identify vorticity-dominated regions associated with coherent vortices, whereas negative values correspond to strain-dominated regions. The particle positions are projected onto the two-dimensional x-y plane.  Since particle dynamics are strongly influenced by vortices generated by the cylinder, subsequent analysis focuses on particles located within the region $Y/D = [-3,3]$. 

The unsteady vortex shedding governs the particle dispersion near the cylinder wake. In the gravity-free case, particles are ejected from the region of high vorticity and preferentially accumulate in the region of low vorticity and high strain, similar findings reported by \citet{eaton1994, maxey1987gravitational} for wall-bounded turbulent flows. The wake is characterized by the presence of coherent vortices, resulting in a regular, alternating pattern of particle clusters and voids downstream. In the immediate wake of the cylinder ($Y/D\lesssim4$), the flow is dominated by the formation and shedding of the vortices, leading to large velocity gradients associated with strong rotation and strain-dominated regions. The particles sample these regions where the magnitude of $Q$ is largest. Further downstream ($Y/D>4$), the particles migrate under the influence of vortices. Owing to their finite inertia, particles are progressively expelled from the rotation-dominated regions and accumulate preferentially in the surrounding strain-dominated regions. As a result, the largest Voronoi cells, corresponding to void regions, are located within the vortex cores, whereas the boundaries of these voids are populated by relatively dense particle clusters. Similar behavior is observed for both the gravity-free and gravity-driven cases, indicating that the vortices continue to govern the spatial organization of particles, while gravity primarily modifies the overall geometry of the void region. It is to be noted that most of the particles are settled in the region of $Q\sim 0$, far away from the vortex core, while the particles that are expelled from the vortex cores settle in the neighboring negative $ Q$ region, indicating a strong correlation between particle distribution and wake dynamics. 

\begin{figure}[htbp]
\centering
\includegraphics[width=0.35\textwidth]{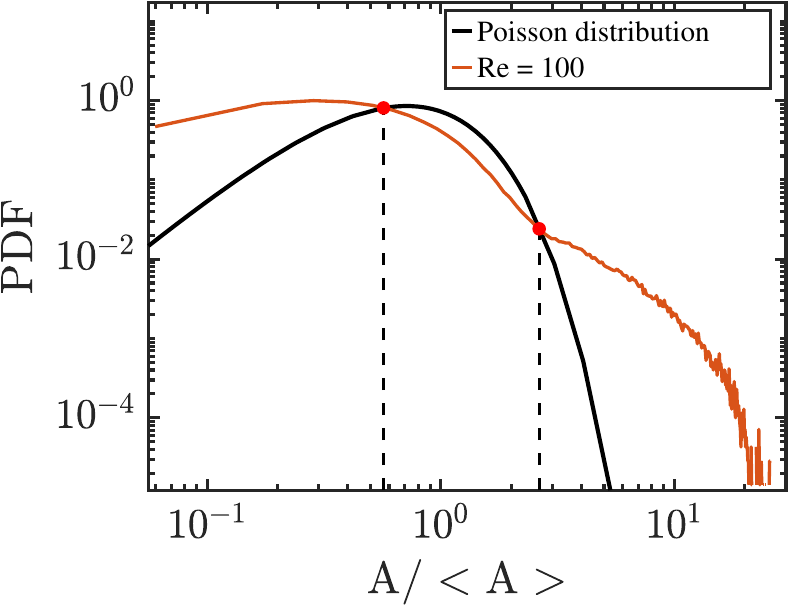}
\caption{Probability density function (PDF) of the normalized Voronoi cell area, $A/\langle A\rangle$, for particle distribution at $St=1$, $\phi_v = 2 \times 10^{-5}$ and $Re=100$ (orange), compared with the Poisson distribution corresponding to a spatially random particle field (black). The intersections between the two distributions define the clustering threshold (left) and the void threshold (right). Voronoi cells with $A/\langle A\rangle$ greater than the void threshold are identified as void cells,  whereas those with $A/\langle A\rangle$ less than the clustering threshold have high local particle concentration and are identified as clusters.}
\label{sample_pdf}
\end{figure}

Figure~\ref{sample_pdf} shows the probability density function (PDF) of the normalized Voronoi cell area, $A/\langle A\rangle$, for particle-laden flow at $St=1$, $\phi_v = 2 \times 10^{-5}$ and $Re=100$, together with the corresponding Poisson distribution representing a spatially random particle field. The intersection between the measured PDF and the Poisson distribution,  marked by the red circles, defines a threshold area, above which Voronoi cells occur more frequently than expected for a random distribution. These cells after the intersection at right  side are therefore identified as void cells and are used to characterize the particle-depleted regions. Similarly, the intersection at small normalized cell areas defines the clustering threshold, below which particles are considered to belong to preferentially concentrated clusters. The threshold values are subsequently used to identify individual void regions and quantify their morphology through measures such as the normalized void area.

The preferential sampling of particles in the high strain region results in a leaf-like structure forming at the central void region, which is aligned with the vortical structures originated from the cylinder. Each leaf-like structure is arranged at the channel center behind cylinder, as seen in Figure~\ref{cylinder void leaf}(a), comprising of large voronoi cells. Here, the individual void cells are formed by particles located at the boundary of the void structure, resulting in well-defined void-cluster patterns in the wake. It is to be noted that the void structures observed in the center are due to the expulsion of particles by the vortical structures, whereas individual void cells can occur anywhere in the domain and are dependent on flow and particle parameters. In finite Froude cases, particle dynamics are influenced not only by the vortical structures in the wake but also by gravity effect on particle momentum. As a result, more particles may enter into a region of high vorticity without getting expelled. The leaf-like void structures become less sharply defined in comparison, with reduced clustering intensity along their boundaries. As a result, the particles arrange themselves around a central wavy void zone, which closely resembles a snake-like voronoi structure, instead of individual leaf-like pattern, as can be seen in Figure~\ref{cylinder void leaf}(b). The Figure shows that particles are able to reside in a high vorticity region without being expelled, indicating that inertial particles are less sensitive to fluid flow and that their distribution is modified under the influence of gravity. The void pattern is strongly influenced by the Stokes, Froude, and Reynolds numbers, while exhibiting a weak dependence on the inlet particle loading.

The structure of the void region modifies due to particle response to local unsteady fluid fluctuations, and is dependent on particle inertia and the presence of gravity. The particle inertia is described by Stokes number St, which is the ratio of the particle relaxation time ($\tau_p = \frac{\rho_p d_p^2}{18\mu}$) and fluid time scale ($\tau_f = \frac{D}{U}$).  Effect of gravity can be described by Froude number ($Fr$), which is the ratio of inertial force to gravitational force ie $Fr^2 = \frac{U^2}{gD}$. Here a low Froude number indicates a stronger influence of gravity on particle motion. The equation of motion for a small, heavy particle in a viscous flow at low particle Reynolds number is given by
\begin{equation}
m_p \frac{d\boldsymbol{v}}{dt}
=
3\pi \mu d_p \left( \boldsymbol{U} - \boldsymbol{v} \right)
+
m_p \boldsymbol{g},
\end{equation}
where $m_p = \frac{\pi}{6}\rho_p d_p^3$ is the particle mass, $\boldsymbol{v}$ is the particle velocity, $U$ is the fluid velocity, $d_p$ is the particle diameter, $\mu$ is the dynamic viscosity, and $\rho_p$ is the particle density.  The equation can be written in non-dimensional form as 
\begin{equation}
\frac{d\boldsymbol{v}^*}{dt^*}
=
\frac{\boldsymbol{U}^* - \boldsymbol{v}^*}{\mathrm{St}}
+
\frac{1}{\mathrm{Fr}^2},
\label{eqn_nd_slip}
\end{equation}

where non-dimensional velocity and time parameters are defined in terms of inlet fluid velocity and cylinder diameter as
\begin{equation}
\boldsymbol{v}^* = \frac{\boldsymbol{v}}{U_0}, \quad
\boldsymbol{U}^* = \frac{\boldsymbol{U}}{U_0}, \quad
t^* = \frac{t U_0}{D}.
\end{equation}
In equation ~\ref{eqn_nd_slip}, first term from left represent non-dimensional particle acceleration, second term is acceleration due to fluid drag and third term represents contribution of gravity on particle acceleration.
The settling velocity relative to the flow at steady state ($d\boldsymbol{v}/dt = 0$) is  $v_t ^*= \boldsymbol{v}^* - \boldsymbol{U}^* $ . Thus, the equation reduces to
\begin{equation}
\frac{St}{Fr^2}
=
v_t^*.
\end{equation}

 Therefore, $v_t^*$ is the ratio of gravitational settling speed to the advective speed of the fluid. Here, $v_t^*$ determines the relative contribution of the interaction of the particle with the vortex and gravitational drift. It is expected it when $v_t^* << 1$, gravitational settling is negligible and particle distribution will be majorly governed by the particle-vortex interaction. On the other hand, for $v_t^* >> 1$, particle dynamics will majorly governed by gravity and their is a possibility of piercing of the vortices by the particle. Since at higher $v_t^*$, particles do not follow the curved streamline, it is expected that the pattern of the void structure will also change for different $v_t^*$, which has been qualitatively presented in Figures~\ref{Scatter_plot_vp_LSt},~\ref{Scatter_plot_vp_HSt} and~\ref{scatter relative velocity}. Also, PDFs of relative particle velocity for different values of $v_t^* (= \frac{v-u}{U_0})$ are shown in Figure~\ref{PDF of relative velocity}.
 
\begin{table}[!t]
\centering
\caption{Simulation cases for the study of varying dimensionless settling velocity($v_t^*$) at $Re=100$. Here, $St$,$Fr$,$\phi_v$ are the Stokes number, Froude number and inlet particle volume fraction, respectively.}
\begin{tabular}{c@{\hspace{10mm}}c@{\hspace{10mm}}c@{\hspace{10mm}}c}
\hline
Case & $St$ & $St/Fr^2$ & $\phi_v$ \\
\hline
A & 1   & 0   & $2\times10^{-5}$ \\
B & 6.5 & 0   & $2\times10^{-5}$ \\
C & 1   & 0.1 & $2\times10^{-5}$ \\
D & 3   & 0.3 & $2\times10^{-5}$ \\
E & 1   & 0.6 & $2\times10^{-5}$ \\
F & 6.5 & 0.6 & $2\times10^{-5}$ \\
G & 1   & 1   & $2\times10^{-5}$ \\
H & 6.5 & 6   & $2\times10^{-5}$ \\
I & 1   & 0   & $2\times10^{-4}$ \\
J & 6.5 & 0   & $2\times10^{-4}$ \\
K & 1   & 0.1 & $2\times10^{-4}$ \\
L & 6.5 & 0.6 & $2\times10^{-4}$ \\
M & 1   & 1   & $2\times10^{-4}$ \\
N & 6.5 & 6   & $2\times10^{-4}$ \\
\hline
\end{tabular}
\label{St_Fr2_cases}
\end{table}

 To quantify the change in particle distribution, Voronoi tessellation analysis is performed for the cases listed in Table ~\ref{St_Fr2_cases}. The probability density functions (PDFs) of the normalized Voronoi cell area corresponding to void cells are compared for different values of $v_t^*=St/Fr^2$ at $Re=100$.   In the present study, variations in $v_t^*$ are carried out by changing both particle density and gravitational acceleration, allowing the study on the combined influence of particle inertia and gravity on void statistics. 

\begin{figure*}[!t]
\centering

% Row 1
\begin{subfigure}[t]{0.23\textwidth}
\centering
\includegraphics[height=6cm,keepaspectratio]{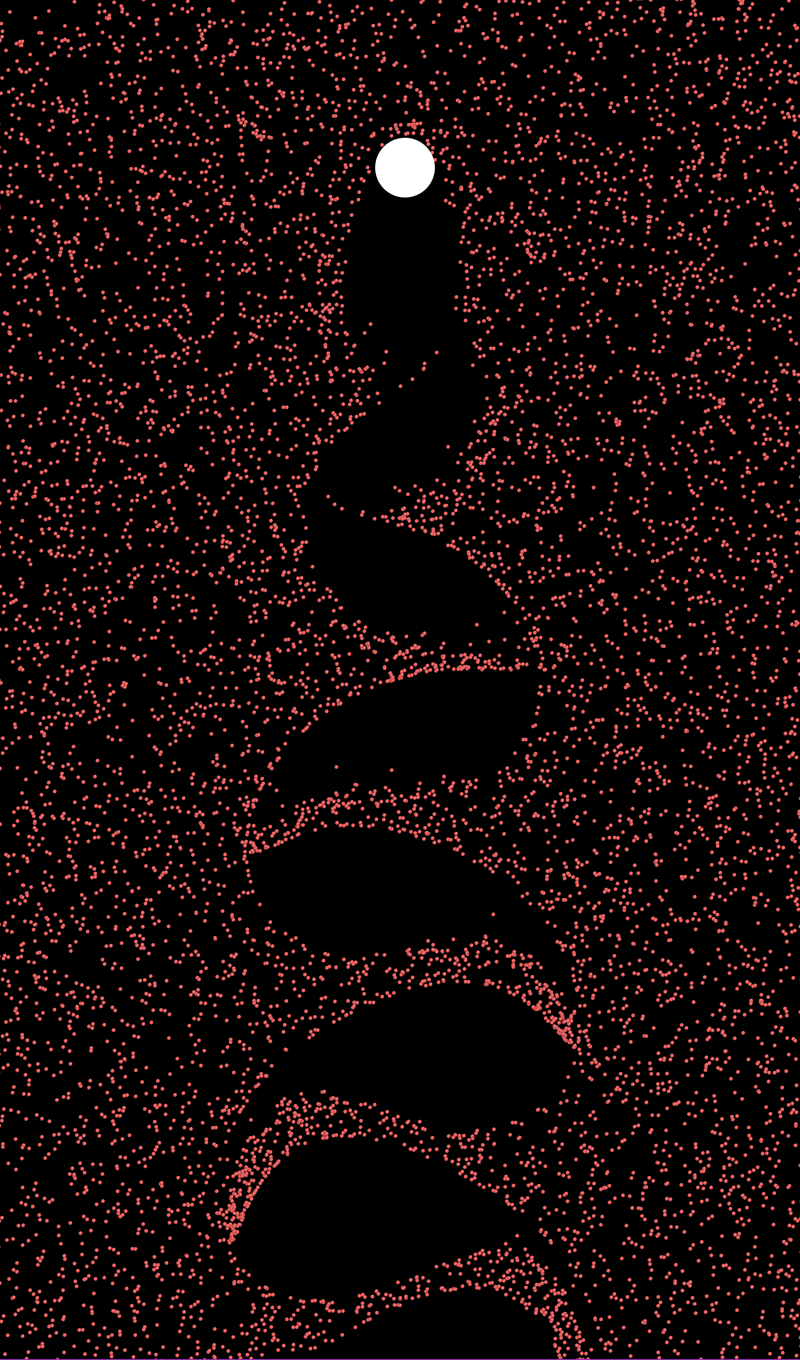}
\caption{}
\end{subfigure}\hfill
\begin{subfigure}[t]{0.23\textwidth}
\centering
\includegraphics[height=6cm,keepaspectratio]{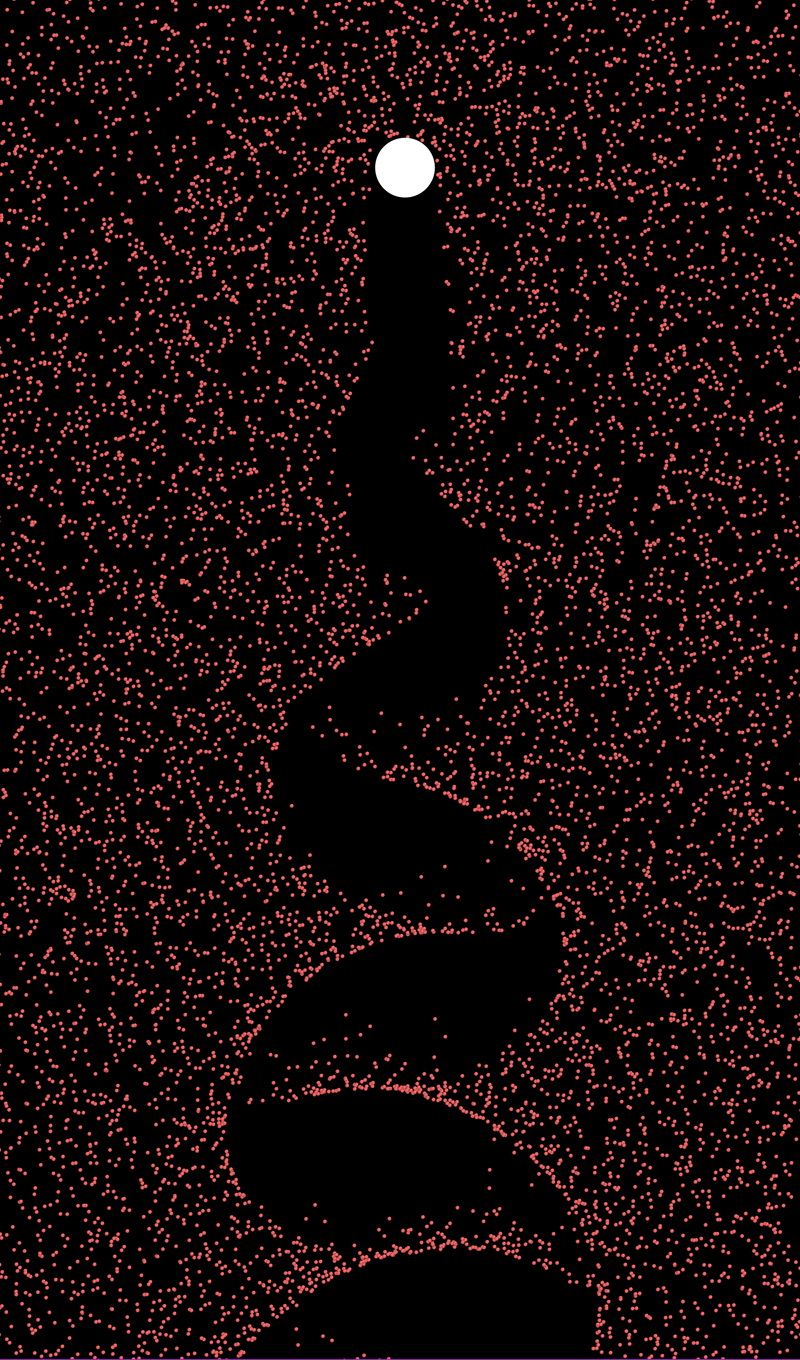}
\caption{}
\end{subfigure}\hfill
\begin{subfigure}[t]{0.23\textwidth}
\centering
\includegraphics[height=6cm,keepaspectratio]{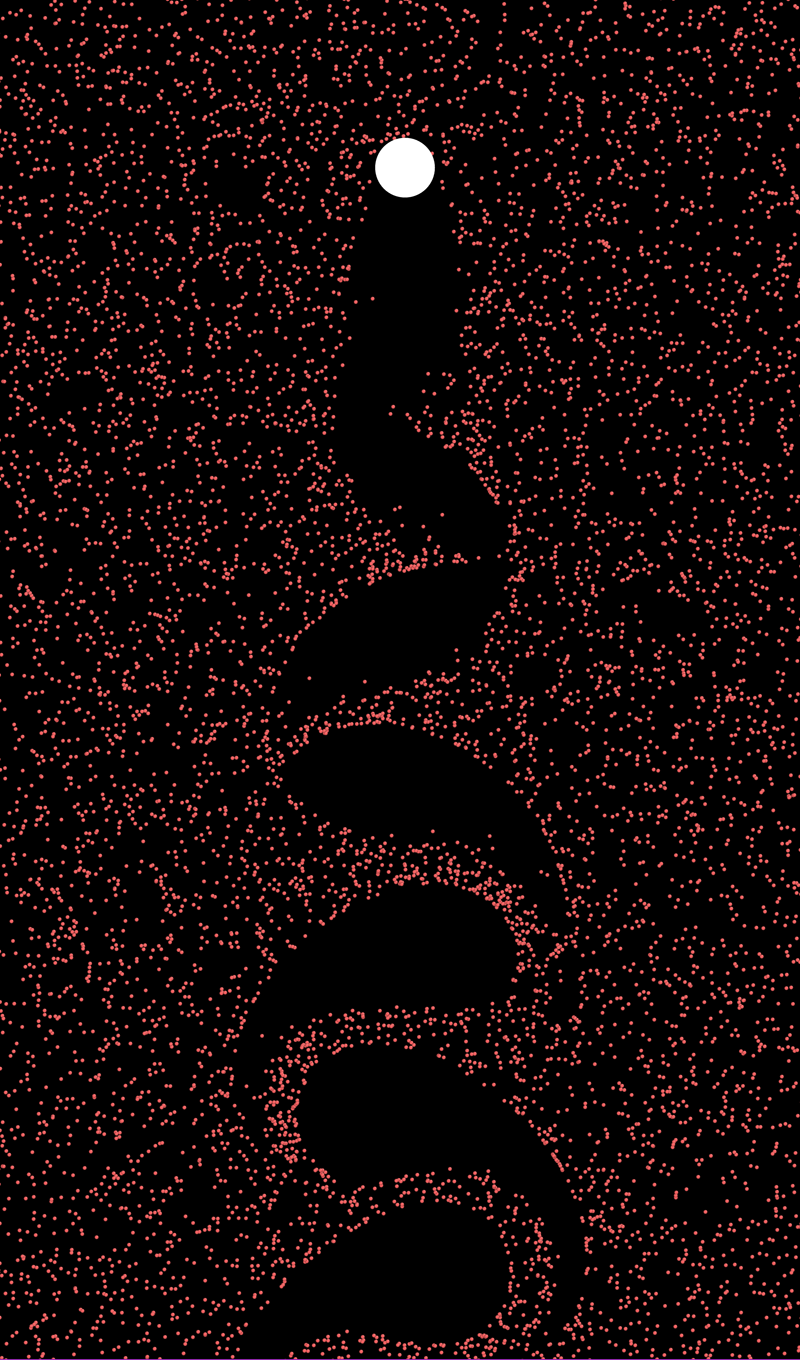}
\caption{}
\end{subfigure}\hfill
\begin{subfigure}[t]{0.23\textwidth}
\centering
\includegraphics[height=6cm,keepaspectratio]{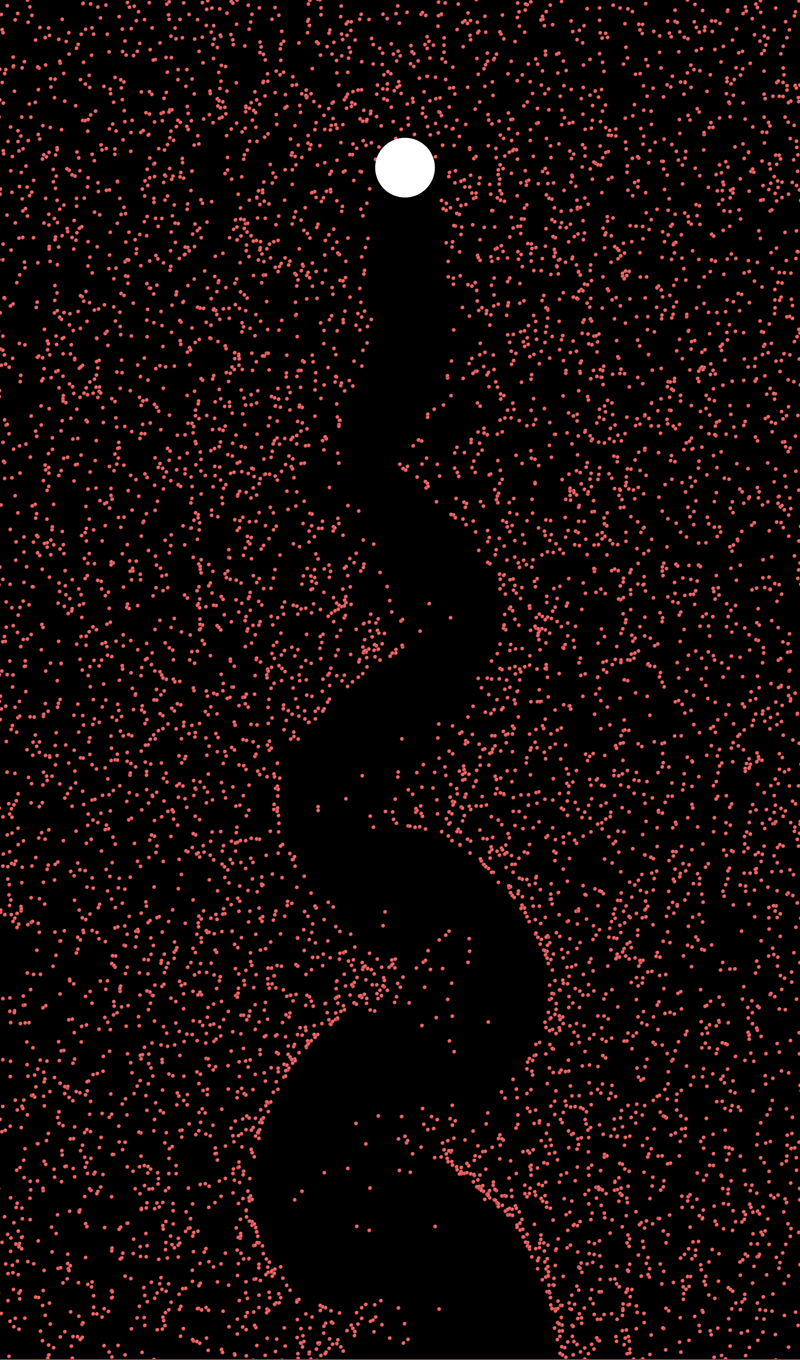}
\caption{}
\end{subfigure}

\vspace{0.8cm}

% Row 2
\begin{subfigure}[t]{0.23\textwidth}
\centering
\includegraphics[height=6cm,keepaspectratio]{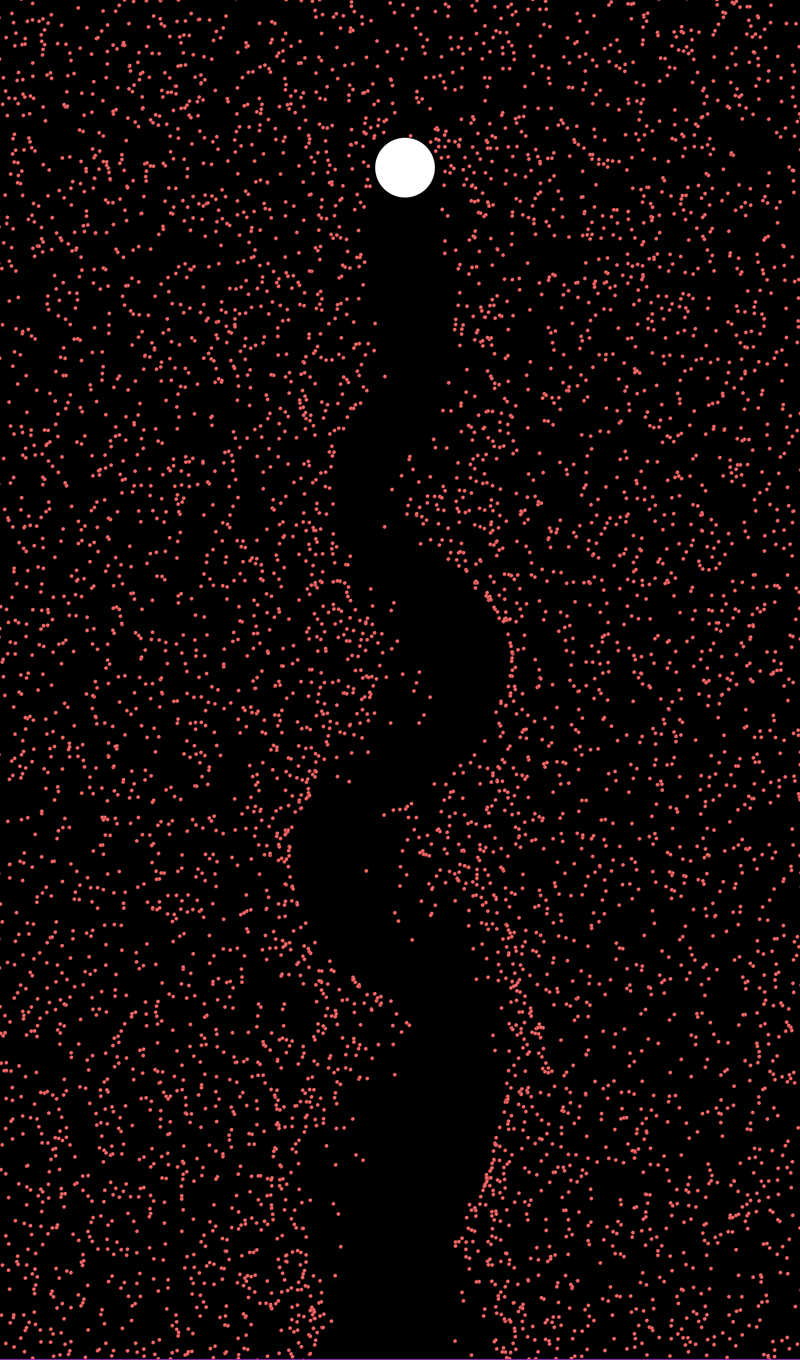}
\caption{}
\end{subfigure}\hfill
\begin{subfigure}[t]{0.23\textwidth}
\centering
\includegraphics[height=6cm,keepaspectratio]{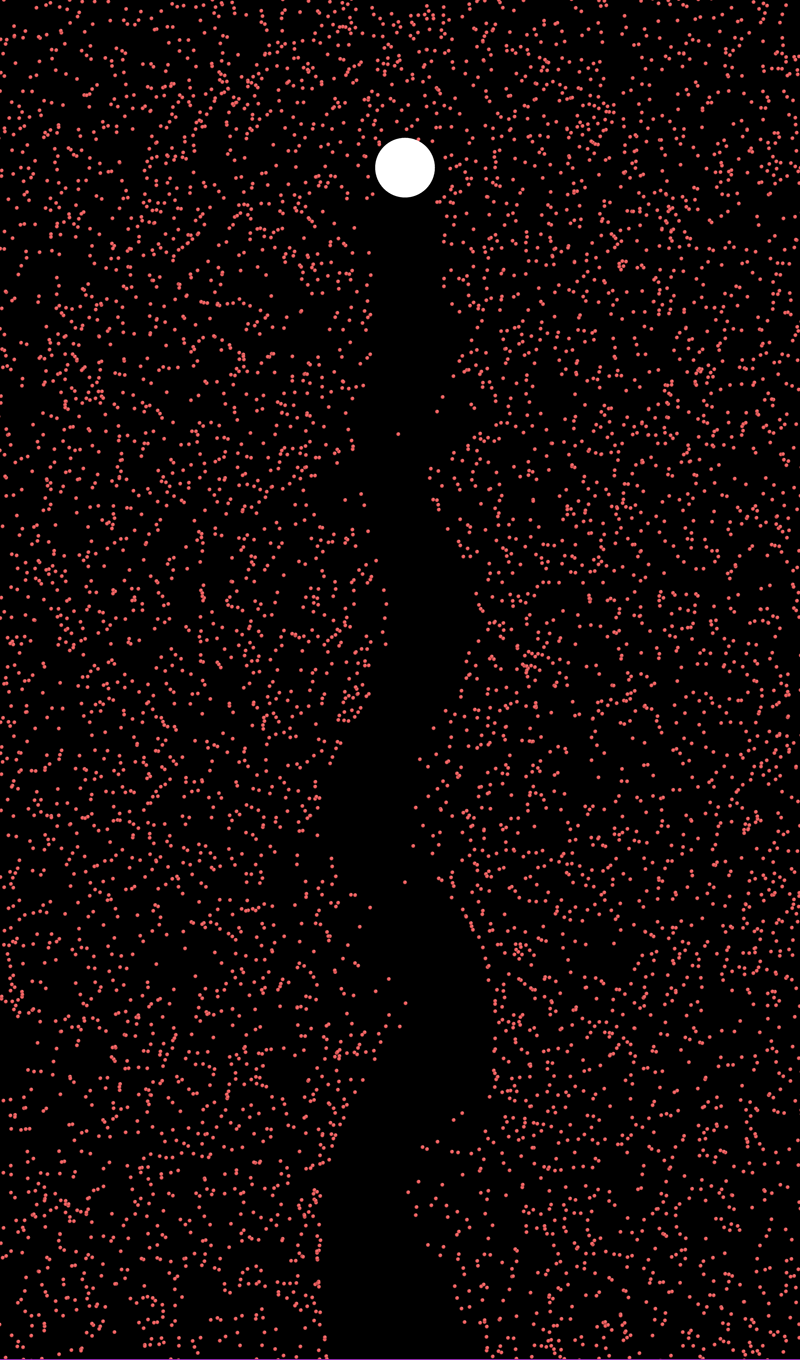}
\caption{}
\end{subfigure}\hfill
\begin{subfigure}[t]{0.23\textwidth}
\centering
\includegraphics[height=6cm,keepaspectratio]{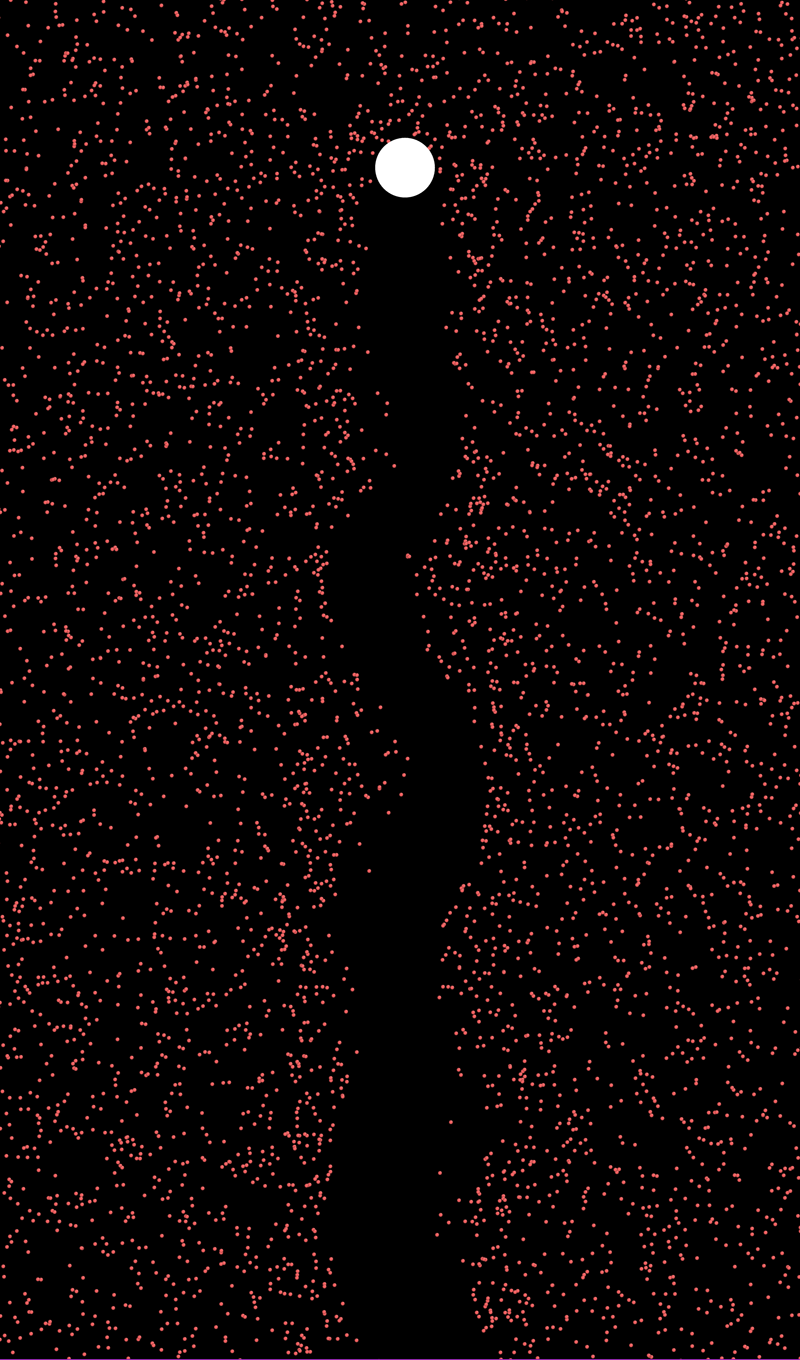}
\caption{}
\end{subfigure}\hfill
\begin{subfigure}[t]{0.23\textwidth}
\centering
\includegraphics[height=6cm,keepaspectratio]{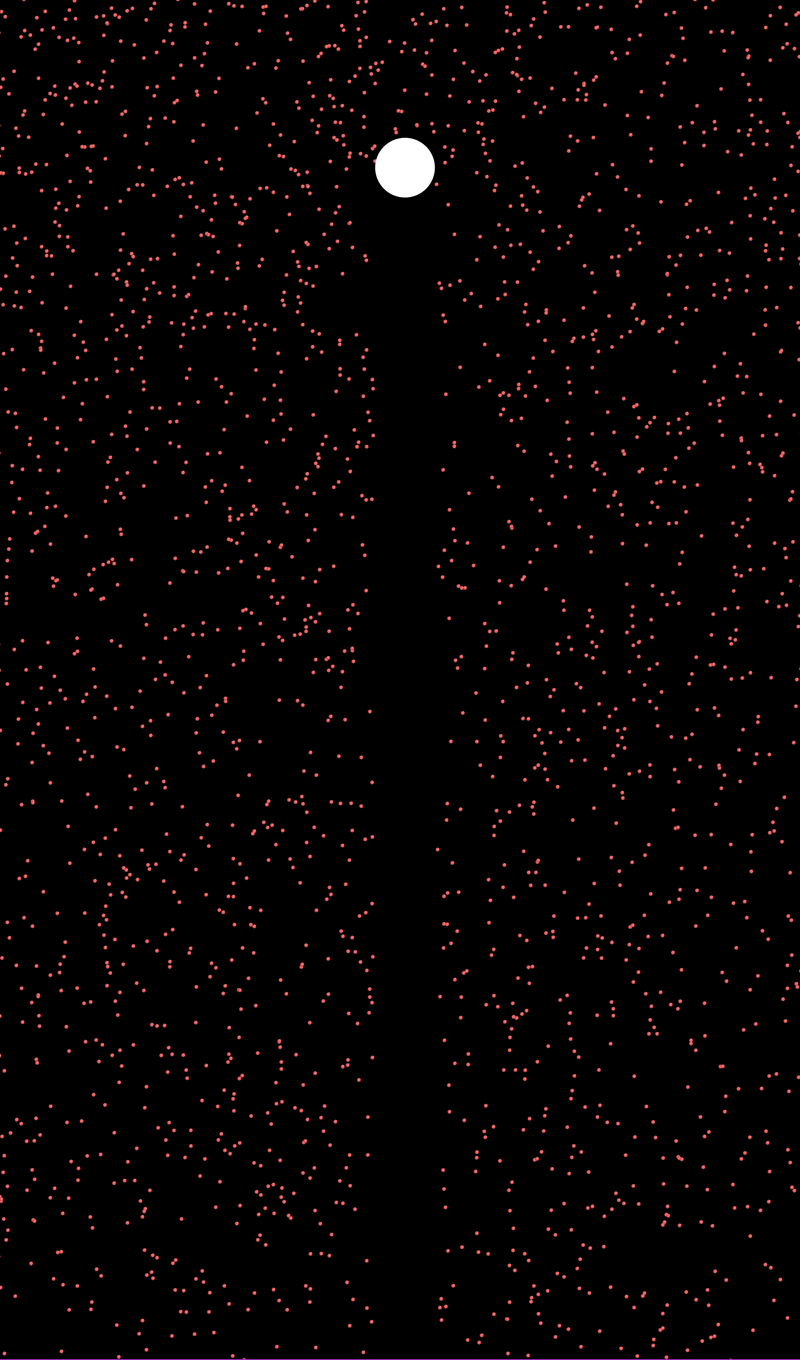}
\caption{}
\end{subfigure}

\caption{Instantaneous snapshots of particle-laden flow over a circular cylinder at $Re=100$ and low particle loading ($\phi_v=2\times10^{-5}$) for cases (a) $St=1$, $St/Fr^2=0$, (b) $St=6.5$, $St/Fr^2=0$, (c) $St=1$, $St/Fr^2=0.1$, (d) $St=3$, $St/Fr^2=0.3$, (e) $St=1$, $St/Fr^2=0.6$, (f) $St=6.5$, $St/Fr^2=0.6$, (g) $St=1$, $St/Fr^2=1$ and (h) $St=6.5$, $St/Fr^2=6$. The white circle denotes the cylinder. The flow is from top to bottom, with the top and bottom boundaries representing the inlet and outlet, respectively.}
\label{ovito_lvf}
\end{figure*}

 Figure~\ref{ovito_lvf} shows the instantaneous snapshot of particles for increasing values of the dimensionless settling velocity, $v_t^*=St/Fr^2$ at $Re=100$ and low particle loading of $\phi_v=2\times10^{-5}$. The cylinder is represented by the large white circle, while the red markers denote the instantaneous particle positions. At $St/Fr^2=0$ and $St=1$, the particle-void region follows the oscillatory wake generated by vortex shedding, resulting in a leaf-like void pattern, as shown in Figure~\ref{ovito_lvf}(a). As the dimensionless settling velocity increases to $St/Fr^2=0.1$ (Figure~\ref{ovito_lvf}(c)), the leaf-like void pattern begins to align with the streamwise direction, resulting in the tilting of the individual void structure. The tilt reflects the increasing influence of gravitational settling of particles on particle motion in comparison to the streamwise displacement caused by the vortices. A further increase to $St/Fr^2=0.6$ (Figure~\ref{ovito_lvf}(e)) suppresses the vortex-induced oscillatory void pattern more significantly, resulting in a narrower and more vertically aligned void region. As gravitational settling increases the streamwise particle velocity, particles travels through the wake region more rapidly and experience less cross-stream deflection by the vortices. As a result, the leaf-like void pattern transitions into a snake-like void structure. At $St/Fr^2=1$ (Figure~\ref{ovito_lvf}(g)), the large gravitational settling velocity causes the particles to descend in the vertical direction with less deflection, producing a nearly straight void region. Here, the gravitational settling dominates the particle motion and substantially weakens the influence of the wake vortices. Figure~\ref{ovito_lvf} (d) presents the intermediate case at $St=3$ and $St/Fr^2=0.3$, where the leaf-like void pattern merges in the streamwise direction to form a single snake-like void pattern. Unlike the $St=1$ case, where increasing gravitational settling primarily distorts the leaf-like void structures, the larger particle inertia at $St=3$  and the same Froude condition as Figure~\ref{ovito_lvf} (c) promotes the merging of adjacent void structures. This demonstrates that increasing particle inertia accelerates the transition from a leaf-like to a snake-like void pattern. At $St/Fr^2=0$, the particle respond less to the underlying fluid fluctuations with the increase in particle inertia from $St=1$ to $St=6.5$, allowing particles to penetrate the void region. As a result, the boundaries of the leaf-like void  become less distinct and the spacing between successive void structures is reduced as shown in Figure~\ref{ovito_lvf}(b). The void region transitions from leaf-like pattern to snake-like structure with increase in $St/Fr^2$, similar to $St=1$ case. However, the void region becomes narrower for the higher Stokes number case (Figure~\ref{ovito_lvf}(f)), as the larger particle inertia and increase in gravitational settling reduces the cross-stream particle migration caused by the vortices. At $St/Fr^2=6$ (Figure~\ref{ovito_lvf}(h)), the particles travels in the vertical direction with minimal cross-stream displacement, resulting in a straight void region downstream to the cylinder. This is due to the particle motion dominated by the combined effects of particle inertia and gravitational settling, with a weaker influence of the vortices on the particle distribution.

\begin{figure*}[!t]
\centering

% Row 1
\begin{subfigure}[t]{0.28\textwidth}
\centering
\includegraphics[height=6cm,keepaspectratio]{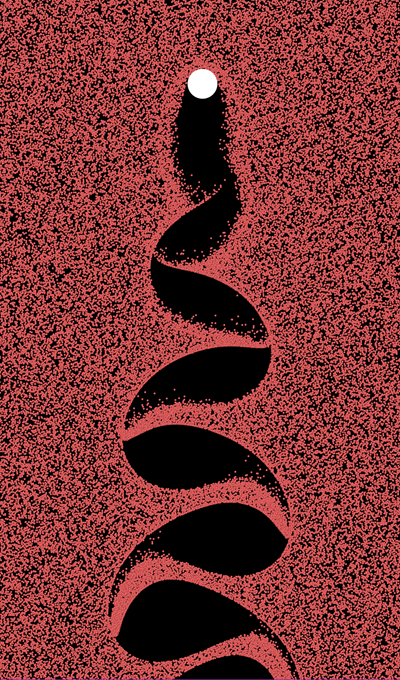}
\caption{}
\end{subfigure}\hfill
\begin{subfigure}[t]{0.28\textwidth}
\centering
\includegraphics[height=6cm,keepaspectratio]{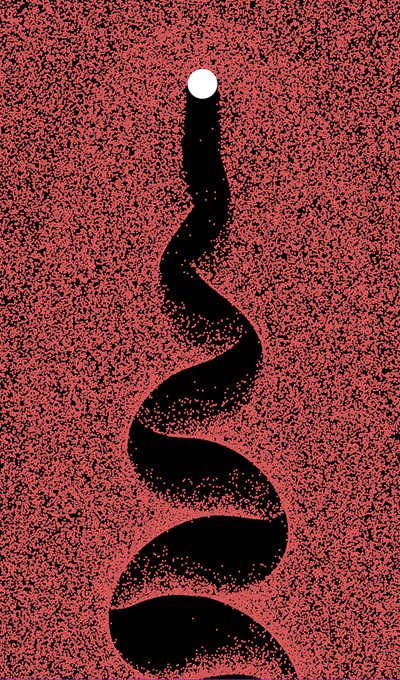}
\caption{}
\end{subfigure}\hfill
\begin{subfigure}[t]{0.28\textwidth}
\centering
\includegraphics[height=6cm,keepaspectratio]{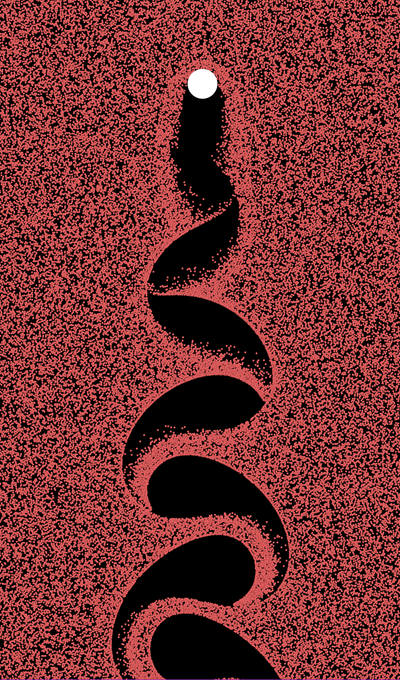}
\caption{}
\end{subfigure}

\vspace{0.8cm}

% Row 2
\begin{subfigure}[t]{0.28\textwidth}
\centering
\includegraphics[height=6cm,keepaspectratio]{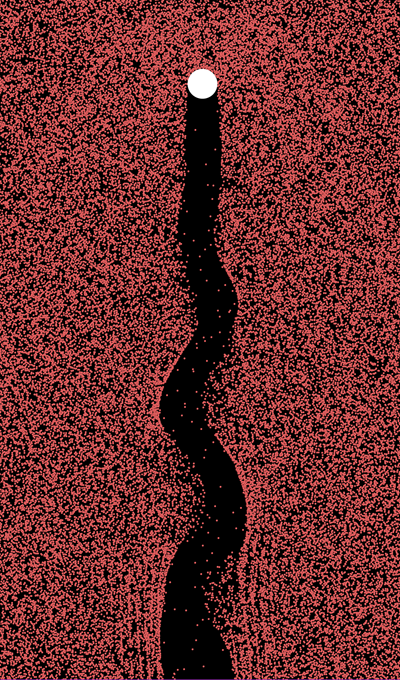}
\caption{}
\end{subfigure}\hfill
\begin{subfigure}[t]{0.28\textwidth}
\centering
\includegraphics[height=6cm,keepaspectratio]{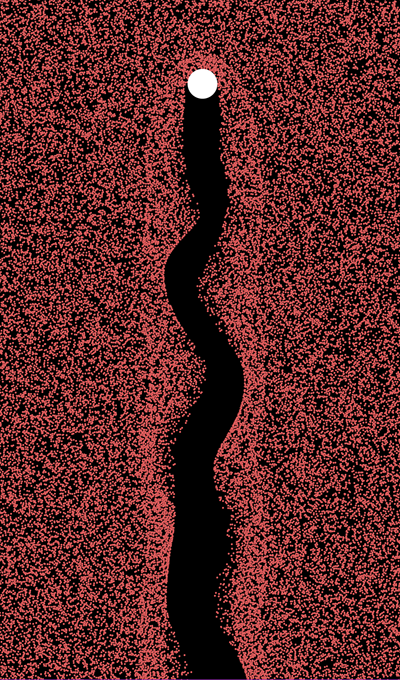}
\caption{}
\end{subfigure}\hfill
\begin{subfigure}[t]{0.28\textwidth}
\centering
\includegraphics[height=6cm,keepaspectratio]{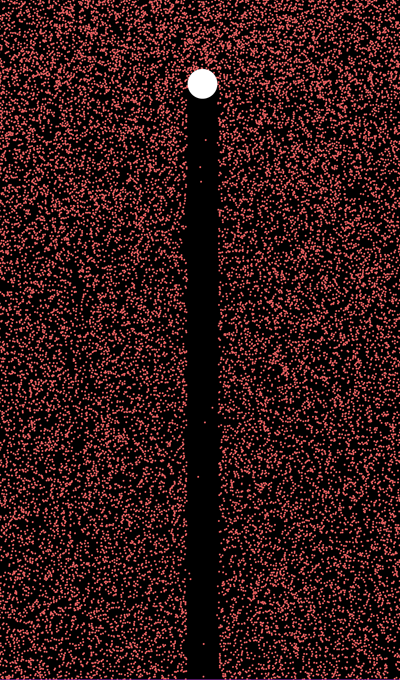}
\caption{}
\end{subfigure}

\caption{Instantaneous snapshots of particle-laden flow over a circular cylinder at $Re=100$ and high particle loading ($\phi_v=2\times10^{-4}$) for cases (a) $St=1$, $St/Fr^2=0$, (b) $St=6.5$, $St/Fr^2=0$, (c) $St=1$, $St/Fr^2=0.1$, (d) $St=6.5$, $St/Fr^2=0.6$, (e) $St=1$, $St/Fr^2=1$, and (f) $St=6.5$, $St/Fr^2=6$. The white circle denotes the cylinder. The flow is from top to bottom, with the top and bottom boundaries representing the inlet and outlet, respectively.}
\label{ovito_hvf}
\end{figure*}

Figure~\ref{ovito_hvf} shows the instantaneous particle snapshots for increasing values of the dimensionless settling velocity, $v_t^*=St/Fr^2$, at $Re=100$ and a high particle loading of $\phi_v=2\times10^{-4}$. The particle-void structures are qualitatively similar to those observed for the low loading case (Figure~\ref{ovito_lvf}). For both $St=1$ and $St=6.5$, the particle-void region evolves from a leaf-like pattern at $St/Fr^2=0$ to a snake-like structure with increasing $St/Fr^2$, and eventually becomes nearly straight when gravitational settling dominates the particle motion. The close agreement between the low particle load and high particle load cases indicates that particle loading has a negligible influence on the void structure under one-way coupling, with the particle distribution governed primarily by particle inertia and gravitational settling.

\begin{figure*}[!t]
    \centering

    \begin{subfigure}{0.45\textwidth}
        \centering
        \includegraphics[width=0.9\linewidth]{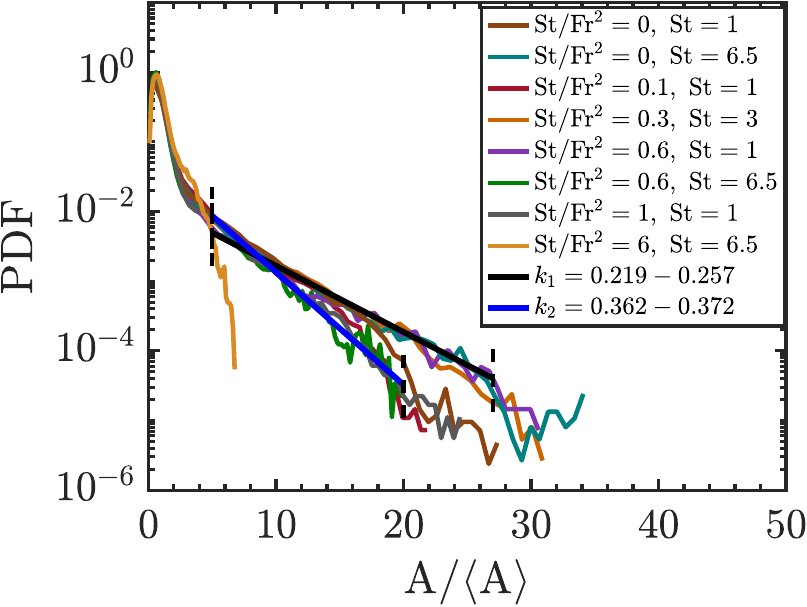}
        \caption{}
    \end{subfigure}
    \hfill
    \begin{subfigure}{0.45\textwidth}
        \centering
        \includegraphics[
            width=0.9\linewidth]{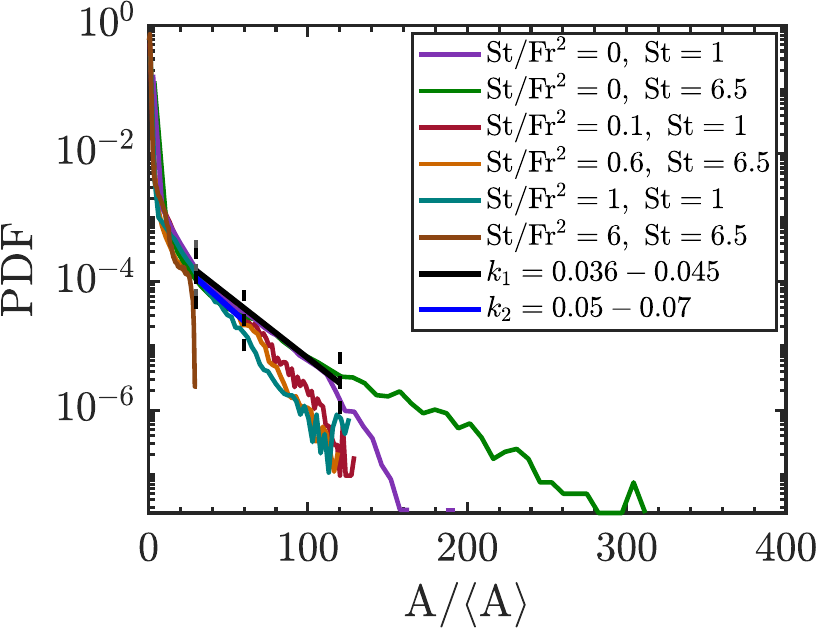}
        \caption{}
    \end{subfigure}

   \caption{Probability density function (PDF) of normalized voronoi cell area 
($A/\langle A \rangle$) for varying $St/Fr^{2}$ at (a) low particle loading ($\phi_v=2\times10^{-5}$) and (b) high  particle loading ($\phi_v=2\times10^{-4}$). 
The solid black and blue lines denote exponential
fits to the tails of the PDFs with decay constants $k_1$ and $k_2$,
respectively. The vertical dashed lines indicate the range of normalized
Voronoi cell areas over which the exponential fits are performed.
}
    \label{pdf_exponential_decay}
\end{figure*}

Although the instantaneous void structures remain qualitatively similar for both particle loadings (Figures~\ref{ovito_lvf} and \ref{ovito_hvf}), the PDF of the normalized voronoi cell area exhibits noticeable differences, as shown in Figure~\ref{pdf_exponential_decay}. Here, we focus on the tail of the PDF, which captures the statistics of large, normalized voronoi cells associated with voids. It is observed that the PDF at low particle loading exhibits a relatively narrow spread and the distribution develops a broader tail at high particle loading, indicating an increased probability of formation of void cells with large normalized voronoi cell areas.

At low particle loading, PDF collapses reasonably well for cases A-G over the intermediate normalized cell-area range $(A/\langle A \rangle \leq 10)$ and exhibits a similar decay. However, as void structure transitions with increasing $\mathrm{St}/\mathrm{Fr}^2$, the tail end of PDF changes due to change in shape of larger voronoi void cells existing in these void patterns. For $St=1$, the PDF is showing decay of $k_2$ at $\mathrm{St}/\mathrm{Fr}^2 =0$, where leaf-like void pattern is clearly visible. As $\mathrm{St}/\mathrm{Fr}^2$ increases to 0.1, only a  slight streamwise elongation  of the leaf-like voids is observed, and the tail of the PDF remains largely unchanged. With a further increase in $\mathrm{St}/\mathrm{Fr}^2$ to 0.6, the leaf-like void structure merge to form an elongated snake-like structure, with  a reduced probability of extremely larger normalized voronoi cell areas (Figure~\ref{pdf_exponential_decay}(a)) . As a result, the PDF tail broadens with a slower exponential decay $k_1$. As the void structure transitions to almost straight vertical region at $\mathrm{St}/\mathrm{Fr}^2=1$, the occurrence of extremely large voronoi void cells is substantially reduced. The PDF tail becomes shorter in this case, with its tail showing decay with exponential decay constant $k_2$. For $St=3$ and $St/Fr^2=0.3$, the PDF is characterized by the exponential decay, $k_1$, where the particle-void region has already transitioned to a snake-like structure, similar to the $St=1$ and $St/Fr^2=0.6$ case. For $St=6.5$ and $St/Fr^2=0$,  the leaf-like void pattern has blurred boundaries with many particle penetrating the void structure, resulting in partially merged void pattern. The PDF for this case shows similar decay to that of transition snake-like void pattern, with a decay constant of $k_1$. With a further increase in $\mathrm{St}/\mathrm{Fr}^2$ to 0.6, the decay constant increases to k2 as void pattern changes to almost vertical snake-like structure, similar to the $St=1$ and $St/Fr^2=1$ case. As the particle-void structure collapses into a nearly straight vertical region at $St/Fr^2=6$, the variation in the sizes of the large voronoi void cells is substantially reduced. Consequently, the PDF develops a shorter tail with a much steeper exponential decay. 

For the high particle loading cases shown in Figure~\ref{pdf_exponential_decay}(b), the PDF of the normalized voronoi cell area exhibits a broader range, extending up to$(A/\langle A \rangle \leq 300)$ compared to the low particle loading cases. The PDF collapses reasonably well for cases I-M over the lower normalized cell-area range $(A/\langle A \rangle \leq 30)$. However, as the void structures transition with varying $St/Fr^2$, the tail of the PDF shifts significantly to reflect the transition of the void region. For $St/Fr^2=0$, both $St=1$ and $St=6.5$ case exhibits a leaf-like particle-void structure and the PDF tail is well described by the  exponential decay, $k_1$, over more than one order of magnitude of the PDF. The extreme end of the PDF varies as particle penetrate void structure at higher inertia, leading to a wider distribution of large voronoi cell areas. PDF decay exponentially at much faster rate ($k_2$) with increase in $St/Fr^2$, as void structure transitions to snake-like structure. As the void structure become narrow, vertical region at $St/Fr^2=6$, the tail end of PDF is short with steep decay, similar to low particle loading case for  $St/Fr^2=6$. Therefore, the dimensionless settling velocity, $St/Fr^2$, governs the spatial distribution of particles by quantifying the balance between gravitational settling and vortex-induced particle migration.

\begin{figure*}[!t]
    \centering

    \begin{subfigure}{0.45\textwidth}
        \centering
        \includegraphics[width=0.9\linewidth]{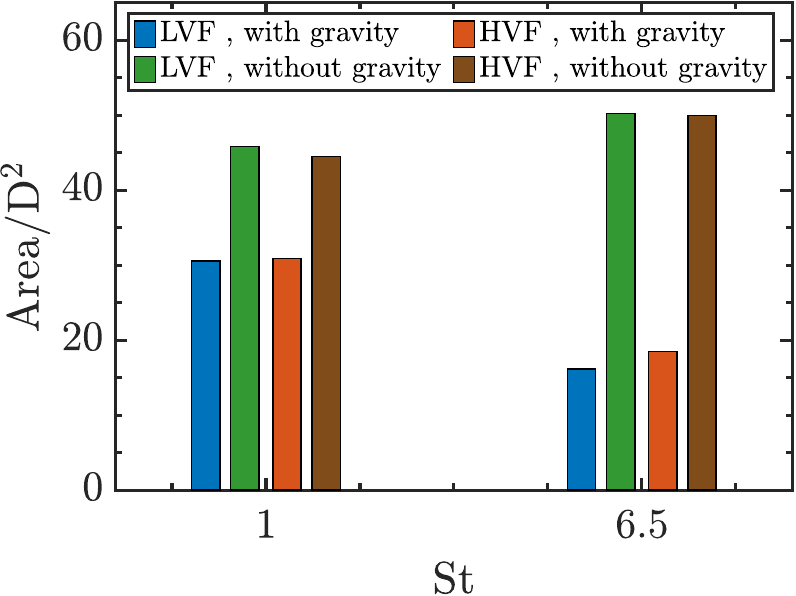}
        \caption{}
    \end{subfigure}
    \hfill
    \begin{subfigure}{0.45\textwidth}
        \centering
        \includegraphics[
            width=0.9\linewidth]{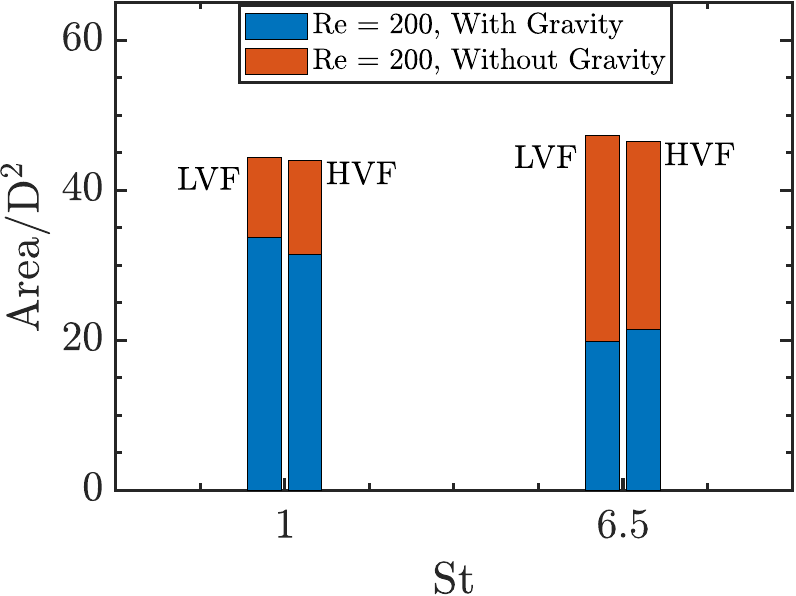}
        \caption{}
    \end{subfigure}

   \caption{Variation of the normalized void area, $Area/D^{2}$, in the wake of a circular cylinder for particles with $St=1$ and $6.5$ at (a) $Re=100$ and (b) $Re=200$. Results are shown for low particle loadings ($\phi_v=2\times10^{-5}$) and high particle loadings ($\phi_v=2\times10^{-4}$) for cases with gravity (Finite Froude number) and without gravity ($Fr\rightarrow\infty$) conditions. The void area is evaluated over a measurement window of size ($L_x \times L_y \sim 25D \times 20D$), where $D$ denotes the cylinder diameter.}
    \label{void_Area_total}
\end{figure*}

The clustering and void statistics observed in the PDF reflect the influence of the vortices on particle distribution in the flow. As seen in Figure~\ref{cylinder void leaf} , smaller voronoi cells are typically located surrounding larger void cells, indicating local accumulation of particles driven by the local vorticity field. This behavior is consistent with preferential accumulation. In addition to these local effects, a distinct large-scale void region is observed along the wake centerline, as discussed in Figures~\ref{ovito_lvf} and \ref{ovito_hvf}. Figure~\ref{void_Area_total} shows the variation of the normalized area of void region, $Area/D^2$, downstream to the cylinder for at $Re = 100$ and $200$, for finite and infinite Froude conditions. 
The most prominent feature across both Reynolds numbers is the stark contrast between the finite Froude number and $Fr\rightarrow\infty$ conditions. In the absence of gravity, represented by the green and brown bars, the normalized void area remains consistently large ($Area/D^2 \approx 45-50$) regardless of the Reynolds number or inlet particle volume fraction and slight increase with increase in Stokes number. Under these conditions, the particle distribution is majorly governed by the oscillatory wake dynamics. The vortices deflect particles in the cross-stream direction, producing well-defined leaf-like void regions that occupy a substantial portion of the channel. On the introduction of gravity, represented by the blue and orange bars, the total void area decreases sharply. This reduction is directly dependent on the dimensionless settling velocity, $ v_t^* = \frac{St}{Fr^2}$. When the gravity effect is present, particles accelerates in downward streamwise direction with gravity, so that they are able to penetrate vortices without deflection, causing the leaf-like structures to narrow, elongate, and ultimately collapse. The significance of particle inertia and gravity becomes highly evident when examining the transition from $St = 1$ to $St = 6.5$ under the finite Froude condition. At $St=1$, the relatively small settling velocity allows the vortices to form moderately wide void regions, yielding $Area/D^2 \approx30$ at both Reynolds number. Increasing the particle inertia to $St=6.5$ increases the dimensionless settling velocity, leading to formation of narrow void region, with the normalized void area to approximately $16-20$. In contrast, the particle volume fraction has only a minor influence on the void area. For both Reynolds numbers, the LVF and HVF cases exhibit similar values of $Area/D^2$, indicating that the void regions are governed primarily by the combined effects of particle inertia and gravitational settling rather than particle loading. Although the PDF of the normalized voronoi cell area exhibits noticeable differences between the two volume fractions, these differences reflect changes in the distribution of local void-cell sizes rather than the overall size of the void structure. The comparison of Figure~\ref{void_Area_total} (a) and (b) shows that these trends remain qualitatively unchanged with increasing Reynolds number. Although the void area at $St=6.5$ is slightly larger for $Re=200$ in comparison to $Re =100$ case, the dimensionless settling velocity is the primary parameter that dominates the particle motion.

The unsteady wake dynamics are studied by examining their instantaneous velocity fields and the spanwise vorticity,
$\omega_z = (\partial v /  \partial x) - (\partial u/\partial y)$, which highlights the shear layers and coherent vortical structures present in the wake. Vortex cores are identified using the $Q$-criterion suggested by \cite{hunt1988eddies}, $
Q = \tfrac{1}{2}\left(\|\boldsymbol{\Omega}\|^2 - \|\boldsymbol{S}\|^2\right)$. Here, positive values of local Q correlate to the presence of rotation-dominated regions, and shear-dominated areas are identified with negative Q. Evaluating Q values at particle positions measures the effect of local wake dynamics on particle dispersion. Here, the analysis is restricted to a three-dimensional slice of cross-stream width  $L_y(slice) = 6D$ symmetric across the channel centerline in order to analyze particles majorly affected by the cylinder wake.

For a dilute suspension of small, heavy particles, the Maxey-Riley reduces to a singular perturbation problem in the small parameter $\mathrm{St}$, given by

\begin{equation}
\mathrm{St}\,\frac{d\bm{v}^*}{dt^*}
=
\bm{u}^*
-
\bm{v}^*.
\label{eq_A}
\end{equation}

where $\bm{v}^*$ , $\bm{u}^*$ and ${t}^*$ is the non-dimensional particle velocity, fluid velocity and time, respectively. Here, inter-particle collisions, particle volume, added mass, pressure gradient, history force, and lift are neglected and only drag force is considered.

\begin{equation}
\frac{d\bm{v}^{*}}{dt^{*}}
=
\frac{d\bm{u}^{*}}{dt^{*}}
-
\mathrm{St}\,
\frac{d^{2}\bm{v}^{*}}{dt^{*2}} .
\label{eq_B}
\end{equation}

Using ~\eqref{eq_A} and \eqref{eq_B} for small St, we obtain

\begin{equation}
\bm{v}^*
=
\bm{u}^*
-
\mathrm{St}
\left(
\frac{d\bm{u}^*}{dt^*}
\right)
+
O(\mathrm{St}^2).
\label{eq:div_vp_11}
\end{equation}

Preferential concentration is governed by the divergence of the particle velocity field,

\begin{equation}
\nabla^* \cdot \bm{v}^*
=
-
\mathrm{St}
\nabla^* \cdot
\left(
\frac{d\bm{u}^*}{dt^*}
\right)
+
O(\mathrm{St}^2).
\label{eq:div_vp_1}
\end{equation}
This relation shows that  the compressibility of the particle phase is dependent on the divergence of the fluid acceleration field.
The non-dimensional material derivative of the fluid velocity and the incompressibility condition $\nabla^* \cdot \mathbf{u}^* = 0$, gives,

\begin{equation}
\nabla^* \cdot
\frac{d\bm{u}^*}{dt^*}
=
\partial_i^*(u_j^* \partial_j^* u_i^*)
=
(\partial_i^* u_j^*)(\partial_j^* u_i^*),
\end{equation}

The velocity-gradient tensor can be defined as,
\begin{equation}
A_{ij}^* = \partial_j^* u_i^*.
\end{equation}

This tensor can be decomposed into its symmetric (strain-rate $S_{ij}^*$) and antisymmetric (rotation $\Omega_{ij}^*$) tensor,\\
\begin{equation}
A_{ij}^* A_{ji}^*
=
S_{ij}^* S_{ij}^*
-
\Omega_{ij}^* \Omega_{ij}^*.
\end{equation}
\begin{align}
S_{ij}^* &= \frac{1}{2}(A_{ij}^* + A_{ji}^*), \\
\Omega_{ij}^* &= \frac{1}{2}(A_{ij}^* - A_{ji}^*).
\end{align}

Substituting the decomposed terms into the  Equation~\eqref{eq:div_vp_1}, the equation becomes
\begin{equation}
\nabla^* \cdot \bm{v}^*
=
-
\mathrm{St}
\left(
S_{ij}^* S_{ij}^*
-
\Omega_{ij}^* \Omega_{ij}^*
\right).
\end{equation}

Rewriting the divergence in terms of non-dimensional Q criterion $Q^*
=
\frac{1}{2}
\left(
\Omega_{ij}^* \Omega_{ij}^*
-
S_{ij}^* S_{ij}^*
\right)$  gives the relation between compressibility of particle phase to non-dimensional Q values and Stokes number, as seen in the relation,
\begin{equation}
\nabla^* \cdot \mathbf{v}^*
= 2\,\mathrm{St}\,Q^*
\label{eqn_Q}
\end{equation}

Equation \ref{eqn_Q} shows that regions with $Q^* < 0$ (strain-dominated) relate to negative divergence of the particle velocity field and promote particle clustering, while regions with $Q^* > 0$ (vorticity-dominated) correspond to particle depletion zones or voids. In the low Stokes number range, inertial particle clustering depends on the balance between strain and rotation rate in the fluid flow, as quantified by the Q-criterion. To analyze this behavior, a study of the relation between local Q values and particle clustering in the region for both St$=$1 and St$=$6.5 at $Re=100$ and $Re=200$ is carried out. It is worth noting that although the Q based analysis which is supposed to be valid for $St<1$, has been extended upto higher St cases to obtain detailed picture about the particle distribution in the vortical field.

\begin{figure*}[!t]
	\begin{subfigure}[b]{1\textwidth}
	\minipage{0.45\textwidth}		
	\includegraphics[width=0.9\textwidth]{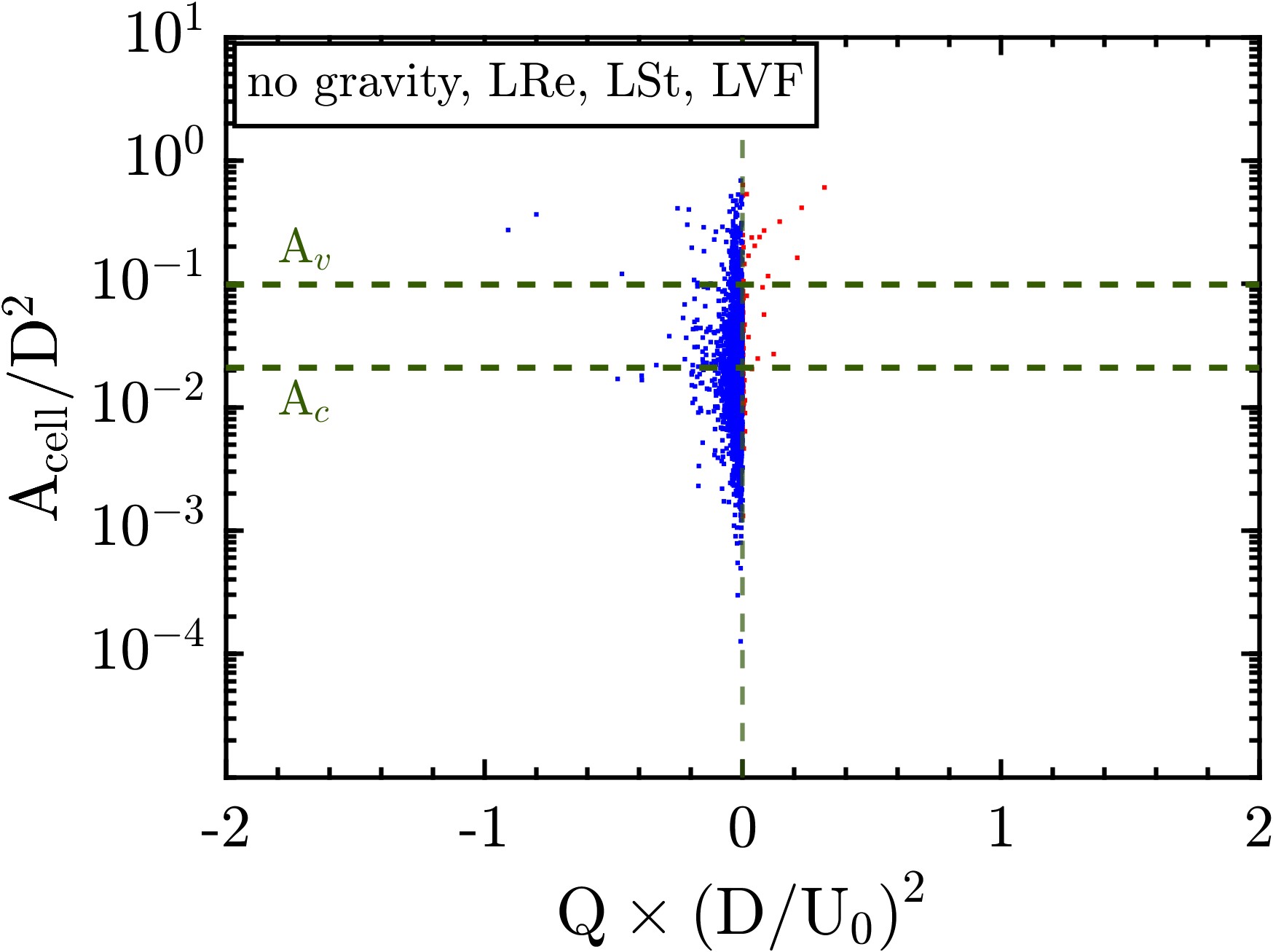}
	\caption{}
	\endminipage
	\minipage{0.45\textwidth}		
	\includegraphics[width=0.9\textwidth]{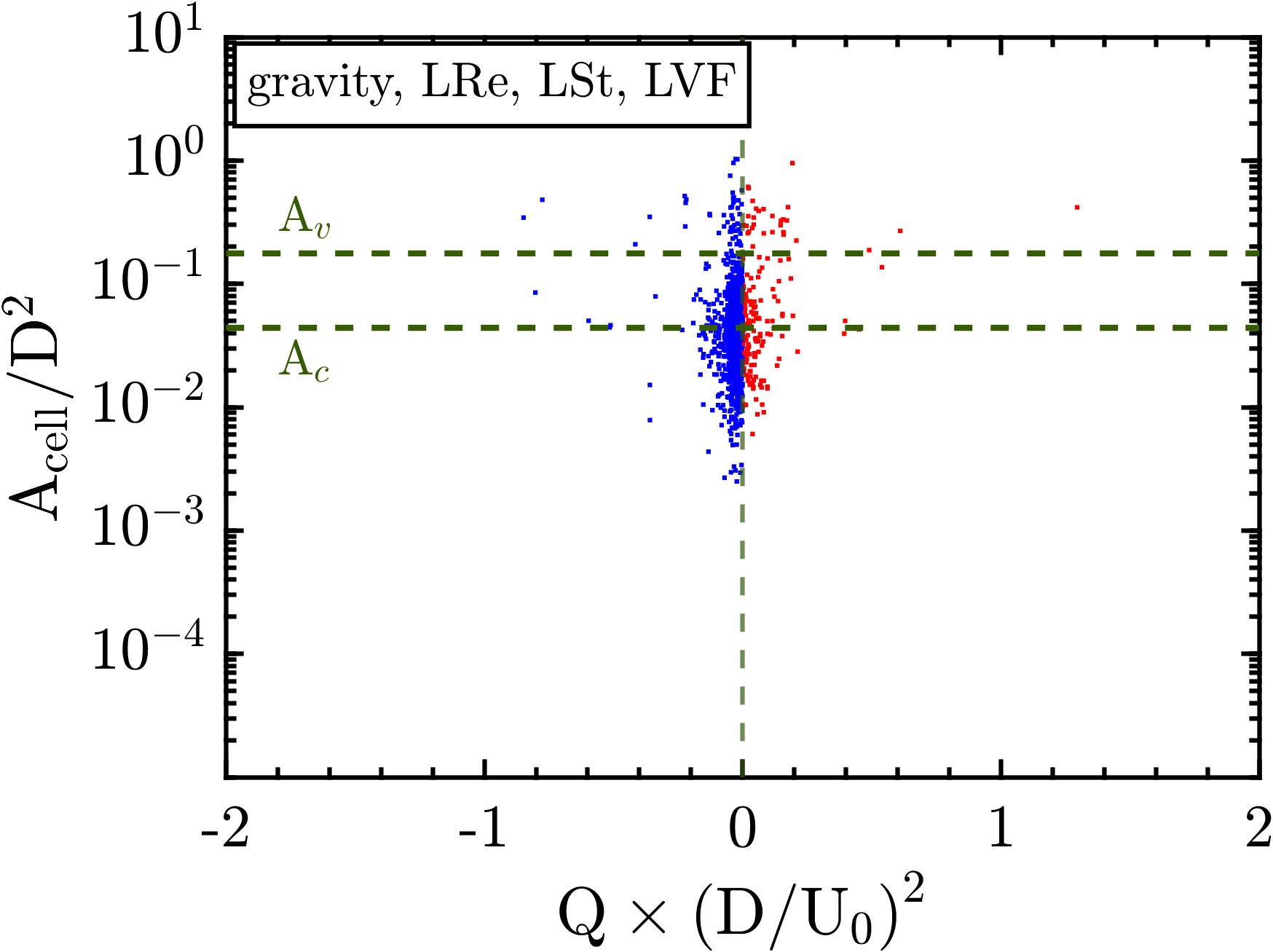}
	\caption{}
	\endminipage \\
	\minipage{0.45\textwidth}		
	\includegraphics[width=0.9\textwidth]{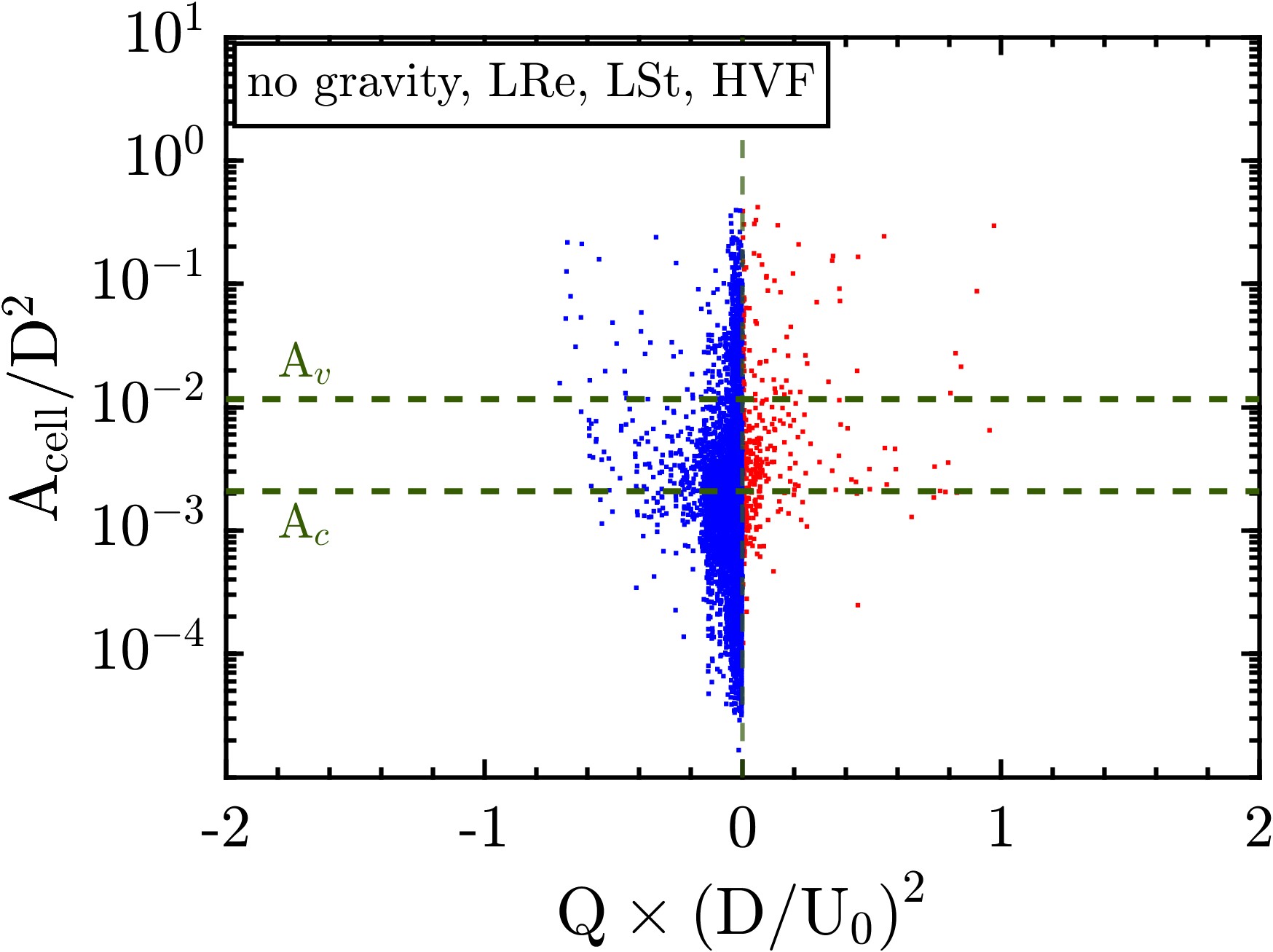}
	\caption{}
	\endminipage
	\minipage{0.45\textwidth}		
	\includegraphics[width=0.9\textwidth]{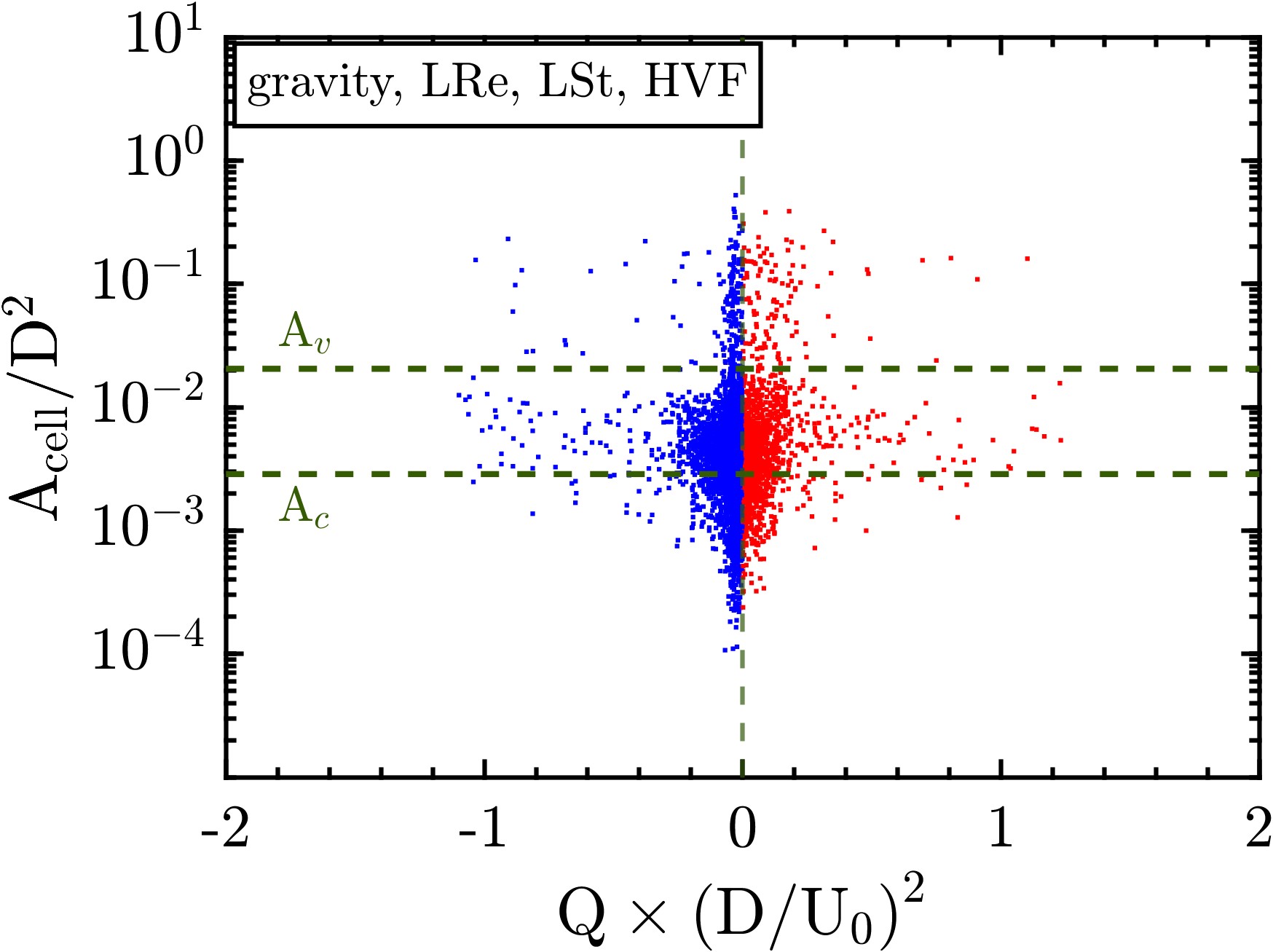}
	\caption{}
	\endminipage			
	\end{subfigure}
\caption{Scatter plots of the normalized Voronoi cell area ($A_{\mathrm{cell}}/D^{2}$) versus fluid $Q$ values evaluated at particle locations for low inertia particles ($St=1$) at $Re=100$. Results for the low particle loadings ($\phi_v=2\times10^{-5}$) and high particle loadings ($\phi_v=2\times10^{-4}$) are shown in (a,b) and (c,d), respectively. Cases without gravity ($Fr\rightarrow\infty$) are shown in (a,c), whereas cases with gravity ($Fr=1$) are shown in (b,d).}
\label{slice_Qre100_Lst}
\end{figure*}

\begin{figure*}[!t]
	\begin{subfigure}[b]{1\textwidth}
	\minipage{0.45\textwidth}		
	\includegraphics[width=0.9\textwidth]{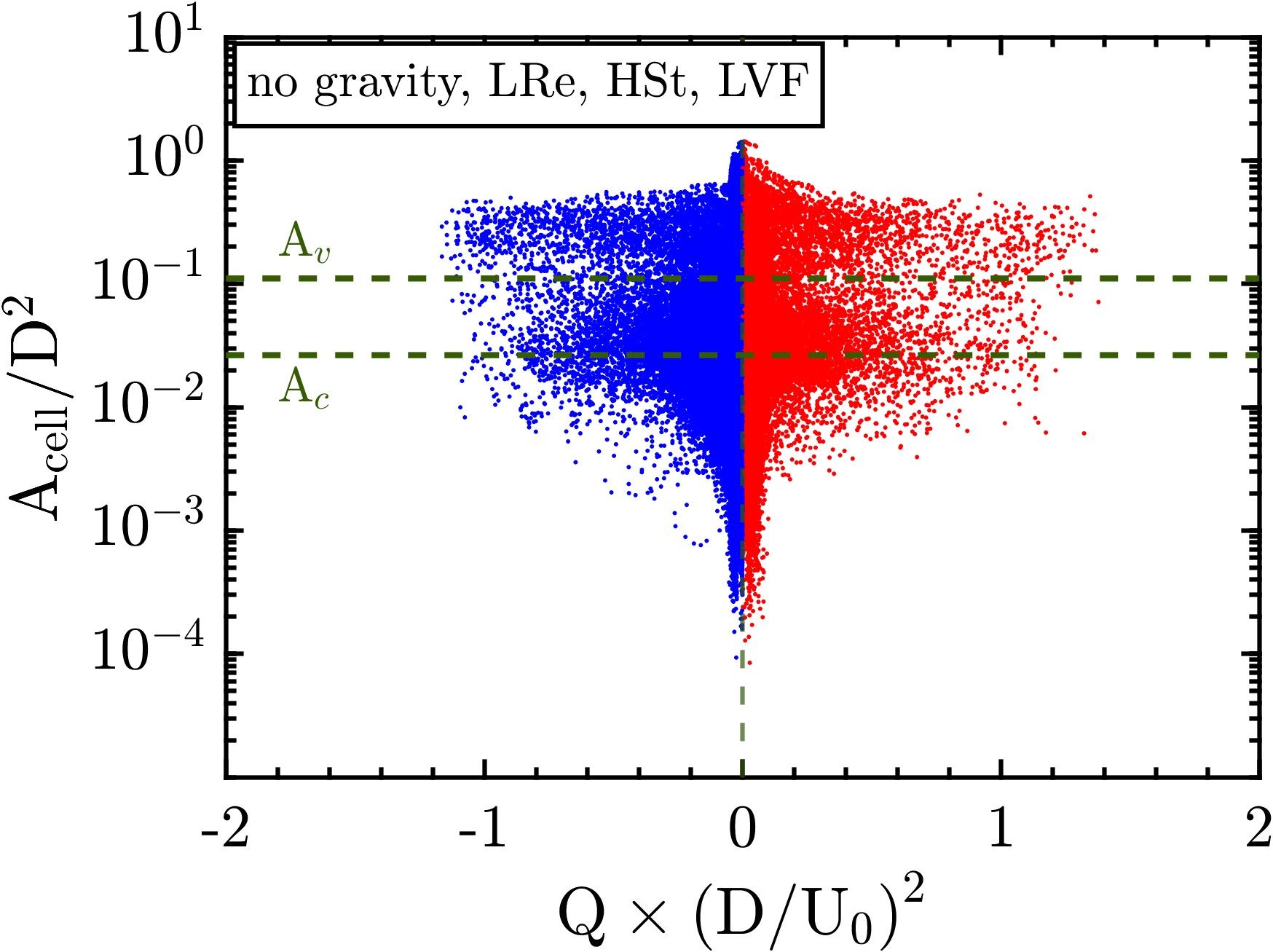}
	\caption{}
	\endminipage
	\minipage{0.45\textwidth}		
	\includegraphics[width=0.9\textwidth]{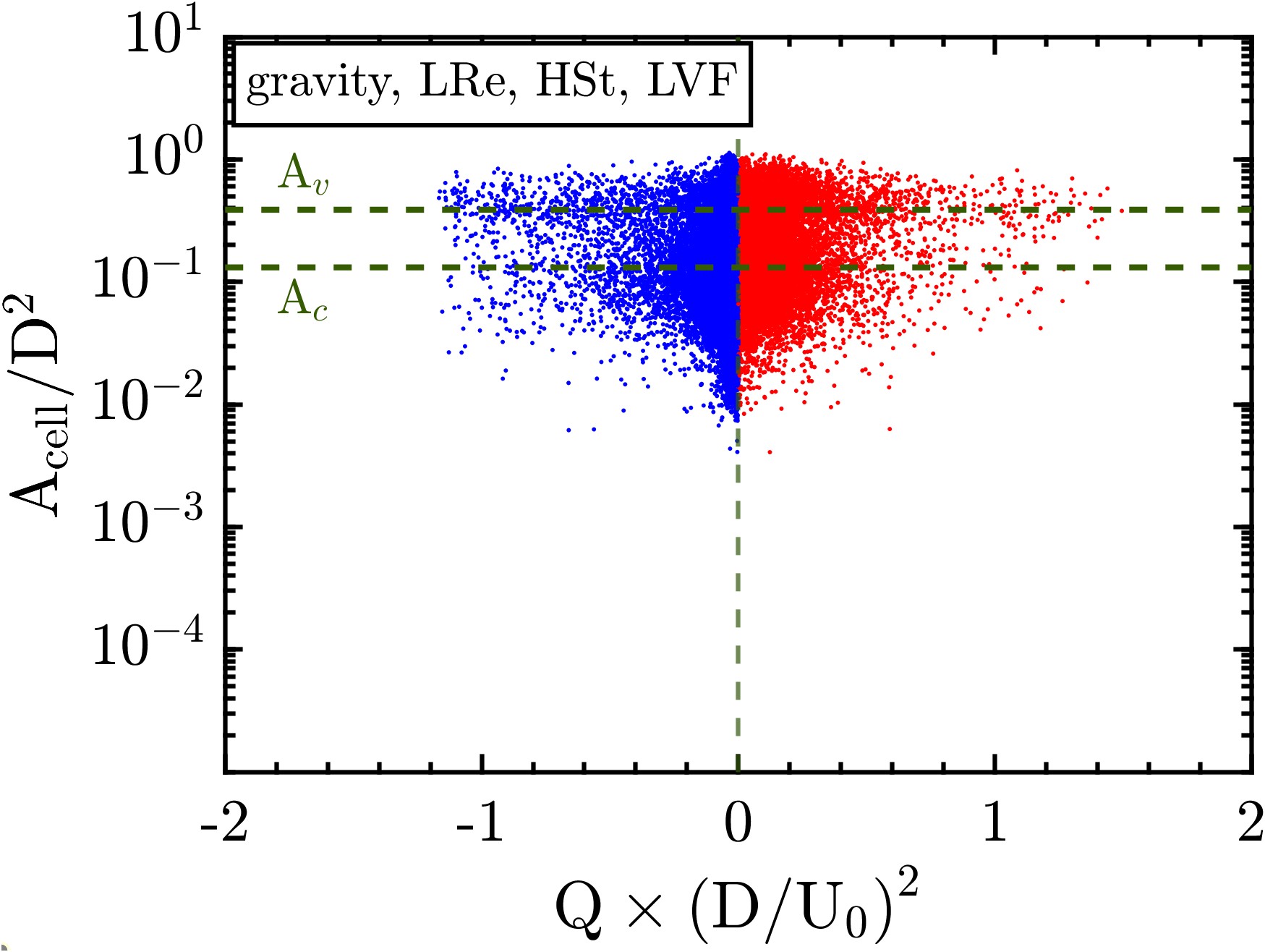}
	\caption{}
	\endminipage \\
	\minipage{0.45\textwidth}		
	\includegraphics[width=0.9\textwidth]{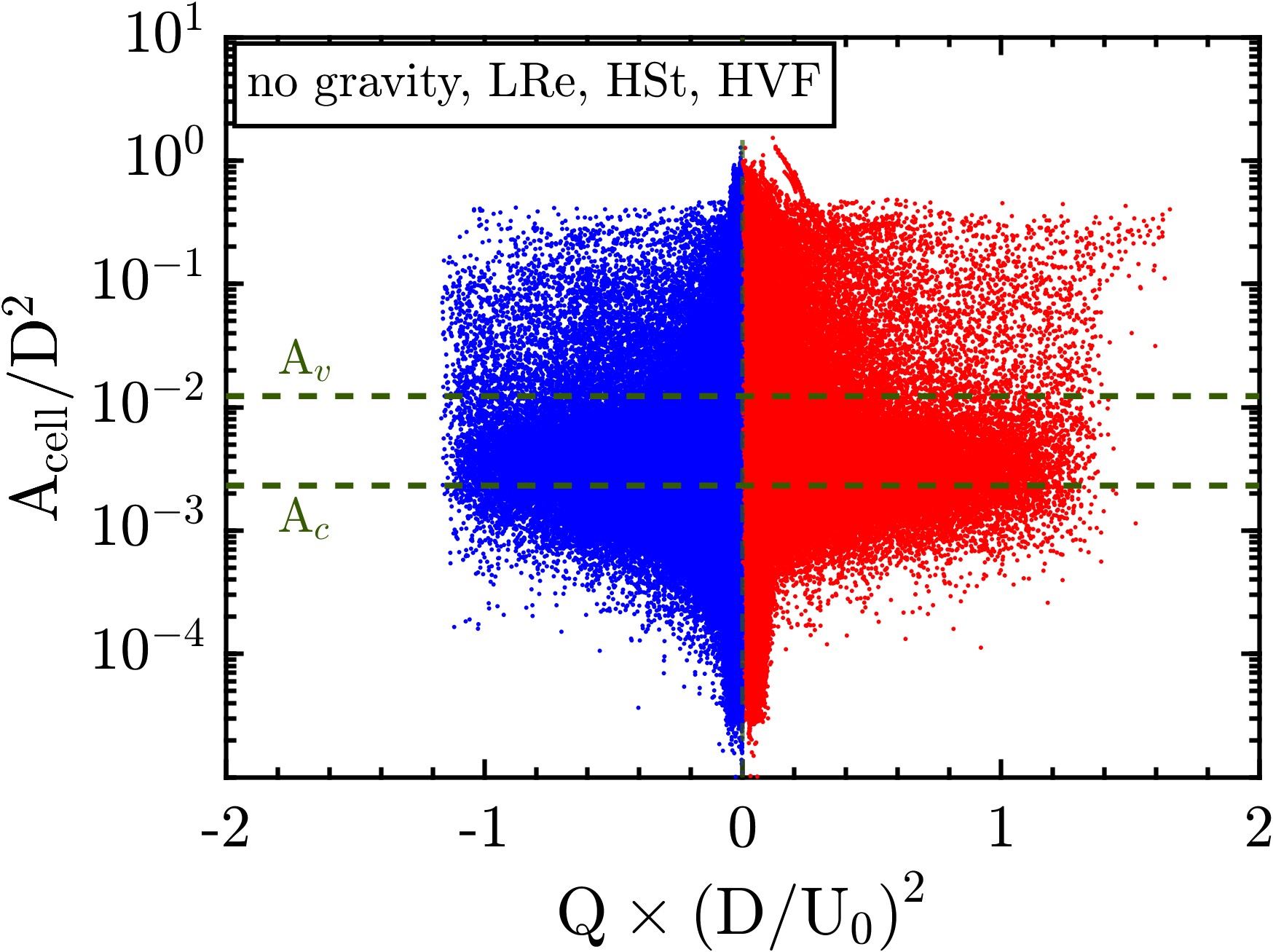}
	\caption{}
	\endminipage
	\minipage{0.45\textwidth}		
	\includegraphics[width=0.9\textwidth]{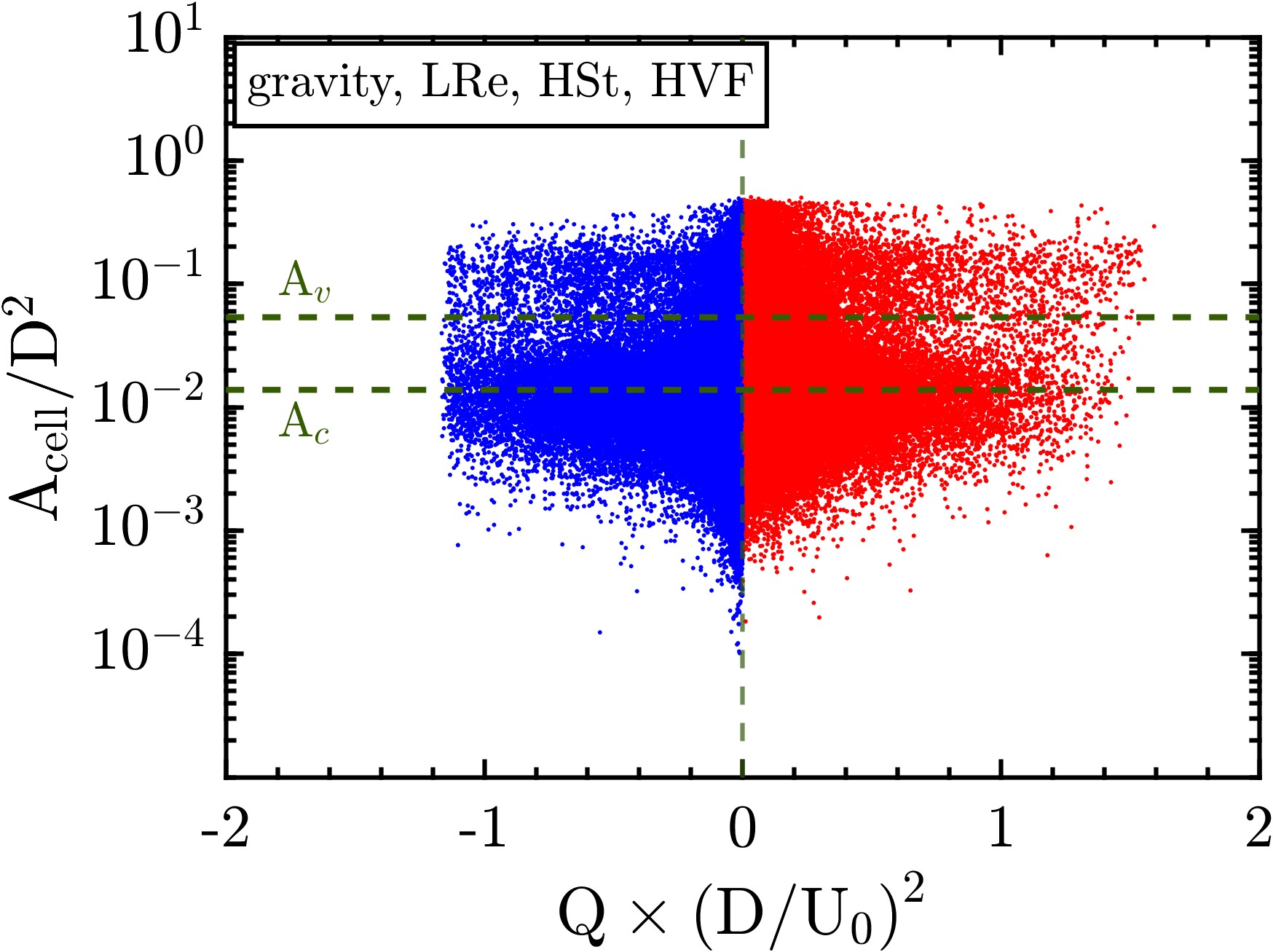}
	\caption{}
	\endminipage			
	\end{subfigure}
\caption{Scatter plots of the normalized Voronoi cell area ($A_{\mathrm{cell}}/D^{2}$) versus fluid $Q$ values evaluated at particle locations for high inertia particles ($St=6.5$) at $Re=100$. Results for the low particle loadings ($\phi_v=2\times10^{-5}$) and high particle loadings ($\phi_v=2\times10^{-4}$) are shown in (a,b) and (c,d), respectively. Cases without gravity ($Fr\rightarrow\infty$) are shown in (a,c), whereas cases with gravity ($Fr=1$) are shown in (b,d).}
\label{slice_Qre100_Hst}
\end{figure*}

\begin{figure*}[!t]
	\begin{subfigure}[b]{1\textwidth}
	\minipage{0.45\textwidth}		
	\includegraphics[width=0.9\textwidth]{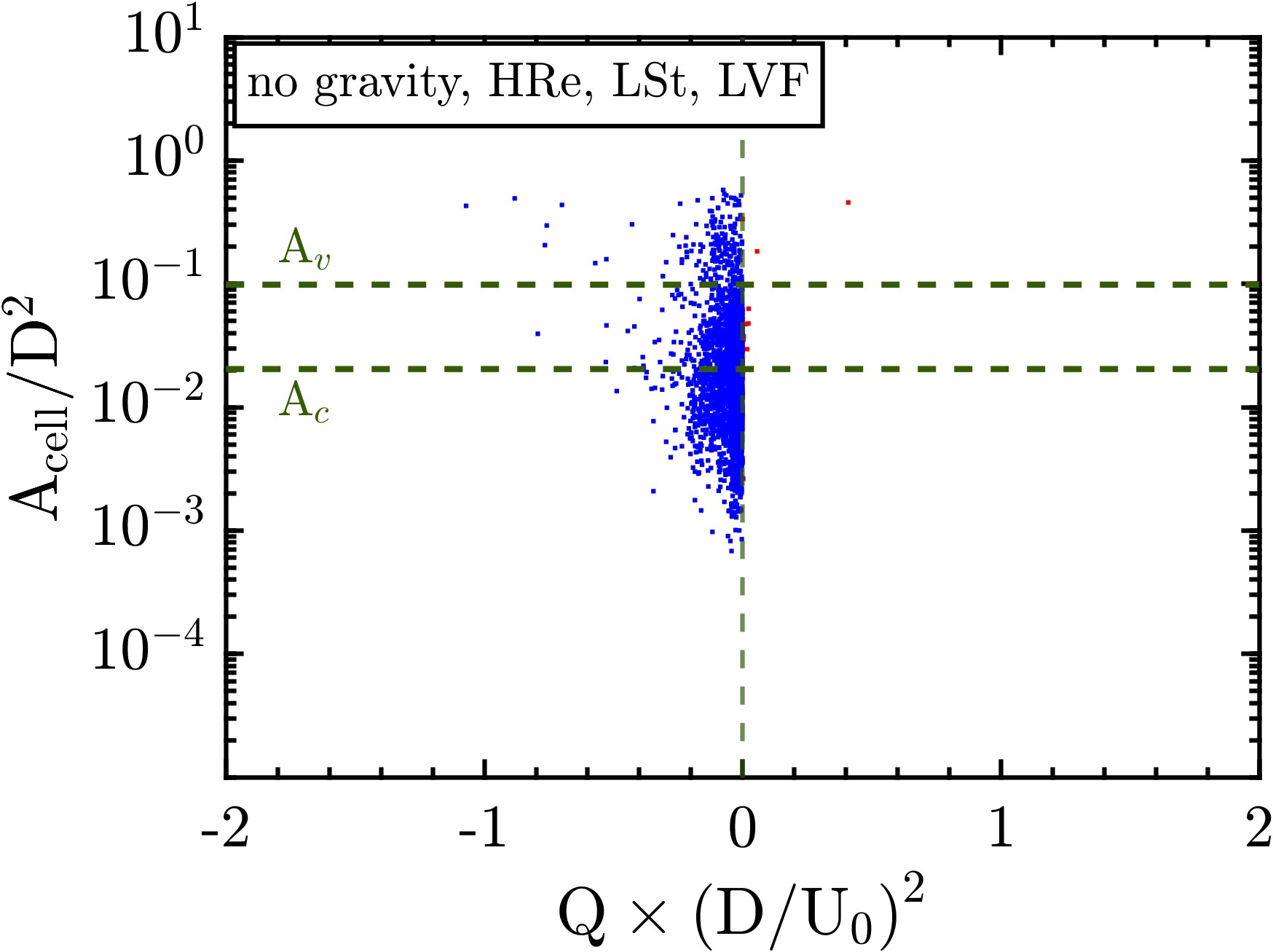}
	\caption{}
	\endminipage
	\minipage{0.45\textwidth}		
	\includegraphics[width=0.9\textwidth]{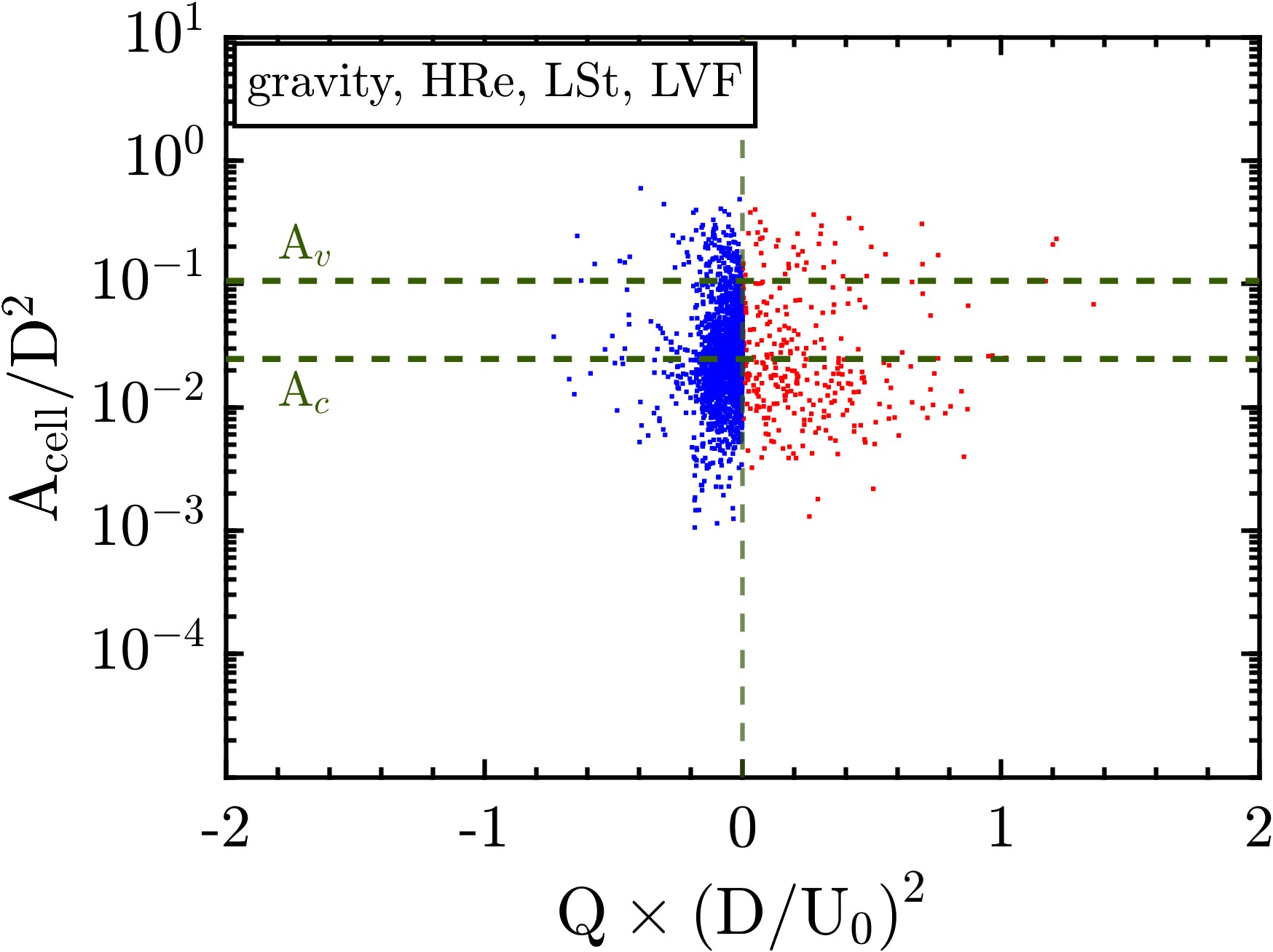}
	\caption{}
	\endminipage \\
	\minipage{0.45\textwidth}		
	\includegraphics[width=0.9\textwidth]{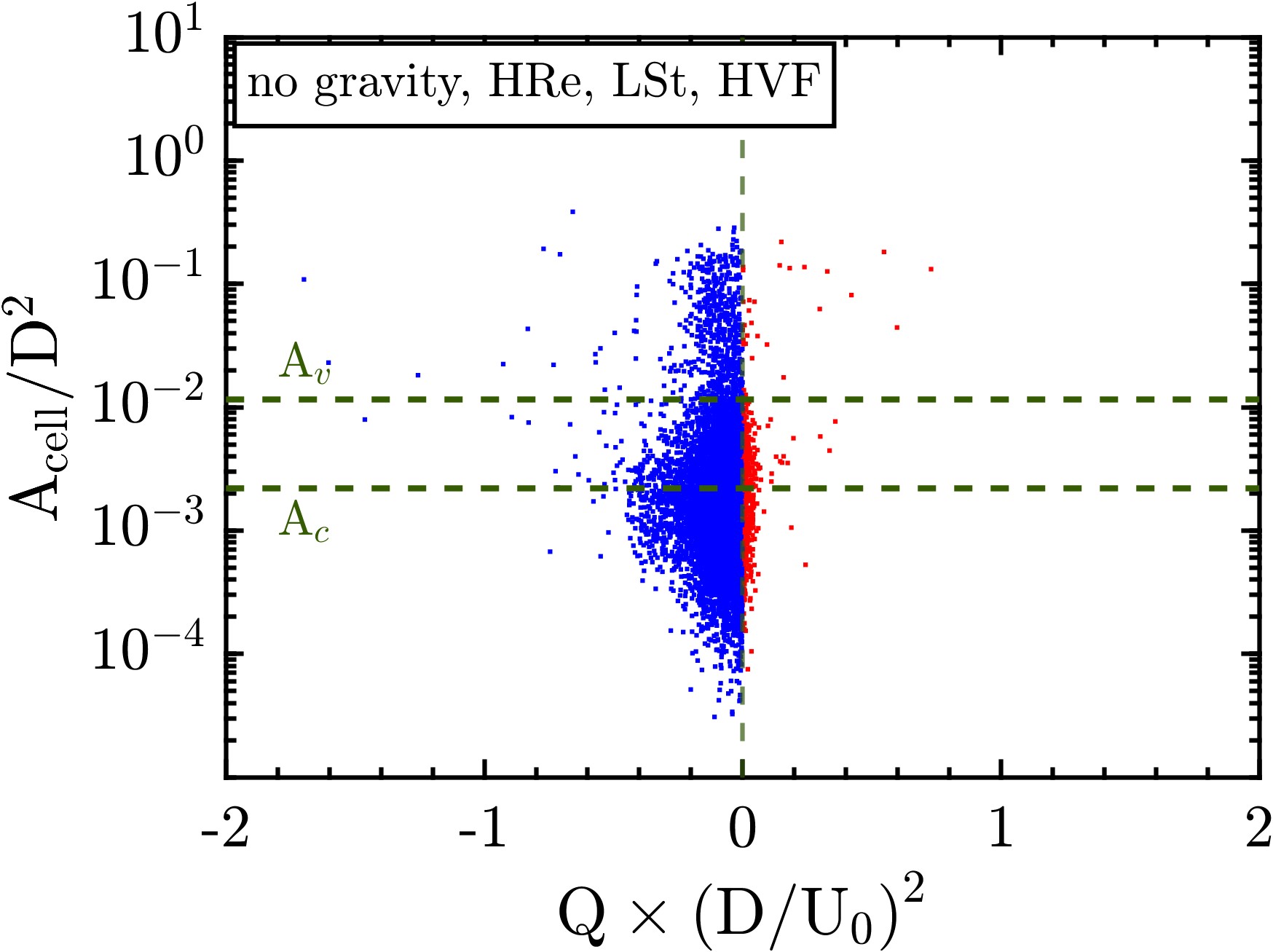}
	\caption{}
	\endminipage
	\minipage{0.45\textwidth}		
	\includegraphics[width=0.9\textwidth]{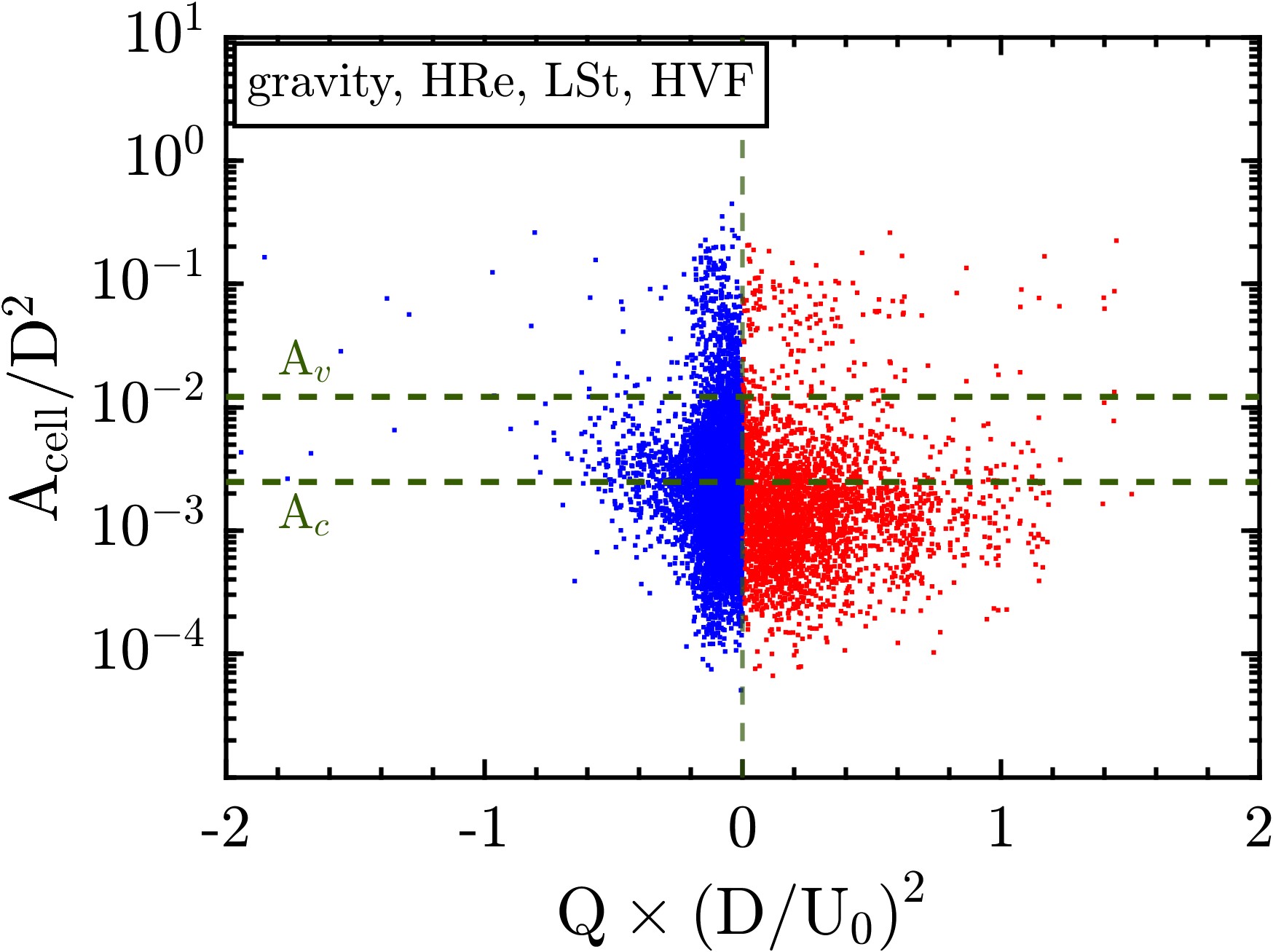}
	\caption{}
	\endminipage			
	\end{subfigure}
\caption{Scatter plots of the normalized Voronoi cell area ($A_{\mathrm{cell}}/D^{2}$) versus fluid $Q$ values evaluated at particle locations for low inertia particles ($St=1$) at $Re=200$. Results for the low particle loadings ($\phi_v=2\times10^{-5}$) and high particle loadings ($\phi_v=2\times10^{-4}$) are shown in (a,b) and (c,d), respectively. Cases without gravity ($Fr\rightarrow\infty$) are shown in (a,c), whereas cases with gravity ($Fr=2$) are shown in (b,d).}
\label{slice_Qre200_Lst}
\end{figure*}

\begin{figure*}[!t]
	\begin{subfigure}[b]{1\textwidth}
	\minipage{0.45\textwidth}		
	\includegraphics[width=0.9\textwidth]{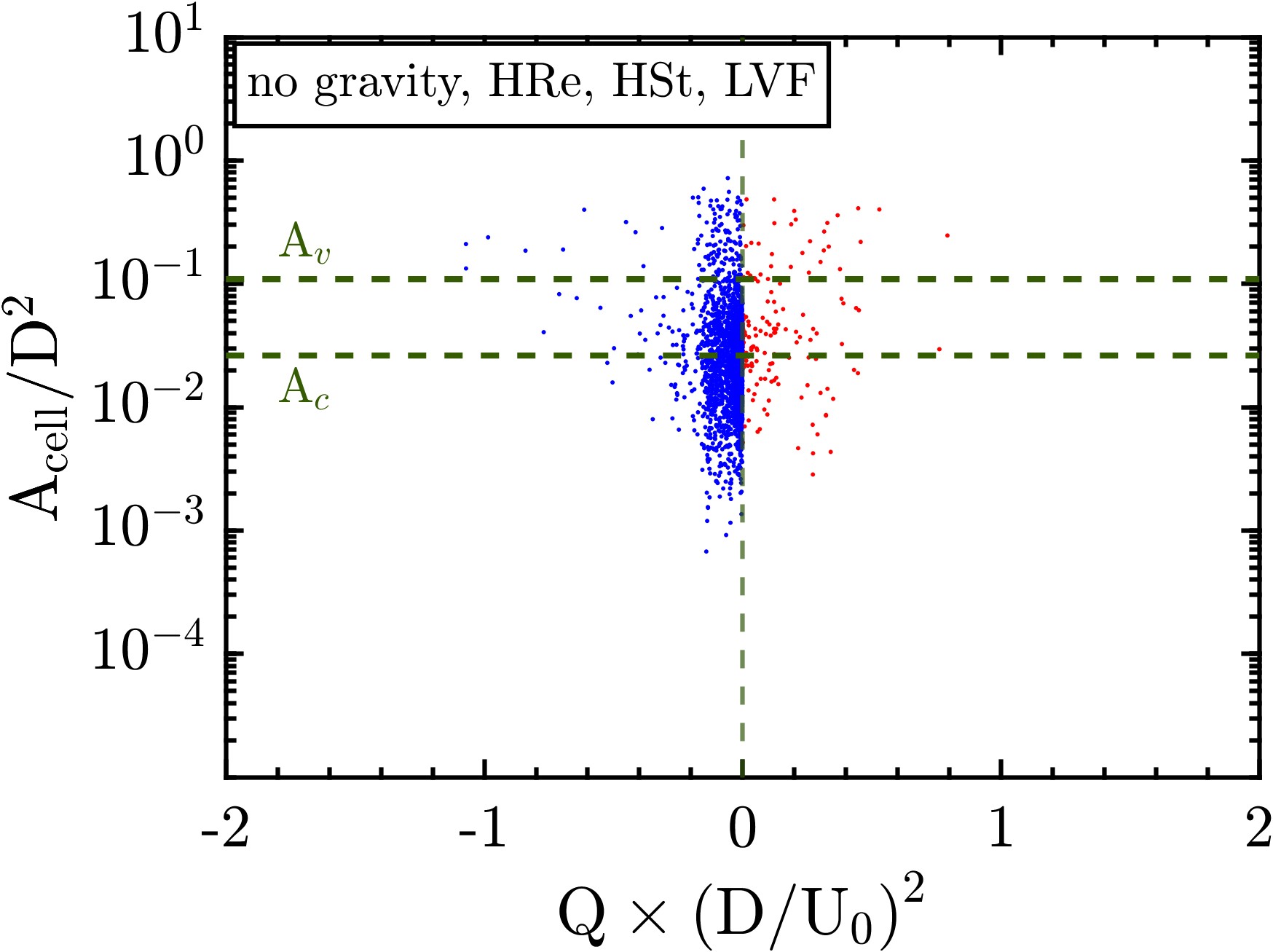}
	\caption{}
	\endminipage
	\minipage{0.45\textwidth}		
	\includegraphics[width=0.9\textwidth]{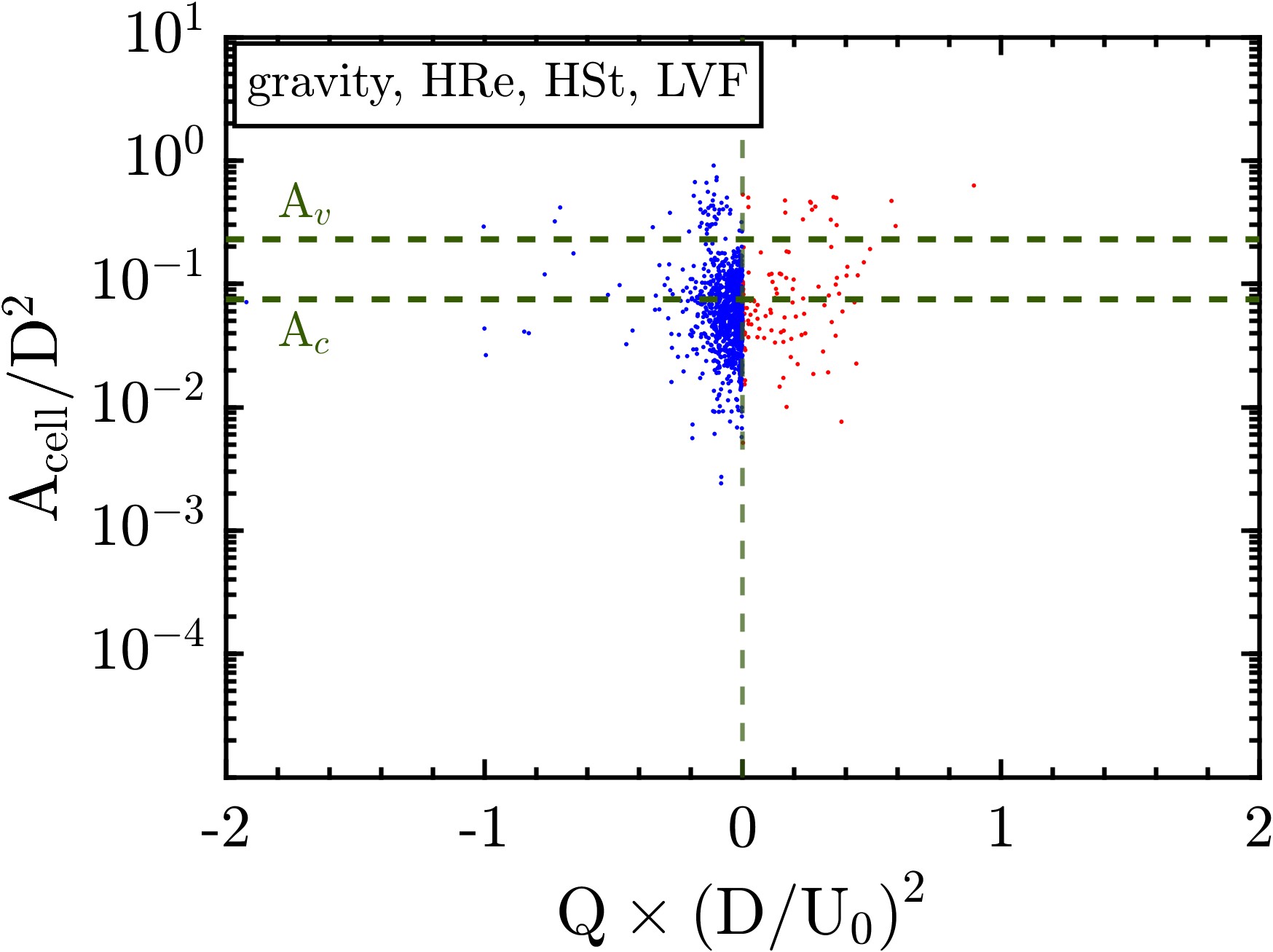}
	\caption{}
	\endminipage \\
	\minipage{0.45\textwidth}		
	\includegraphics[width=0.9\textwidth]{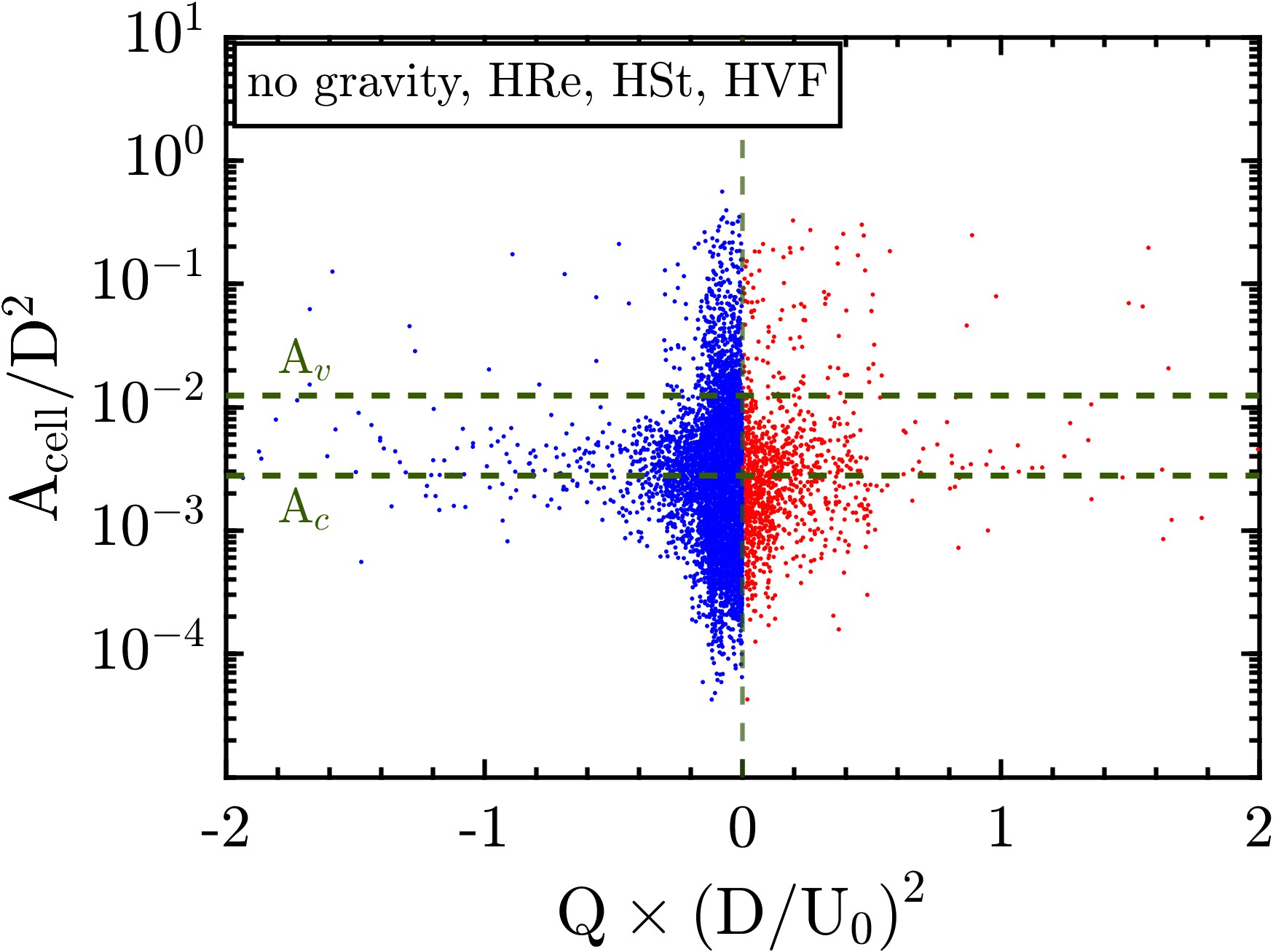}
	\caption{}
	\endminipage
	\minipage{0.45\textwidth}		
	\includegraphics[width=0.9\textwidth]{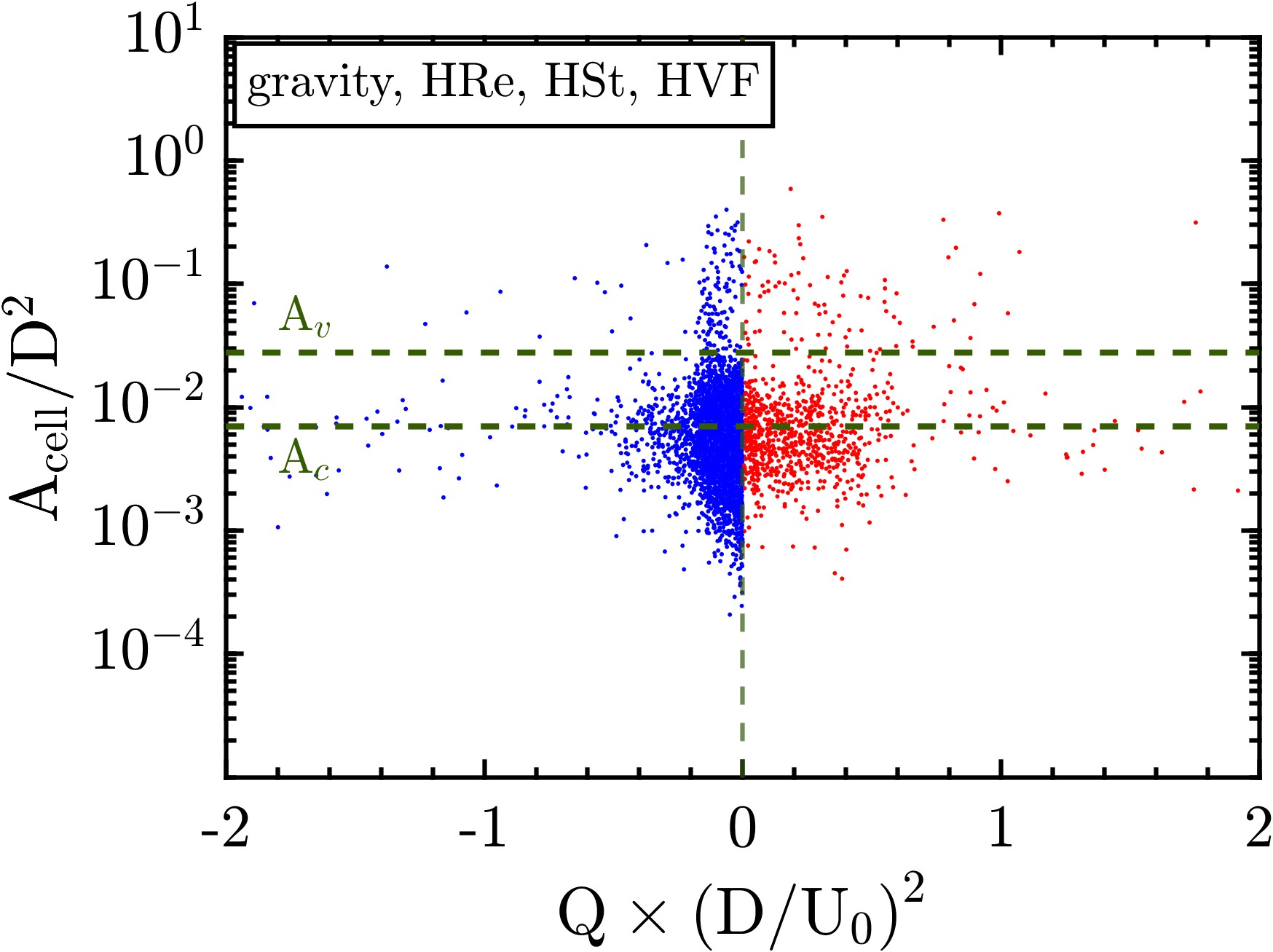}
	\caption{}
	\endminipage			
	\end{subfigure}
\caption{Scatter plots of the normalized Voronoi cell area ($A_{\mathrm{cell}}/D^{2}$) versus fluid $Q$ values evaluated at particle locations for high inertia particles ($St=6.5$) at $Re=200$. Results for the low particle loadings ($\phi_v=2\times10^{-5}$) and high particle loadings ($\phi_v=2\times10^{-4}$) are shown in (a,b) and (c,d), respectively. Cases without gravity ($Fr\rightarrow\infty$) are shown in (a,c), whereas cases with gravity ($Fr=2$) are shown in (b,d).}
\label{slice_Qre200_Hst}
\end{figure*}

Figures~\ref{slice_Qre100_Lst}  to ~\ref{slice_Qre200_Hst} present the scatter plots of the normalized voronoi cell area, $A_{\mathrm{cell}}/D^{2}$, as a function of $Q$, evaluated at particle locations within the near-wake region of the cylinder. The scatter plots for finite $Fr $ and $Fr\rightarrow\infty$  case shows particle dispersion and its correlation with fluid phase dynamics for dataset sampled over 200 frames. As discussed above, the voronoi cell area provides a quantitative measure of local particle concentration, where a smaller cell area indicates particles are closely packed together in a local high concentration region (clusters), and a larger cell area corresponds to particle-depleted or void zones.
The vertical axis in the plot represents the non-dimensional second invariant of the fluid velocity gradient term, $Q \times (D/U_0)^2 = 0 $, which distinguishes between vorticity-dominated ($Q>0$) and strain-dominated ($Q<0$) regions. The dashed horizontal lines denote the normalized voronoi area thresholds, $A_c$ and $A_v$, separating clustered ($A_{\mathrm{cell}}/D^{2} < A_c$), randomly distributed ($A_c < A_{\mathrm{cell}}/D^{2} < A_v$), and dilute particle distribution ($A_{\mathrm{cell}}/D^{2} > A_v$) zones.

In Figure~\ref{slice_Qre100_Lst}, scatter plots are presented for $Re=100$ and $St=1$.  In the absence of gravity ($Fr\rightarrow\infty$) for low particle loading cases, particles are located mainly in the strain dominated region, with negative Q values. The scatter plot is relatively compact across Q range, which suggests that the particle dispersion is mainly driven by large scale wake dynamics. Particles are largely excluded from regions with higher vorticity ($Q>0$). In these zone, few voids are present with larger voronoi areas ($A_{\mathrm{cell}}/D^{2} > A_v$) , indicating the presence of coherent void structures associated with vortex cores. In case of finite Froude number, as shown in Figure~\ref{slice_Qre100_Lst}(b)(d), particles are still present prominently in  negative-$Q$ regions. However, a noticeable broadening of the distribution is observed in the scatter plot with more prominent presence of particles in the positive-$Q$ regime. This indicates that particle motion is no longer governed purely by unsteady vortical dynamics, as particle acceleration under gravity effect, which enhances particle-fluid slip and piercing of the vortices by the particles, and promotes the particles to sample regions with positive $Q$ values. As particle loading increases, the  distributions extend further into the positive Q regime, indicating that inter-particle interactions disrupt the preferential sampling of vortical regions. In the presence of gravity ($Fr =1$), the voronoi cell area scatter becomes more symmetric around $Q=0$, signifying a more balanced sampling of vorticity and strain-dominated flow regions.

Figure~\ref{slice_Qre100_Hst} shows the scatter plot of the voronoi cell area as a function of the local Q values for high inertia particles at $St =6.5$ and $Re=100$. For Figure~\ref{slice_Qre100_Hst}(a) to (d), the distribution of particles is approximately symmetric about $Q=0$, indicating that particles are present in both strain-dominated and vorticity-dominated regions of the wake and there is no strong preference for particles to occupy negative $Q$ regions. The range of the normalized voronoi cell area($A_\text{cell}/D^2$) is observed to decrease with inclusion of gravity effect in the flow. A wider spread in $A_\text{cell}/D^2$ observed at $Fr\rightarrow\infty$ indicates stronger variations in local particle concentration, with the presence of clustered regions having smaller voronoi cell areas. In presence of gravity, the scatter remains symmetric in $Q$, but the spread in voronoi cell area decreases, showing that gravity influences the inter-particle spacing within the flow. Increasing particle loading further increase the range of distribution of voronoi cell area, with lower cell size in the clustered zone.

At higher Reynolds number ($Re=200$), the particle samples a larger range of Q while still showing more prominent presence in high strain - low vorticity region, as shown in Figure~\ref{slice_Qre200_Lst} and Figure~\ref{slice_Qre200_Hst}. The preferential clustering is observed even for high inertia particles at $St=6.5$ and $Re=200$, due to stronger vortical field present in the wake as seen in Figure~\ref{slice_Qre200_Hst}. As in case of particles with low Stokes, more particles enter into the region of high vorticity when gravity driven flow($Fr = 2$).

\begin{figure*}[!t]
	\begin{subfigure}[b]{1\textwidth}
	\minipage{0.45\textwidth}		
	\includegraphics[width=0.9\textwidth]{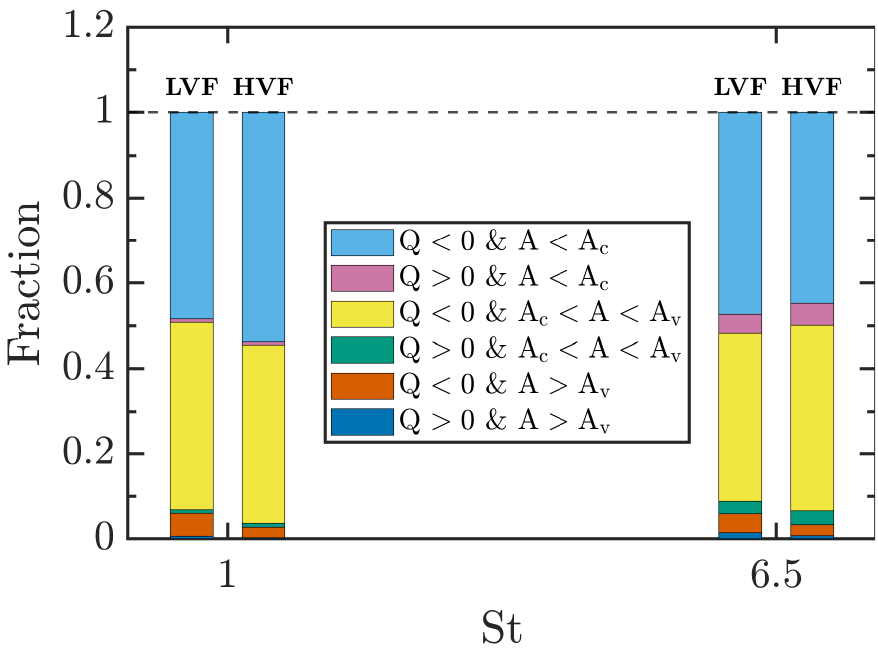}
	\caption{}
	\endminipage
	\minipage{0.45\textwidth}		
	\includegraphics[width=0.9\textwidth]{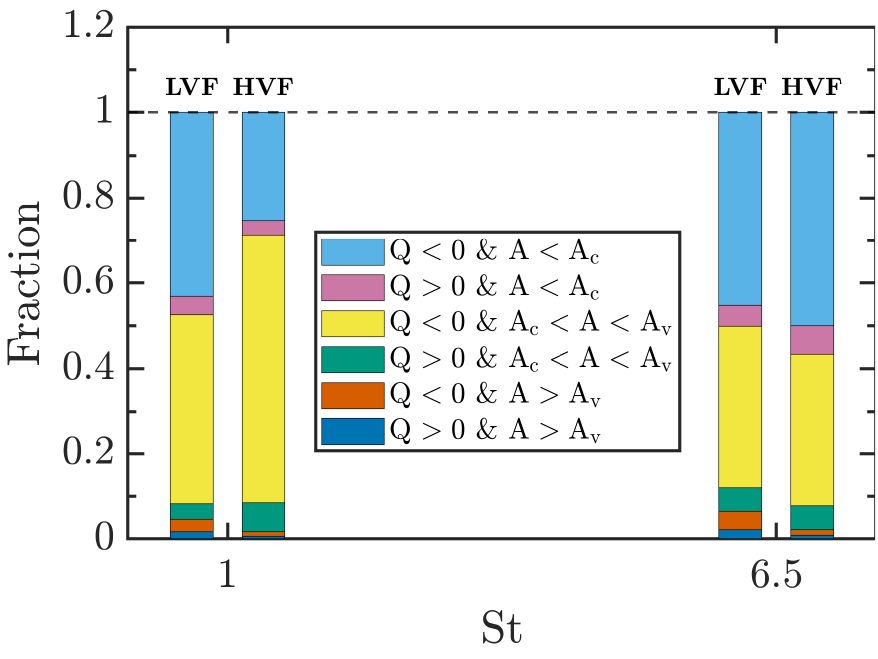}
	\caption{}
	\endminipage \\
	\minipage{0.45\textwidth}		
	\includegraphics[width=0.9\textwidth]{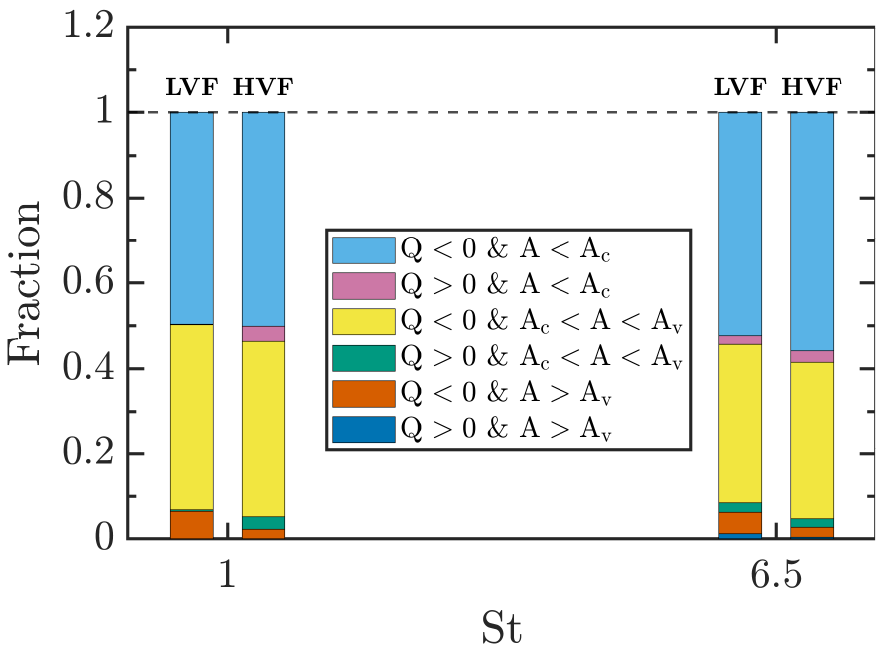}
	\caption{}
	\endminipage
	\minipage{0.45\textwidth}		
	\includegraphics[width=0.9\textwidth]{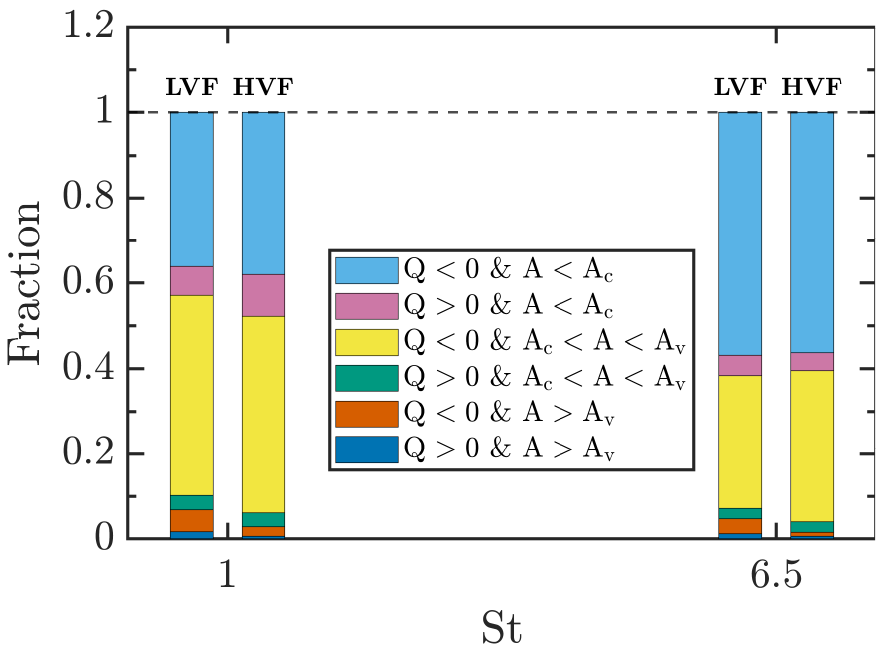}
	\caption{}
	\endminipage			
	\end{subfigure}
\caption{Bar charts showing the fraction of Voronoi cells classified as clusters, intermediate cells, and voids in the vorticity-dominated and strain-dominated regions. Results for $Re=100$ are shown in (a,b), and those for $Re=200$ are shown in (c,d). Cases without gravity ($Fr\rightarrow\infty$) are shown in (a,c), whereas cases with finite Froude number are shown in (b,d).}
\label{slice_frac}
\end{figure*}

In Figure~\ref{slice_frac}, particles are grouped into six zones in terms of the local Q values of fluid and the particle concentration based on voronoi cell area A, where $A_{c}$ , $A_{v}$ denote normalized cell area for clustered cells and void cells respectively. The zone for $Q>0$ and $A>A_{v}$ corresponds to particles located in strong rotational regions of the flow and forms void regions. These are the high vorticity regions from where the particles are expelled. The fraction of particles in this zone is very small in all the cases, indicating that particles rarely occupy high vorticity regions where large voids exist. This is consistent across Reynolds numbers and gravity conditions, reflecting the general tendency of particles to avoid strongly rotating cores.
Particles in the zone $Q<0$ and $A>A_{v}$ lie in strain-dominated regions but still belong to void-like regions. The fraction here is small but non-zero, indicating that void cells are not exclusively associated with vortical regions and can also occur in high strain parts of the wake. These void cells are majorly made up of particles which are expelled from high vorticity regions and are located near the void cells. This fraction remains limited and decreases with inclusion of gravity effect (finite $Fr$ case). This is accompanied by transition of void structure from its characteristic leaf-like void pattern towards a vertical,straight void structure. The zone $Q>0$ and $A_{c}<A<A_{v}$ represents particles in high-rotation, low-strain regions with moderate cell areas, indicating particles in the region are neither strongly clustered nor part of voids. The fraction of particles in this zone increases slightly for finite $Fr$ cases (in presence of gravity), suggesting that gravity enhances particle-fluid slip and allows particles to sample high rotational regions more readily. The zone  $Q<0$ and $A_{c}<A<A_{v}$  is the dominant zone in all the cases. It corresponds to particles located in strain-dominated regions with moderate local particle concentration. A large fraction of particles follow this region, indicating that most particles reside in strain-dominated wake regions without forming clusters or voids. This trend is observed across range of Reynolds numbers, Stokes numbers, particle loading and Froude number investigated here. The zone  $Q>0$ and $A<A_{c}$ corresponds to clustered particles inside regions of high rotation and low strain. The fraction of particles in this zone is very small in gravity-free cases, emphasizing that clustering inside high vorticity regions is negligible. With gravity, a slight increase in this fraction is observed, indicating that gravity increases the particle momentum and particle-fluid slip, causing it to overcome rotational effect of underlying flow structures, mentioned as vortex piercing. The zone $Q<0$ and $A<A_{c}$ represents clustering in strain-dominated regions in the flow. A significant fraction of particles resides in this zone in gravity-free cases. These particle are flung from high vorticity regions and cluster in the region of high strain. In presence of gravity, this fraction decreases for low particle loading case, indicating a weakening of intense clustering as particle has high slip velocity and does not get affected by local fluid fluctuations. This is more prominent for $St=1$ case. Here, particle distribution shows modified void structure and the clustering phenomenon is suppressed. Higher inertia particles, on the other hand, forms large bow-shock like structure which is known to increase local particle concentration. This leads to an increase in fraction in case of $St =6.5$ at high particle loading of $\phi_v=2\times10^{-4}$.

\section{Conclusion}
The present study has investigated the preferential clustering of inertial particles in the wake of a circular cylinder using one-way coupled direct numerical simulations over a range of Reynolds numbers ($Re=100$ and $200$), particle inertia ($St=1$ and $6.5$), particle loading ($\phi_v=2\times10^{-5}$ and $2\times10^{-4}$), and Froude number. This is done in order to analyze the coupled influence of coherent structures in the wake, particle inertia, and gravity on particle transport. The unladen flow exhibits the classical vortex pattern at both Reynolds numbers, with increasing Reynolds number producing a narrower but more energetic wake. The flow at higher Reynolds number is characterized by stronger vortices, faster wake recovery, and enhanced unsteady fluctuating intensity, with the peak normalized streamwise velocity fluctuations increasing from approximately $0.10$ at $Re=100$ to $0.15$ at $Re=200$, while the peak cross-stream fluctuations normalized by inlet velocity increase from approximately $0.17$ to $0.45$. These Reynolds-number-dependent wake characteristics establish the coherent flow structures that govern preferential concentration of inertial particles. It is observed that the dimensionless settling velocity, $
v_t^*=\frac{St}{Fr^2},$ is the primary parameter governing particle distribution in unsteady wake flows. This parameter represents the ratio of the gravitational settling velocity to the characteristic advective velocity of the wake and therefore quantifies the balance between gravitational settling and vortex-induced particle migration. For $St/Fr^2=0$ ( in the absence of gravity), particles are centrifugally expelled from vortices and preferentially accumulate in strain-dominated regions, producing well-defined leaf-like void structures that oscillate with the wake. As $St/Fr^2$ increases and particles accelerate under effect of particle inertia and gravity, enhanced particle-fluid slip velocity weakens vortex trapping, allowing particles to penetrate high vorticity regions and modify the void structure. The void region undergoes a continuous transition from individual leaf-like structures to elongated snake-like voids at intermediate settling velocities and eventually to narrow, nearly vertical void regions when gravitational settling dominates the particle motion. This transition demonstrates that gravity modifies preferential concentration primarily through weakening of vortex-induced particle clustering and promoting  vortex piercing by particles. Voronoi tessellation combined with the local fluid $Q$-values helps to carry out a quantitative study on balance between particle concentration and unsteady flow dynamics. For  $Fr\rightarrow\infty$ (in absence of gravity), particles preferentially sample strain-dominated regions ($Q<0$), leaving the largest Voronoi cells within vortex cores and producing highly organized cluster-void patterns. Increasing $St/Fr^2$ progressively broadens the particle distribution into positive-$Q$ regions, demonstrating that increasing slip velocity enables inertial particles to penetrate vorticity-dominated regions instead of being completely expelled by centrifugal forces. This transition is accompanied by a substantial reduction in the normalized void area. In the absence of gravity, the normalized void area remains approximately $45$-$50\,D^2$, largely independent of the Reynolds number or inlet particle volume fraction and shows slight increase with increase in Stokes number. The introduction of gravity reduces the void area to approximately $30\,D^2$ for $St=1$ and to approximately $16$-$20\,D^2$ for $St=6.5$, reflecting the suppression of cross-stream particle migration by gravitational settling. The probability density functions of the normalized Voronoi cell area further reveal that particle loading has only a secondary influence on particle organization within the one-way coupling regime considered here. Although increasing particle volume fraction broadens the tail of the Voronoi-area distribution, indicating an increased probability of individual void cells with higher normalized cell area, the overall void structure and its total area remains nearly unchanged. This demonstrates that particle inertia and gravitational settling dominate the large-scale migration of particles, whereas particle loading primarily affects the local statistics of voronoi cell areas. An increase in upstream particle spatial inhomogeneity due to particle-cylinder interactions is observed, where particle collisions with the cylinder generate a bow-shock-like region upstream of the stagnation point, whose extent increases with particle inertia. High inertia particles deviate more strongly from the curved streamlines around the cylinder, producing wider bow-shock structures than low inertia particles. For finite Froude number cases, particles accelerate under gravity effect and further enhances the spatial extent of the bow-shock region. These observations demonstrate that particle travels under the combined influence of upstream collision and downstream vortex-particle interactions.

Overall, the present study establishes that preferential clustering in unsteady wakes is governed by the coupled influence of unsteady wake dynamics, particle inertia, and gravitational settling, with the dimensionless settling velocity, $St/Fr^2$, providing a scaling parameter for particle distribution. The combined Voronoi-based and fluid Q value based analyses quantitatively relate coherent particle structures (clusters and voids) to the fluid vortex and provide a unified framework for understanding their formation in wake flows over a range of particle Stokes number, particle loadings, Reynolds number and Froude numbers. These findings helps to further the understanding of particle transport in particle-laden unsteady wake flows and provide useful insight for predicting particulate migration in various fields like sediment transport, aerosol dynamics, and other environmental and industrial multiphase flows involving coherent vortical structures.

% \appendix

% \section{Appendixes}

\section*{References}
% \nocite{*}
\bibliography{aipsamp}% Produces the bibliography via BibTeX.

\end{document}